\documentclass[]{jfm}

\usepackage{graphicx}
\usepackage{newtxtext}
\usepackage{newtxmath}
\usepackage{natbib}
\usepackage{hyperref}
\usepackage{xcolor}
\usepackage{subcaption}

\hypersetup{
    colorlinks = true,
    urlcolor   = blue,
    citecolor  = black,
}
\definecolor{rltgreen}{rgb}{0,0.5,0}

\newcommand{\RomanNumeralCaps}[1]
\linenumbers

\title{Modal analysis of the triadic interactions in the dynamics of a
	transitional shock wave boundary layer interaction}

\author{Ismaïl Ben Hassan Saïdi\aff{1},
	Stéphane Wang\aff{2}, Guillaume Fournier\aff{3},
	Christian Tenaud\aff{2}
	\and Jean-Christophe Robinet\aff{1}\corresp{\email{jean-christophe.robinet@ensam.eu}} }

\affiliation{\aff{1}École Nationale Supérieure des Arts et Métiers, Laboratoire DynFluid, 75013, Paris, France
	\aff{2}Université Paris-Saclay, CNRS, CentraleSupélec, Laboratoire EM2C, 91190, Gif-sur-Yvette, France
	\aff{3}Université Paris-Saclay, Université d'Évry, LMEE, 91020, Évry, France}

\usepackage{float}

\begin{document}
\maketitle

\begin{abstract}
This work is a numerical study of a transitional shock wave boundary layer interaction (SWBLI). The main goal is to improve our understanding of the well known low-frequency SWBLI unsteadiness and especially the suspected role of triadic interactions in the underlying physical mechanism. To this end, a Direct Numerical Simulation (DNS) is performed using high-order finite volume scheme equipped with a suitable shock capture procedure. The resulting database is then extensively post-processed in order to extract the main dynamical features of the interaction zone (involved characteristic frequencies, characteristics of the vortical structures, etc.). The dynamical organisation and space-time evolution of the flow at dominant frequencies are then further characterised by mean of a Spectral Proper Orthogonal Decomposition (SPOD) analysis. In order to study the role of triadic interactions occurring in the interaction region, a BiSpectral Mode Decomposition (BSMD) analysis is applied to the data base. It allows us to extract the significant triadic interactions, their location and the resulting physical spatial modes. Strong triadic interactions are detected in the downstream part of the separation bubble whose role on the low-frequency unsteadiness is characterised. All the results of the various analyses are then discussed and integrated to formulate a possible mechanism fuelling low-frequency SWBLI unsteadiness.
\end{abstract}

\begin{keywords}
Shock wave boundary layer interactions (SWBLI), SPOD, BSMD, non-linear dynamics, SWBLI unsteadiness, Direct Numerical Simulation (DNS)
\end{keywords}


\section{Introduction}
\label{sec:Introduction}

Situations in which a shock-wave interacts with a boundary layer (SWBLI: Shock Wave Boundary Layer Interaction) are numerous in the aeronautical and space industries. These interactions can exist on external surfaces (transonic profiles, junctions of surfaces, etc.) or in internal aerodynamics devices (supersonic air intakes, cascade of turbine blades, nozzles, etc.). Under certain conditions (high Mach number, large shock wave angle...), these interactions can generate a transient separation bubble (compressible separation) causing increased drag force, heat fluxes, and pressure fluctuations. It is also known since decades that the separation bubble and reflected shock wave are subject to low-frequency motion, known as "SWBLI unsteadiness". For turbulent interactions, with a turbulent incoming boundary layer, the characteristic frequency of these oscillations is two order of magnitude lower than the characteristic frequencies of the incoming boundary layer. The SWBLI unsteadiness can be detrimental to engineering system performances and can expose structures to oscillating loads, potentially damaging the solid structure's integrity (\cite{dolling_2001,delery_dussauge_2009,Babinsky_Harvey_2011,Gaitonde_2013,Clemens_2014}).  During the last decades, attention
has been focused on numerical and experimental studies of unswept SWBLIs (where the shock impingement line is
orthogonal to the incident boundary layer), employing various analysis techniques like Fourier analysis of signals (e.g. pressure or velocity probe's signal), modal
decomposition (POD (\cite{sirovich_1987,Shinde_2019}), SPOD (\cite{Towne_Spectral_2018}), DMD (\cite{Schmid_2010,Priebe_Tu_Rowley_Martin_2016})), and stability analysis (\cite{Theofilis_2003,theofilis_2011,robinet_2007,Sartor_Mettot_Bur_Sipp_2015,guiho_alizard_robinet_2016,Song_2023}).

The research community identifies two main mechanisms for SWBLI unsteadiness: i) upstream, caused by the advection of large-scale structures of the incoming boundary layer (\cite{Ganapathisubramani_2007,Ganapathisubramani_2009}) and ii) downstream, in which the dynamics of the separation bubble generates disturbances that drive the oscillations of the reflected shock. As stated in \cite{Clemens_2014}, following \cite{Souverein_2010}, a consensus view emerged in the community that strong interactions, which exhibit large separation bubbles, are primarily driven by a downstream instability (mechanisms of type ii)), whereas weakly separated interactions (also called incipient interactions) can be strongly influenced by fluctuations in the upstream boundary layer (mechanisms of type i)). 

\begin{figure}
 	\centering{\includegraphics[width=15cm]{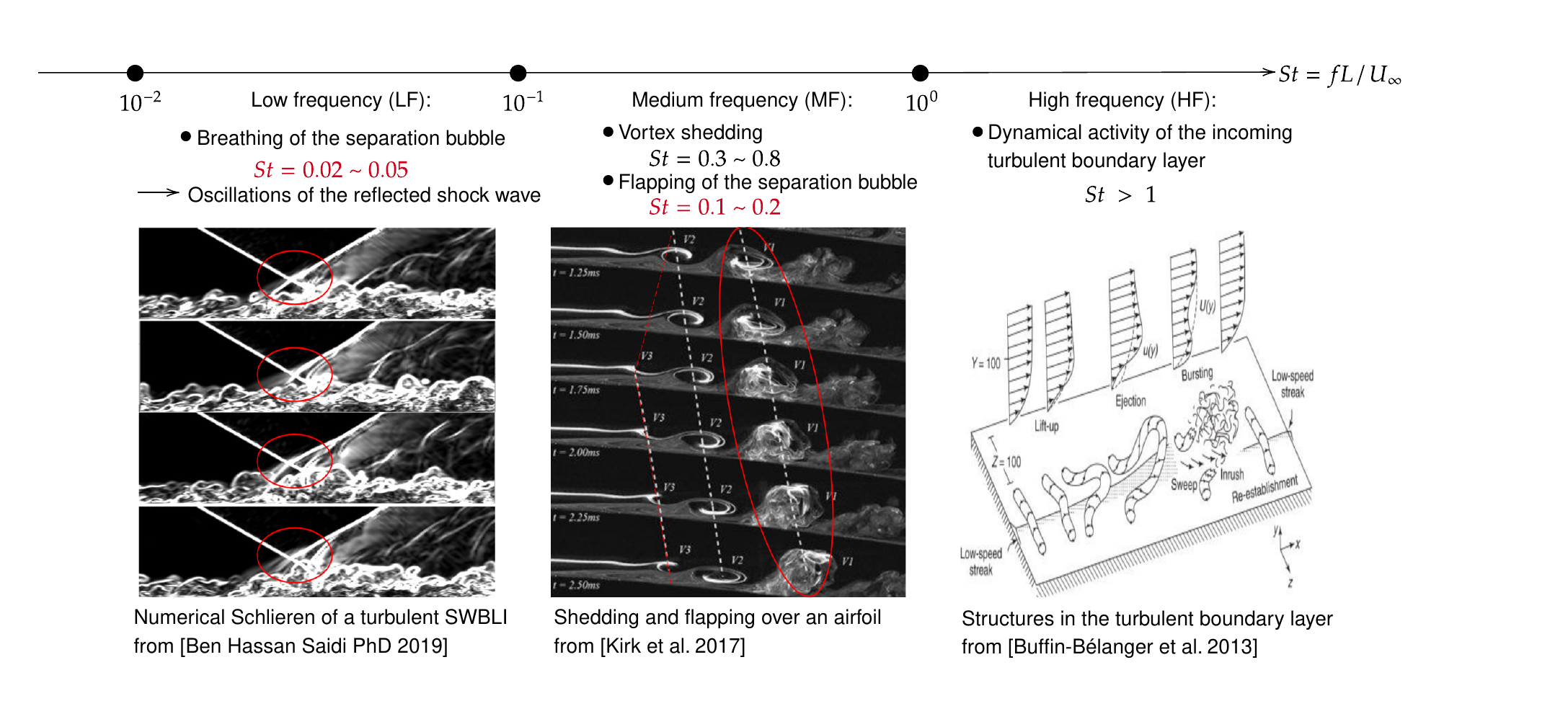}}
 	\caption{Illustration of the documented dynamical features of SWBLIs. The figures are from \cite{benhassansaidi_2019}, \cite{kirk_2017} and \cite{buffinbelanger_2013}}
 	\label{fig:frequencies_of_SWBLIs}
 \end{figure}

Regarding strong interactions a large number of studies have been undertaken in the last decades to document the low-frequency dynamics of the separation bubble and to identify the exact physical mechanism of type ii) underlying the SWBLI unsteadiness.  It is now well established that the dynamics of the flow around this mean configuration is characterised by several unsteady phenomena whose characteristic scales spread over a large broadband spectrum range. As illustrated in figure \ref{fig:frequencies_of_SWBLIs}, the main dynamical features of the flow can be classified according to three frequency ranges spreading over two decades: high, medium and low-frequencies. 
For turbulent interactions, the incoming boundary layer is turbulent with the most energetic fluctuations at high frequencies characterised by a Strouhal number $St_\delta=f\delta/U_e\sim 1$ (where $f$, $\delta$ and $U_e$ are respectively the main frequency of the fluctuations, the boundary layer thickness before the interaction and the free stream velocity). Various studies carried out to characterise the dynamics of subsonic turbulent separated and reattached flows have highlighted a now well documented medium-frequency dynamics of the separation bubble (\cite{Weiss_2015,Wu_Meneveau_Mittal_2020}). In this range of frequencies the shear layer, bounding the upper part of the bubble, is subjected to a global flapping motion of
the separated shear layer at a Strouhal number $St_L=fL/U_e \sim 0.1-0.2$ (where $L$ is the mean separation length) and a higher medium-frequency instability linked to the shedding of
large-scale vortical structures at $St_L \sim 0.3-0.8$ near the reattachment. The flapping mode consists of successive enlargement and shrinkage of the recirculation bubble, the shrinkage being associated with the shedding of a large vortex downstream of the recirculation bubble. These two medium-frequencies unsteadinesses correspond to the modes that have long been documented for subsonic fixed-separation separation bubbles (\cite{Cherry_1984,Kiya_Sasaki_1985}). For these fixed separation bubbles, the flapping frequency was reported as the lowest frequency present in the flow. Nevertheless, in turbulent SWBLIs, a low-frequency flapping mode was also observed at a Strouhal number of $St_L=fL/U_e=0.02-0.05$. In the following, we will refer to this low-frequency mode as the "breathing" of the separated zone. This mode is accompanied by the oscillations of the reflected shock wave, in the same frequency range, which constitute the so-called SWBLI unsteadiness. Thus, the low-frequency separation bubble oscillations documented for SWBLIs were one order of magnitude lower than those documented for fixed-separation subsonic separation bubbles. This led to a series of research projects aimed at explaining the low-frequency oscillations of SWBLIs as being a specificity of these flows, namely dependent on the compressible nature of the flow or the presence of the shock wave \citep{Pirozzoli_2006,piponniau_2009,touber_2011,Aubard_2013,Priebe_Tu_Rowley_Martin_2016,adler_gaitonde_2018}. This view of the problem has been challenged in the recent work of \citet{Weiss_2015} studying the dynamics of pressure induced subsonic turbulent separation bubbles on flat test surfaces in which a breathing mode at a frequency of the same order of magnitude than the breathing of SWBLI induced separation bubbles have been documented. This result tends to demonstrate that the low-frequency breathing mode is an intrinsic characteristic of the dynamics of separation bubbles, where the position of the separation point is not imposed by the geometry, independently of the compressible nature of the flow. 

Another furrow dug for decades to identify the physical mechanism behind the low-frequency SWBLI unsteadiness is the simulation and analysis of transitional SWBLIs with forced or non forced laminar incoming boundary layers in order to remove the presumed influence of large-scale turbulent structures in the incident boundary layer. This approach allows to carefully analyse the dynamics of the separation bubble that is suspected to be the source of the low frequency SWBLI unsteadiness \citep{robinet_2007,Sansica_2014,guiho_alizard_robinet_2016,Diop_2019,Mauriello_2022,bugeat_2022,Mahalingesh_2023,Song_2023}. This approach allows to carefully analyse the dynamics of the separation bubble that is suspected to be the source of the low frequency SWBLI unsteadiness. Part of these works involved characterising the linear dynamics of the flow by means of stability analyses. The global (self sustained dynamics) linear stability analysis performed by \citet{robinet_2007} on a transitional incident oblique shock-wave SWBLI on a flat plate at $M=2.15$ demonstrated that the fixed point of this flow is stable for 2-D perturbations. This result have then been confirmed in \citet{guiho_alizard_robinet_2016}. For 3-D perturbations, the flow exhibited a global instability when the shock angle exceeded a certain critical value. This result has then been confirmed by several studies \citep{Song_2023}. This mode is however stationary and therefore not directly related to the self-sustained low frequency dynamics. As the low-frequency unsteadiness could not be directly linked to any global mode around the fixed point the receptivity of the flow has been studied by mean of resolvent analysis. In particular, the results obtained by \citet{bugeat_2022} show that the global resolvent presents no frequency selectivity at low frequency and therefore behaves as a low-pass filter with respect to external disturbances and no clear evidence of quasi-resonance has been found. In light of the results of both global stability and resolvent analyses cited above, potential purely linear mechanisms explaining the low-frequency unsteadiness rely on the existence of low-frequency external perturbations of the flow. This brings us to the limits of a purely linear analysis of the low-frequency dynamics of SWBLIs. Indeed, the low-frequency SWBLI unsteadiness has been documented flow configurations with laminar incoming boundary layer, without any imposed low-frequency forcing \citep{sansica_2016,benhassansaidi_2019,Mauriello_2022,Mahalingesh_2023}. The search for non-linear mechanisms playing a role in the generation of low frequencies is therefore a promising avenue. This approach is encouraged in particular by the work of \citet{sansica_2016}, who studied a case of transitional interaction forced upstream by unstable high-frequency modes. In this SWBLI, analysis of the Fourier spectra of the parietal pressure showed that the low-frequencies appear to be generated in the transition zone towards turbulence (suggesting the non-linear nature of the mechanism), close downstream of the reattachment and be travelling upstream in the separation bubble. In a more recent study of another upstream forced (with high frequencies) transitional SWBLI \citep{Mauriello_2022}, non-linear spectral analyses (bispectrum and bicoherence) have been performed that pointed out quadratic coupling between frequencies as a possible mechanism involved in the low-frequency dynamics of the separation bubble. This work highlights the appearance of energy at low-frequency near the reattachment point as a result of quadratic coupling with linear unstable modes developing within the shear layer. This energy at low-frequency is then shown to be convected upstream through a feedback mechanism. These recent works point out the possible importance of triadic interactions in the mechanism driving the breathing of the separation bubble. However, these previous studies were based on local analyses of the signals, which did not allow to determine precisely the spatial location of the non-linear interactions in the interaction zone. In addition, the methods used did not allow the physical structure of the modes resulting from these non-linear interactions to be determined.

The objective of the present work is to go further in the analysis of the triadic interactions in the separation bubble of SWBLIs and their involvement in the mechanisms underlying the low-frequency unsteadiness of SWBLIs. To this end,  Direct Numerical Simulation (DNS) of a non-forced transitional oblique shock-wave SWBLI is performed. We study non-linear coupling between modes at different frequencies using the Bi-spectral Mode Decomposition (BSMD) \citep{Schmidt_Bispectral_2020} which is a modal decomposition method specially designed to study triadic interactions in flows. The use of this method is of major interest compared with previous studies because it is designed to analyse triadic interactions in 3-D signals. As a result, this algorithm allows to highlight quadratic coupling between modes at different frequencies. This approach allows to associate a spatial mode to each frequency resulting from a triadic interaction. Moreover, the location of the interactions between two frequencies in the flow field can be represented in a so-called interaction map. These two outputs of this method represent a major advantage with respect to local methods. The main objective of this analysis is to document the suspected non-linear link between the dynamical activity at medium and low frequencies, given that the presence of medium frequency modes of the shear layer are already strongly documented. The underlying objective is to demonstrate that the non linear evolution of the medium-frequency shear layer dynamics create the frequencies characterising the SWBLI unsteadiness. The Spectral Proper Orthogonal Decomposition (SPOD) \citep{Towne_Spectral_2018} has also been used as a support for the physical interpretation of the involved modes. Indeed, for each frequency, the SPOD performs a decomposition of the flow in spatio-temporal modes ordered by their energy content, highlighting the structure of the flow for a given frequency.

In our study, we have deliberately chosen a transitional configuration in which the incident boundary layer is not forced. In this respect, the present case study represents a limit case in which the mechanisms at the origin of the low-frequency dynamics appear to be present at low intensity. Moreover, the flow dynamics are not disturbed by the natural frequencies of the incident turbulence. The characteristic frequencies and mechanisms associated with the low and medium frequencies can therefore be studied independently of the mechanisms and frequencies specific to the turbulence in the upstream boundary layer. We therefore expect, so to speak, to find in the flow the main features and mechanisms of the dynamics we are interested in, but no more. We believe that this characteristic makes the present flow configuration a good candidate for identifying the mechanisms underlying the low-frequency of the separation bubble by modal analysis; in particular, modal analysis associated with the search for triadic interactions in the interaction zone. Indeed, a more complex dynamics induced by the presence of the turbulence or the forcing of the incident dynamics would imply spurious triadic interaction, not relevant for present study.

In the following, we first present the numerical simulation and database. The studied flow configuration is described as well as the governing equations and numerical strategy used for the simulation. The salient features of the mean flow and its dynamics are documented and the database used for both the SPOD and BSMD analysis are precisely described. We then present the SPOD analysis. We particularly describe the dominant modes of the flow at medium and low-frequencies. The BSMD analysis follows, in which the triadic interactions in the interaction region are documented. The physical relevance of the results regarding the mechanism at the origin of the SWBLI unsteadiness is the discussed. The paper ends with concluding remarks and perspectives.

\section{Numerical simulation and database}
\label{sec:Numerical_simulation_and_database}
   \subsection{Flow configuration}
   \label{subsec:Flow_config}
   
\begin{figure}
 	\centerline{\includegraphics[width=15cm]{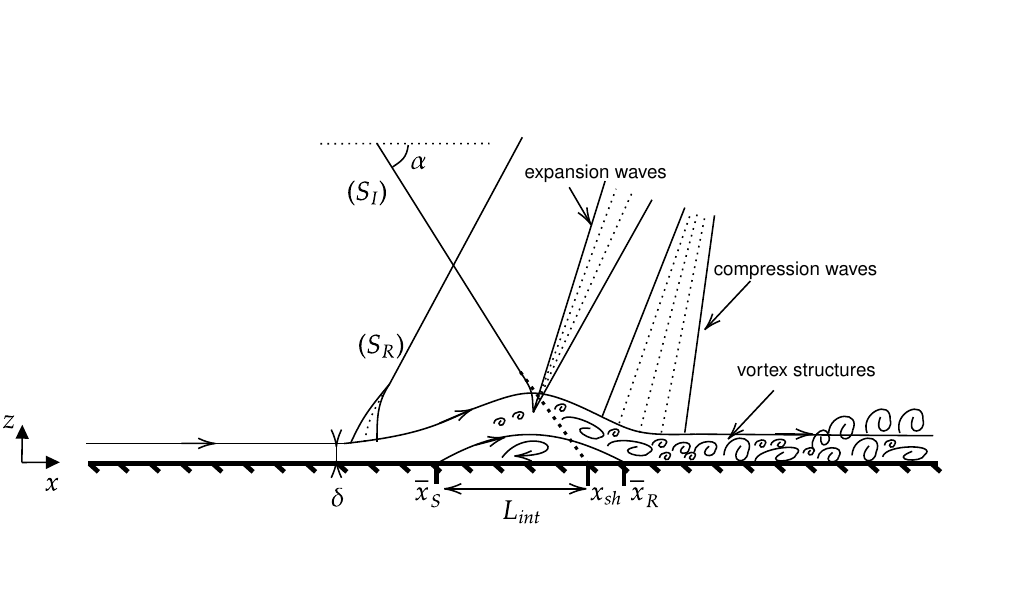}}
 	\caption{Sketch of a 2D slice of the flow, where $(S_1)$ is the incident shock wave of angle $\alpha$, $\delta$ is the boundary layer thickness just upstream of the interaction zone, $(S_R)$ is the reflected shock-wave, $\overline{x}_S$ is the mean separation point, $\overline{x}_R$ is the mean reattachment point, $x_{sh}$ is the location of the impingement shock-wave (if the flow were non-viscous) and $L_{int}=x_{sh}-\overline{x}_S$ is the interaction length. }
 	\label{fig:sketch_flow}
 \end{figure}

  The flow consists in the interaction between an incident shock wave and a laminar boundary layer developing on a flat plate. A 2D sketch of the flow (2D slice) is shown in Figure \ref{fig:sketch_flow}. The freestream conditions are similar to the experimental and numerical test case documented in \citet{Degrez_1987}, with a freestream Mach number $M=2.15$ and a freestream Reynolds number (based on the distance $x_{sh}$ between the leading edge of the flat plate and the shock impingement point) $\Rey=10^5$. Compared to the configuration of \citet{Degrez_1987}, where the whole flow is steady (fully laminar interaction), the incident shock wave angle $\alpha$ is increased up to $33.8^\circ$. The interaction is thus strengthened, the separation bubble is enlarged and the dynamics of the separation bubble becomes unsteady with transition to turbulence triggered in the bubble \citep{benhassansaidi_2019}. Table \ref{tab:physical_parameters} summarises the physical parameters of the simulation.
   
  \begin{table}
	\begin{center}
		\def~{\hphantom{0}}
		\begin{tabular}{lc}
			Mach number & $M=2.15$  \\
			Reynolds number   & $\Rey=10^5$ \\
			Prandtl number   & $\Pr=0.71$ \\
			Shock wave angle  &  $\alpha =33.8$ \\
			Ratio of specific heats  & $\gamma = 1.4$
		\end{tabular}
		\caption{Flow parameters of the SWBLI.}
		\label{tab:physical_parameters}
	\end{center}
  \end{table}

   \subsection{Governing equations}
   \label{subsec:Governing_equations}
   
     To perform the DNS, we consider the dimensionless compressible Navier-Stokes equations expressed in a Cartesian coordinate system. The length and time scales for nondimensionalisation are respectively $x_{sh}$ and $U_\infty / x_{sh}$, with $U_\infty$ being the freestream velocity. All the primitive variables are nondimensionalised
by the corresponding freestream quantities except the thermodynamic pressure $P^*$ and the total energy per unit volume $\rho^* E^*$ that are nondimensionalised by the freestream dynamic pressure $\rho_\infty U_\infty^2$. The dimensionless Navier-Stokes system is therefore written as:
     \begin{equation}
     	\frac{\partial \boldsymbol{U}}{\partial t} + \nabla \cdot \boldsymbol{F}(\boldsymbol{U}) - \nabla \cdot \boldsymbol{F_v}(\boldsymbol{U}, \nabla \boldsymbol{U}) = 0,
     	\label{eq:N-S}
     \end{equation}
    where $\boldsymbol{U}$ is the dimensionless vector of conservative variables, $\boldsymbol{F(U)}$ the dimensionless convective fluxes,  and $\boldsymbol{F_v(U, \nabla(\boldsymbol{U}))}$ the dimensionless diffusive fluxes that write respectively:
    
\begin{equation}
	\begin{array}{lll}
		\boldsymbol{U}=\left[ \begin{array}{c}
			\rho\\ 
			\rho \boldsymbol{u}\\
			\rho E 
		\end{array}
		\right], \quad	
		&
		\boldsymbol{F}=\left[ \begin{array}{c}
	      \rho \boldsymbol{u}\\ 
	      \rho \boldsymbol{u} \otimes \boldsymbol{u} + P \mathbb{I}\\
	      (\rho E + P) \boldsymbol{u} 	
       \end{array}
       \right],
       \quad \text{and} \quad	
		&
       \boldsymbol{F_v}=\left[ \begin{array}{c}
       	0\\ 
       	{ \frac{1}{\Rey} \mathbf{\sigma}} \\
       	{ \frac{1}{\Rey} \boldsymbol{u}. \mathbf{\sigma} + \frac{\mu}{ (\gamma - 1) \Rey \Pr M^2} \nabla T }
       \end{array}
       \right] .
	\end{array}
	\label{eq:NS_flux}
\end{equation}	

$\mathbb{I}$ stands for the identity matrix, $\rho$ is the dimensionless density, $\boldsymbol{u}=[u,v,w]^T$ is the dimensionless velocity vector, $E$ is the dimensionless total energy per unit of mass, $P$ is the dimensionless thermodynamic pressure and $\lambda$ is the dimensionless thermal conductivity. $P$ relates to the dimensionless conservative variables through the following relationship: 

 \begin{equation}
    P = (\gamma - 1) \left(  \rho E - \frac{1}{2}  \frac{(\rho \boldsymbol{u}) \cdot  (\rho \boldsymbol{u})}{\rho} \right),  
    \label{eq:pression}
 \end{equation}
and to the dimensionless static temperature $T$ through the following ideal gas dimensionless equation of state:
\begin{equation}
	T = \gamma M^2 \frac{P}{\rho}.
	\label{eq:eq_state}
\end{equation}

The dimensionless viscous stress tensor is expressed as
\begin{equation}
   \displaystyle{\mathbf{\sigma}= \mu ( \nabla \boldsymbol{u} + \nabla^T \boldsymbol{u} ) - \frac{2}{3} \mu \left(\nabla \cdot \boldsymbol{u}\right)\boldsymbol{I} }.
\end{equation}

We assume that the dynamic viscosity only depends on the temperature through the Sutherland's law. Given the Prandtl number, the dimensionless thermal conductivity is then deduced from the dimensionless dynamic viscosity: $ \lambda = \mu Cp/\Pr$.

   \subsection{Numerical methods}
   \label{subsec:numerical_methods}
   
   The DNS of this flow is performed using an in-house parallel (MPI) Finite-Volume based DNS/LES solver. The numerical methods employed are presented in detail in \citet{BenHassanSaidi_2020}, where the ability of this code to compute high Reynolds compressible (turbulent and shocked) flows has also already been demonstrated (including the boundary conditions presented in section    \ref{subsec:Computational_domain_and_BC}). The convective fluxes are discretised by Monotonicity-Preserving shock-capturing scheme (OSMP7) (first introduced in \citet{Daru_2004}), based on the Lax-Wendroff approach to obtain a $7^{th}$-order accurate coupled time and space approximation. A $2^{nd}$-order centered scheme is used for the diffusive fluxes. In \citet{BenHassanSaidi_2020}, we showed that the use of a higher-order scheme for computing diffusive fluxes has a negligible effect on the accuracy of the SWBLI calculation and is therefore unnecessary. 
   
   \subsection{Computational domain and boundary conditions}
   \label{subsec:Computational_domain_and_BC}
   
   The geometry of the 3D domain is $\mathcal{D}=[0;250\delta]\times[0;125\delta]\times[0;62.5\delta]$, with $\delta=1.6 \times 10^{-2}~m$ the boundary layer thickness just upstream of the interaction zone. The domain is discretised using a cartesian mesh with non-uniform spacing in the direction normal to the wall ($z$). The mesh employs 800 × 400 × 202 grid points in $(x\times y \times z)$. In the normal-to-the-wall direction, the mesh is tightened close to the wall using a hyperbolic tangent law to obtain a minimum grid spacing over the plate of $\Delta z_{min}=0.0125 \delta$. 
   
   A uniform flow is prescribed as the inlet of the domain as prescribed by the guidelines of the $4^{th}$HiOCFD workshop \citep{Wang_2013,HiOCFD4_2016} to compute the SWBLI in the configuration of \citet{Degrez_1987}, which has been validated in \citet{BenHassanSaidi_2020} against reference experimental and numerical results. The incident shock wave is created by imposing the Rankine-Hugoniot relationships in the inlet plane at a height $z$ chosen so that the shock wave impinges the flat plate at the desired abscissa $x_{sh}=1$. No-slip and adiabatic wall conditions are prescribed at the flat plate location $(z = 0)$. Outlet time-dependent non-reflecting boundary conditions \citep{Thompson_1987} are imposed at the top surface and at the downstream outlet boundary of the computational domain. Periodic conditions are used in the spanwise direction ($y$).   
   
  The numerical strategy have already been validated for this type of simulation. Indeed the same numerical schemes and boundary conditions have already been successfully employed to compute SWBLIs in the flat plate configuration. Please refer to \citet{BenHassanSaidi_2020} for more details.

\subsection{Mean flow organisation and frequency content}
   \label{subsec:Mean_flow_and_frequency}

\begin{figure}
 	\centerline{\includegraphics[width=1.0\linewidth,trim=4 2.5 4 0,clip]{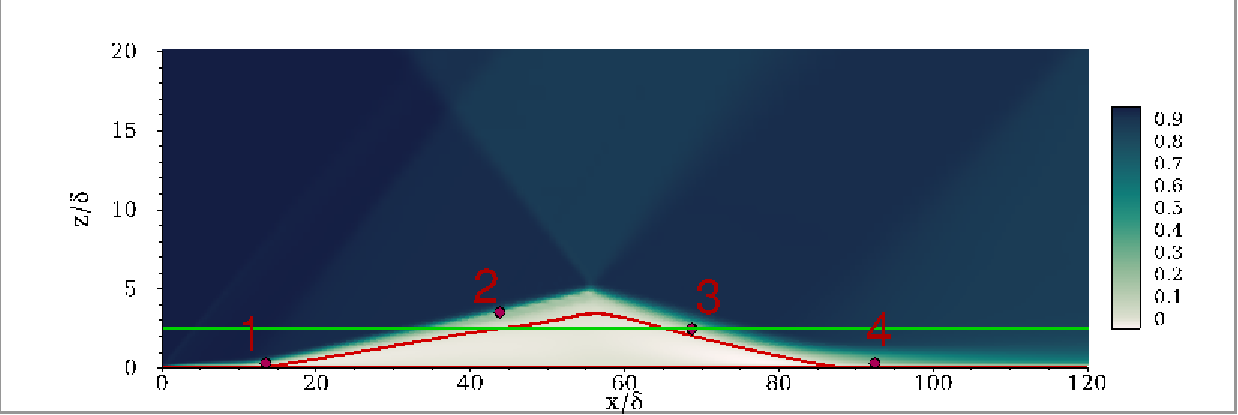}}
 	\caption{Time mean longitudinal velocity field averaged in the spanwise direction $\overline{u}$. The red line shows
the isocontour $\overline{u} = 0$. The green line is parallel to the wall at a height of $z/\delta = 2.5$. Coordinates of probes in the $(x,z)$ basis, from 1 to 4: $(13.44 \delta,0.3\delta)$, $(43.75,3.5\delta)$, $(68.75 \delta, 2.5 \delta)$, $(92.50 \delta,0.3 \delta)$.}
 	\label{fig:mean_flow}
 \end{figure}

\begin{figure}
 	\centerline{\includegraphics[width=10cm]{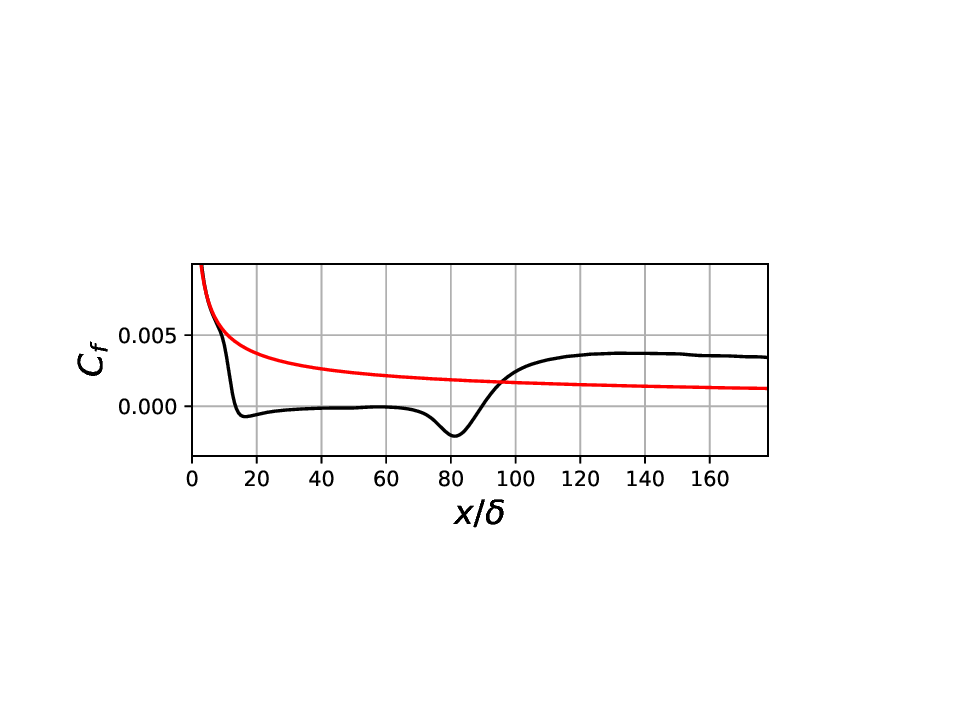}}
 	 \caption{Distribution of the spanwise averaged skin friction along the flat plate ({\color{black}{\rule{0.05\linewidth}{0.3mm}}}). Laminar boundary-layer solution ({\color{red}{\rule{0.05\linewidth}{0.3mm}}}).}
 	\label{fig:skin_friction}
   \end{figure}

   The time mean longitudinal velocity averaged in the spanwise direction $\overline{u}$ is plotted in figure \ref{fig:mean_flow} in the interaction
zone. The velocity is scaled by the free-stream velocity. The red
line shows the isocontour $\overline{u} = 0$. It allows to clearly identify the separation bubble in which the
mean longitudinal velocity is negative. The separation bubble also corresponds to the region in which the mean
skin friction coefficient plotted in figure \ref{fig:skin_friction}
is negative.

As the incident boundary layer is perfectly laminar, it is particularly prone to separation when subjected to the adverse pressure gradient imposed by the incident shock wave, compared to turbulent boundary layers. The resulting mean separation
length is $L=\overline{x}_R-\overline{x}_S\simeq74.68 \delta$, with $\overline{x}_S\simeq13.44 \delta$ and $\overline{x}_R\simeq88.12 \delta$ respectively the time mean separation and reattachment points averaged in the spanwise direction. The interaction length is $L_{int}=x_{sh}-\overline{x}_S=49,06 \delta$. The time mean height of the bubble, averaged in the spanwise direction, is $\overline{h}=3.5 \delta$. Moreover, the mean separation bubble is quasi-symmetric with respect to the vertical axis passing through the apex of the bubble. This length and aspect of the mean bubble is consistent with comparable simulations \citep{robinet_2007,Song_2023}. For similar Mach and Reynolds numbers, the length of separation bubbles for SWBLIs with laminar incident boundary layers is around one order of magnitude bigger than their counterpart for SWBLIs involving turbulent incident boundary layers ; see for example \citet{adler_gaitonde_2018} where $\overline{L}=4.0 \delta$ or \citet{BenHassanSaidi_2020} where $\overline{L}=4.92 \delta$. For turbulent SWBLIs, the symmetry of the separation bubble is also broken, the second part of the bubble being less long than the first part. It is due to increased mixing rate in the shear layer. The same symmetry breaking (less marked) is observed for SWBLIs with sufficiently strongly forced laminar incident boundary layer \citep{Sansica_PhD_2015,Diop_2019,Mauriello_2022}. 

The difference of topology of the mean separation bubble of the present flow configuration with respect to turbulent SWBLIs calls for a comment about the characteristic length used to build non dimensionalised frequencies. Indeed, the main frequencies involved in the interactions have been introduced in the literature, for turbulent interactions, as Strouhal numbers based on the separation length: $St_L=fL/U_e$. For the present flow configuration, as the bubble is symmetric the interaction length $L_{int}$ is more suitable, to be consistent with the characteristic frequencies highlighted in turbulent interactions. In the following, the frequencies will therefore be expressed as Strouhal numbers based on the interaction length: $St_{L_{int}}=fL_{int}/U_e$.   
   
\begin{figure}
 \centerline{\includegraphics[width=13cm,trim=0 4 4 0,clip]{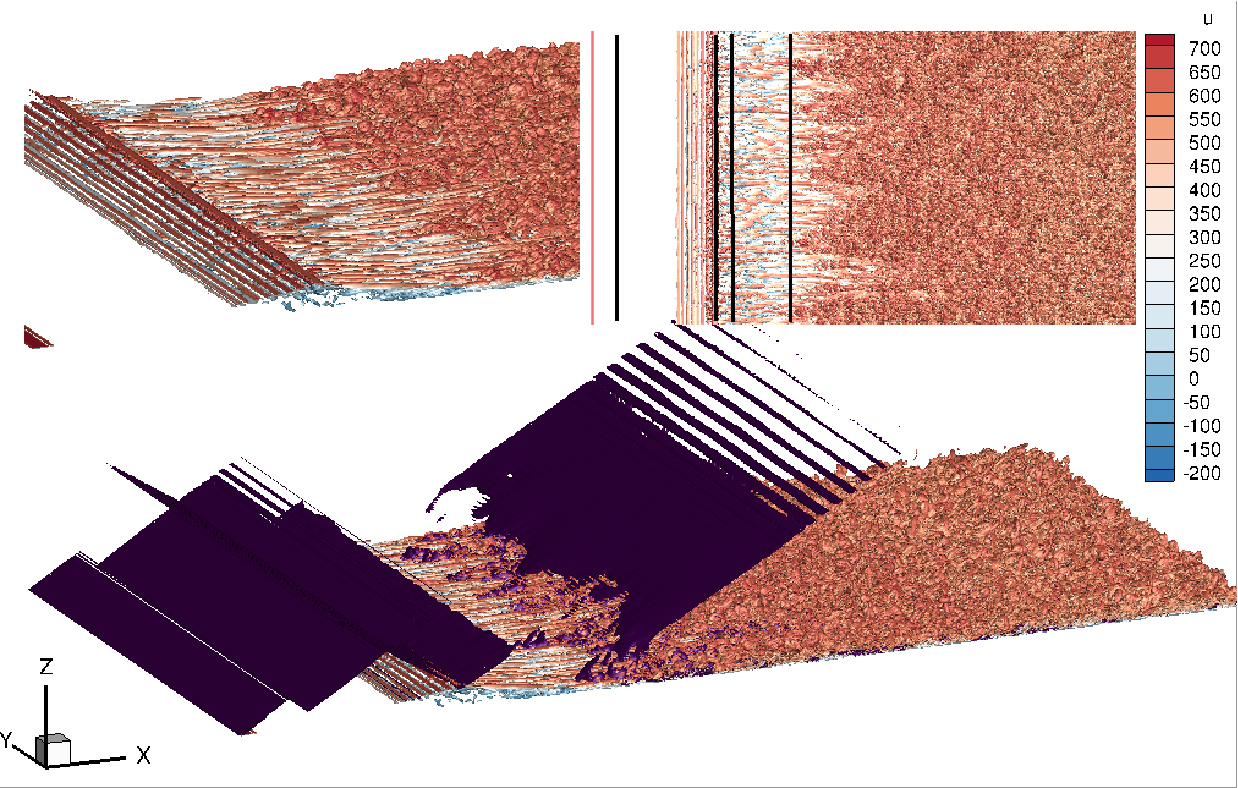}}
 	\caption{Vortical structures highlighted by isosurfaces of the discriminant criterion colored by the magnitude of the longitudinal component of the velocity. Shock waves are highlighted by isosurfaces of $\lVert \nabla P \rVert$. The upper left insert shows a zoomed view of the vortical structures around the separation bubble. The upper right insert shows an overhead view of the vortical structures in the separation bubble and in the downstream boundary layer. On the upper right insert, the vertical black lines indicate from left to right: the mean separation line, the mean line of incident shock impingement on the apex of the separation bubble, the mean line of shock impingement on the flat plate and the mean reattachment line.}
 	\label{fig:champ_instantanne}
 \end{figure}

A snapshot of the flow is shown in figure \ref{fig:champ_instantanne}. The discriminant criterion, introduced in \citet{chong1998turbulence} already used by \citet{Pirozzoli_2006} in the context of SWBLI simulations, is used to identify the vortex structures present in the flow. Shock-waves are highlighted by isosurfaces of $\lVert \nabla P \rVert$ colored in black.
The upstream boundary layer not being artificially forced by modes resulting from a stability problem \citep{Sansica_2014,sansica_2016} or a free-stream turbulence method, the only fluctuations supplied to the flow come from shockwaves (leading-edge, separated or the incident shock waves) which disrupt the flow mainly two-dimensionally although the natural most unstable normal modes are 3-D. Figure \ref{fig:champ_instantanne} shows clearly this characteristic. In what we seek to show, the existence of a non-linear interaction between medium frequency modes generating low frequency modes, this has no consequence whether the convective modes exited are 2-D or 3- D.
Large spanwise vortices, corresponding to Kelvin-Helmholtz
rolls that progressively develop, are clearly visible in the shear layer edging the separation bubble between the reflected and the incident shocks. After the incident shock impingement, the shear layer is populated by elongated structures in the streamwise direction. Inside the separation bubble, 3D structures are visible in the downstream part of the separation zone. For the particular value of the discriminant criterion shown in the figure \ref{fig:champ_instantanne}, no coherent structure is visible in the early part of the interaction. The dynamical activity inside the separation bubble therefore seems to be mainly concentrated in its second part. The dynamic activity of the separation bubble causes the rapid transition of the boundary layer after the reattachment point. 


\begin{figure}
  \begin{subfigure}{.5\textwidth}
\centerline{\includegraphics[width=1.0\linewidth]{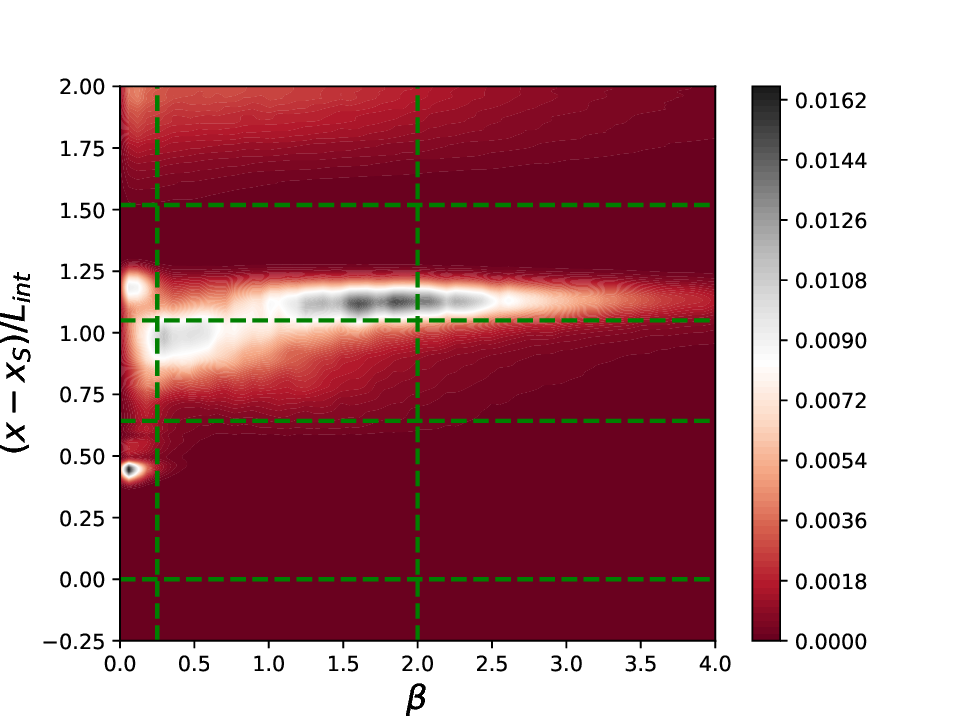}}
 	\caption{Time averaged spanwise wave number of the longitudinal component of velocity $u$ in a plane parallel to the flat plate at a height of $z/\delta = 2.5$. The vertical dashed lines indicate the $\beta=0.25$ and $\beta=2$ wavenumbers. The horizontal dashed lines indicate, from bottom to top: separation point, raising shear layer, descending shear layer and reattachment point.}
 	\label{sub_fig:spectre_spawise_6}
 	  \end{subfigure}%
 	 \begin{subfigure}{.5\textwidth}
\centerline{\includegraphics[width=1.0\linewidth,trim=0 4 4 0,clip]{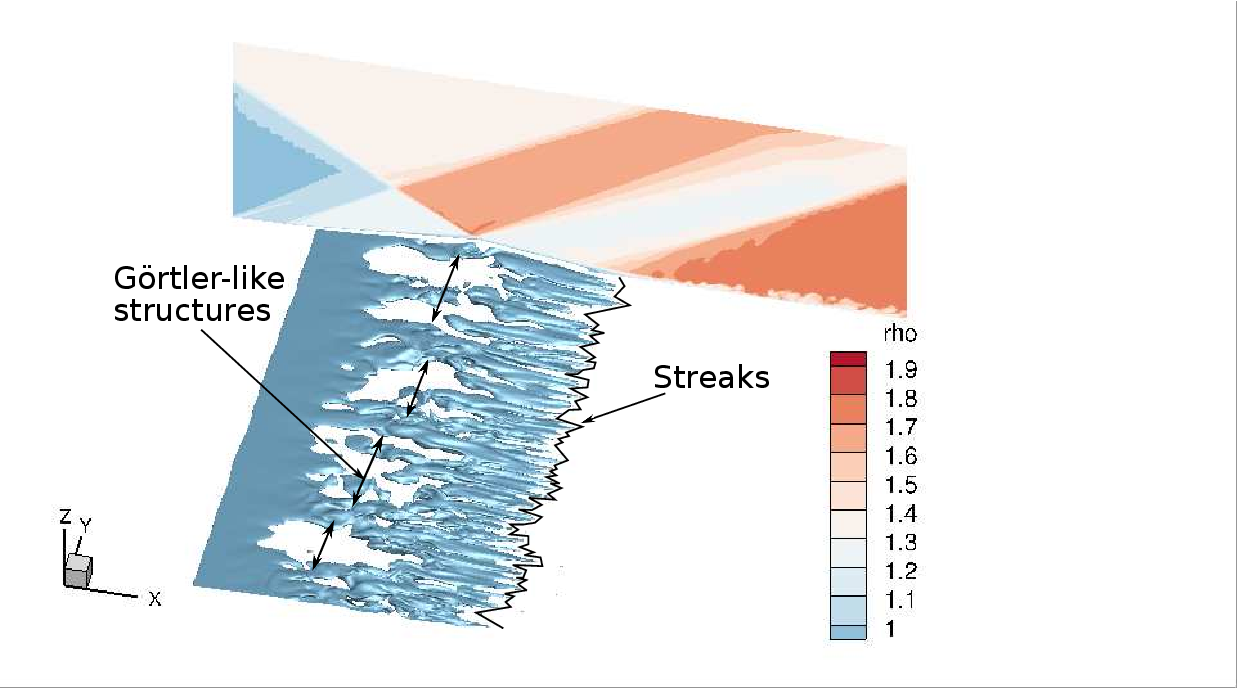}}
 	\caption{Isocontours of the longitudinal component of velocity $u=-0.041$ colored by $u$. Slice ($y$ plane) of the density $\rho$.       }
 	 	\label{sub_fig:iso_u_Gortler_streaks}\vspace{2.5cm}
 	  \end{subfigure}%
 	   	\caption{Spanwise length scales in the interaction zone.}
 	\label{fig:Gortler_streaks_spectre}
 \end{figure}

In order to physically characterise the longitudinal structures identified in Figure \ref{fig:champ_instantanne}, the spectral content of the velocity signal in the transverse direction is studied. To this end, the time averaged 1-D spatial Fourier transform in the spanwise direction of the longitudinal component of the velocity ($u$) was calculated in different planes parallel to the wall, defined by $z=cste$ ($z/\delta=0.3,0.5,1,1.5,2,2.5,3,3.5,4,4.5$). Figure \ref{sub_fig:spectre_spawise_6} shows the evolution of these spectra as a function of the abscissa along the flat plate in the $z/\delta = 2.5$ plane marked by the green line in Figure \ref{fig:mean_flow}.  The wavenumber is scaled using the same scaling as in \citet{bugeat_2022} for comparison. The abscissa is normalised by using the mean interaction length $L_{int}$ and the mean separation point $\overline{x}_S$. In this figure, the two horizontal dashed lines indicate the intersection between the plane under consideration and the shear layer bounding the separation bubble. The line at $(x-\overline{x}_S)/L_{int}=0.445$ indicates the shear layer upstream of the impact of the incident shock, which we will call the "rising shear layer". 
The line at $(x-\overline{x}_S)/L_{int}=1.2$ indicates the shear layer downstream of the impact of the incident shock, which we will call the "descending shear layer". The vertical dashed lines indicate the $\beta=0.25$ and $\beta=2$ wavenumbers that were highlighted as the two maxima of the $G(\beta)$ curve, where $G$ is the optimal gain of a low-frequency resolvent analysis, and that were respectively associated with Görtler-type structures and streaks in \citet{bugeat_2022}. Just upstream of the descending shear layer (i.e. inside the bubble) we can clearly see a peak at $\beta=0.25$. In the descending shear layer, we see a plateau containing the value $\beta=2$. These same peaks are found in the spectra of the other planes (not shown here), indicating the presence of Görtler-type structures inside the bubble and streaks in the descending shear layer. This location of Görtler-type structures and streaks is in agreement with the results of \citet{bugeat_2022}. The optimal modes of low-frequency resolvent analysis are therefore found in the DNS data of the present unforced transitional SWBLI. Analysis of the instantaneous 3-D fields highlights these longitudinal structures and illustrates their location. For example, figure \ref{sub_fig:iso_u_Gortler_streaks} shows the isocontour of the longitudinal velocity field $u=-0.041$ next to a 2-D slice coloured by $\rho$ to locate the separation bubble. The wavelength of the Görtler type structures and the streaks clearly manifest themselves in the isocontour of velocity in the zones identified by the spectral analysis. 

  We now want to document the frequency content of the dynamics of the interaction zone. The dynamic structures mainly visible in Figure  	\ref{fig:champ_instantanne} seem to be concentrated in the shear layer. We therefore begin by examining the frequency content of the mixing layer. To do this, we examine the Power Spectral Density (PSD) of the longitudinal velocity signal measured at different points in the shear layer. These points are shown in a 2D plane in figure \ref{fig:mean_flow}, where their coordinates in the $(x,z)$ plane are specified. In practice, for these coordinates in the $(x,z)$ plane, a probe was placed at each mesh point in the $y$ direction to measure the velocity signal and compute the PSD. The PSDs were then averaged over span. All the PSD of time signals presented in this section have been computed using the periodogram method implemented in the scipy.signal package of python \cite{2020SciPy-NMeth} using Hann windowing. For all signals studied in this section the sampling characteristics are given in table \ref{tab:PSD_details}:
  
   \begin{table}
	\begin{center}
		\def~{\hphantom{0}}
		\begin{tabular}{lc}
			Signal length & $T_{tot}=895 \times L_{int} / U_\infty  $  \\
			Time step   & $\Delta t \simeq 0.282 \times L_{int} / U_\infty $ \\
			Number of samples &  $3000$ \\
		\end{tabular}
		\caption{Sampling characteristics of time signals.}
		\label{tab:PSD_details}
	\end{center}
  \end{table}
 We can see that in the zone of the separation point (probe 1), the energy level of the oscillations in the shear layer is very low and localised in the low and medium frequency ranges characterising the breathing and flapping phenomena. There are clear peaks at frequencies $St_{L_{int}}=0.04,0.0667$ and $0.1$. This frequency content is amplified in the rising mixing layer. Indeed, the signal measured by probe 2 has approximately the same frequency content as probe 1, but at a higher energy level. 
  
  We saw above that the passage through the incident shock is associated with a distortion of the vortex structures, which are mainly oriented in the spanwise direction upstream of the shock and preferentially in the streamwise direction downstream of the shock. These structures are rapidly amplified downstream of the incident shock. Just downstream of the incident shock (probe 3), the energy of the fluctuations is amplified compared with the level observed at probe 2. The peaks measured at low and medium frequencies are still present, but the spectrum is filling out and these frequencies are less predominant. In addition, higher frequencies appear. This signal is massively amplified in the descending shear layer. In the attachment zone (probe 4), the energy of the signal is almost quadrupled compared with probe 3. This amplification is also selective. The frequencies already highlighted: $St_{L_{int}}=0.04,0.0667$ and $0.1$ are particularly peaked in the reattachment zone. There is also a further filling in of the spectrum at high frequencies concomitant with the generation of fine structures associated with the transition to turbulence in this region. 
 
 These measurements in the shear layer seem to indicate that the mixing layer is forced from the separation point by the characteristic SWBLI frequencies at low and medium frequencies. The rising shear layer amplifies these frequencies by a factor of around 10. However, the energy levels in the rising shear layer are low relative to those in the descending shear layer. The latter shows a massive amplification of fluctuations in the low and medium frequency range, and in particular of the frequencies characteristic of the SWBLI unsteadiness $St_{L_{int}}=0.04,0.0667$ and $0.1$. These observations are consistent with various previous works, in particular those of \citet{Sansica_2014} and \citet{Mauriello_2022} who have shown that the low-frequency content of the interaction zone is created by the dynamics of the shear layer and seems to result from the non-linear evolution of the latter in the downstream part of the recirculation bubble. These authors also showed that disturbances at these frequencies then travel upstream inside the recirculation bubble. 
 
  \begin{figure}
 	\begin{subfigure}{.5\textwidth}
 		\centering
 		\includegraphics[width=.9\linewidth]{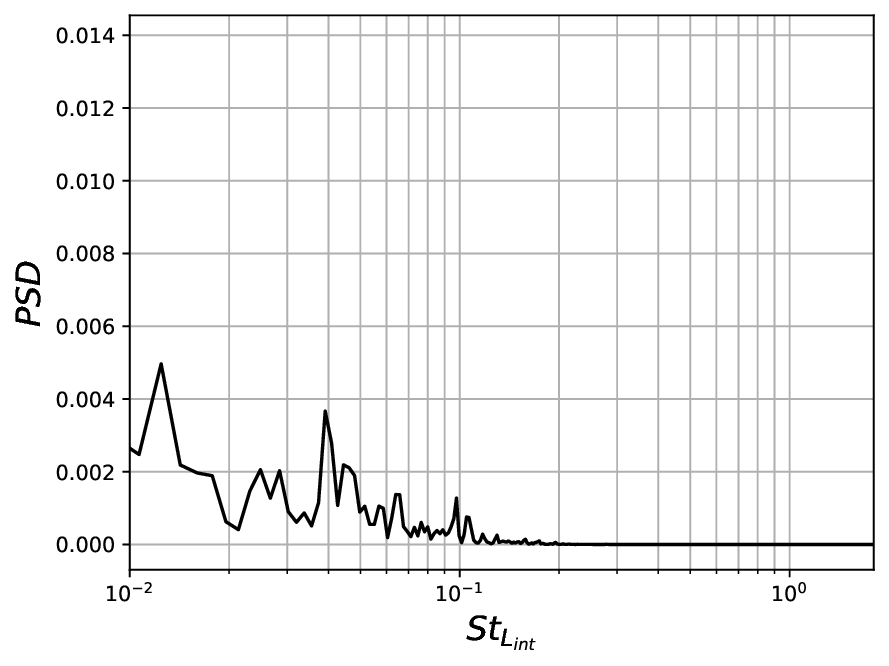}
 		\caption{Probe 1}
 		\label{fig:probe_1}
 	\end{subfigure}%
 	\begin{subfigure}{.5\textwidth}
 		\centering
 		\includegraphics[width=.9\linewidth]{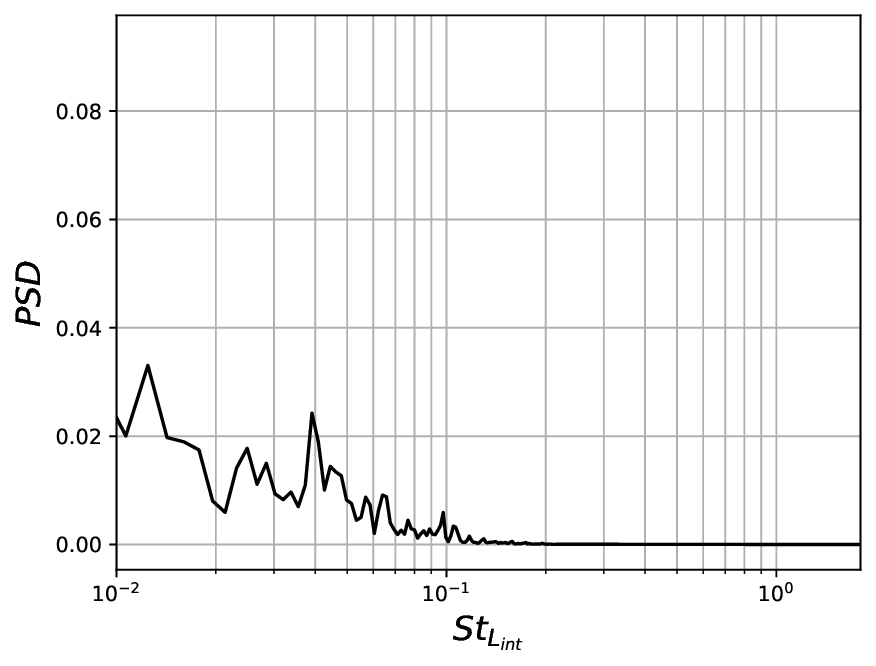}
 		\caption{Probe 2}
 	\end{subfigure}
 	\begin{subfigure}{.5\textwidth}
 		\centering
 		\includegraphics[width=.9\linewidth]{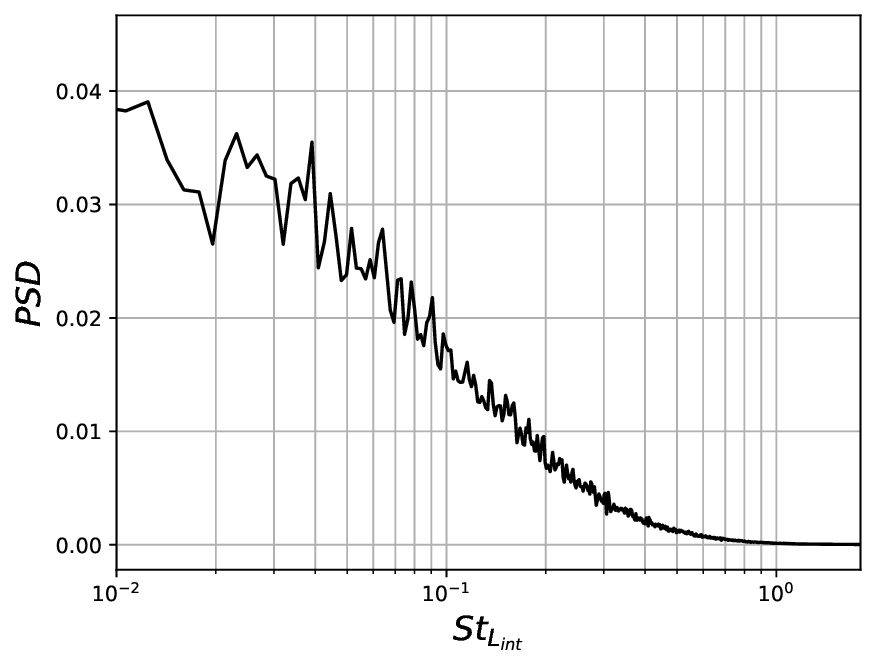}
 		\caption{Probe 3}
 	\end{subfigure}%
 	\begin{subfigure}{.5\textwidth}
 		\centering
 		\includegraphics[width=.9\linewidth]{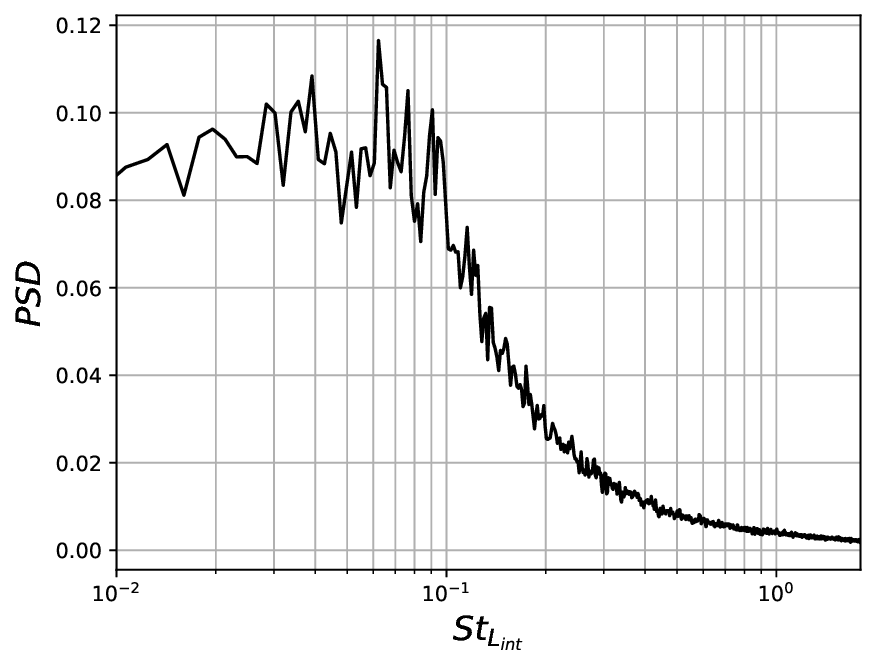}
 		\caption{Probe 4}
 	\end{subfigure}
 	\caption{Power spectral densities of the spanwise averaged longitudinal velocity $u$ signal recorded on probes 1, 2, 3 and 4 as shown in figure \ref{fig:mean_flow}.}
 	\label{fig:PSD_prob}
 \end{figure}
 
\begin{figure}
 	\begin{subfigure}{.5\textwidth}
 		\centerline{\includegraphics[width=1.0\linewidth]{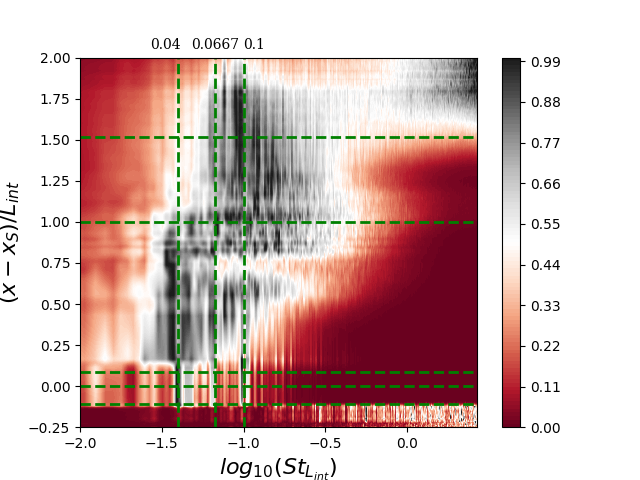}}
 		\caption{Distribution along the flat plate of the spanwise averaged premultiplied and normalised power spectral density of the longitudinal component of velocity $u$ measured in a plane parallel to the flat plate at a height of $z/\delta = 0.3$. The vertical dashed lines indicate significant frequency peaks in the low and medium frequency range. The horizontal dashed lines indicate, from bottom to top: the reflected shock foot, the separation point, the crossing of the rising shear layer, the incident shock impingement location and the reattachment point.}
 		\label{fig:DUPONTRAMA_U_y=0.3}
 	\end{subfigure}
 	\begin{subfigure}{.5\textwidth}
 		\centerline{\includegraphics[width=1.0\linewidth]{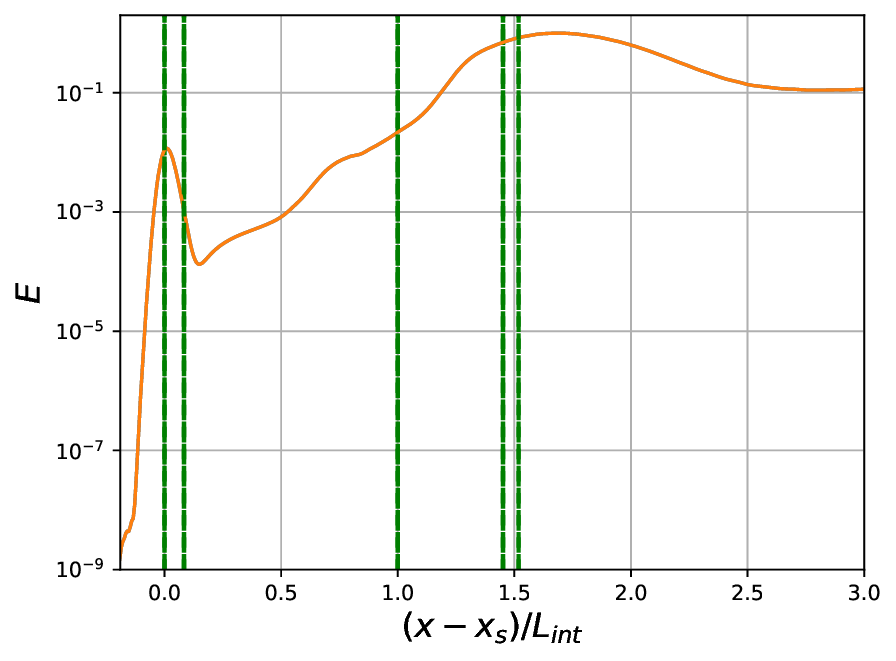}}
 		\caption{Distribution along the flat plate of the expected power in the range $St_{L_{int}}\in [0.01,0.1]$ of the longitudinal component of velocity $u$ measured in a plane parallel to the flat plate at a height of $z/\delta = 0.3$. The vertical dashed lines indicate, from left to right: separation point, the crossing of the rising shear layer, the incident shock impingement location, the crossing of the descending shear layer and the reattachment point.}
 		\label{fig:Energie_U_St_0.01-0.1_y=0.3}\vspace{1cm}
 	\end{subfigure}
 	\caption{Power spectral density and power of the longitudinal velocity signal close to the wall inside the separation bubble.}
 \end{figure}

 In order to study the frequency content of this information feedback in the recirculation bubble, we plot on figure \ref{fig:DUPONTRAMA_U_y=0.3} the evolution of the spanwise averaged premultiplied PSD of the longitudinal velocity signal ($u$) measured in a plane parallel to the flat plate, in very close proximity to the wall $(z/\delta=0.3)$. For each abscissa along the flat plate, the power spectral density at this abscissa is normalised and premultiplied. This makes it possible to determine the predominant frequencies for each abscissa, but the relative energy differences between abscissas are erased by this representation. In this figure, the horizontal dashed lines indicate, from bottom to top, the positions of the separation point, the rising shear layer, the impact of the incident shock on the flat plate and the reattachment. We can see that the spectral content in the attachment zone is present throughout the bubble in the near wall, but with a progressive narrowing of the significant frequency range towards the low and medium frequency range as we move upstream in the bubble. Thus, the disturbances reaching the foot of the rising shear layer have an exclusively low and medium frequency spectral content with a cut-off frequency of approximately $St_{L_{int}}\simeq 0.1$. We can therefore see that the rise of the disturbance inside the bubble selects the low and medium frequencies. This result had already been shown in \citet{Mauriello_2022}. In order to determine whether this feedback into the bubble is damped, we integrate the power spectral density between the $St_{L_{int}}=0.02$ and $0.1$ for each abscissa along the flat plate to obtain the streamwise evolution of the expected power in this frequency range: $E(x)=\int_{St=0.02}^{St=0.1} PSD(u(x)) dSt$. This power distribution is shown in figure \ref{fig:Energie_U_St_0.01-0.1_y=0.3}. This figure clearly shows a large peak of power in the reattachment region, in agreement with the analysis of the spectra shown in Figure \ref{fig:PSD_prob}. We can clearly see an exponential decrease in power inside the bubble. The upwelling of disturbances is therefore damped inside the bubble. The signal reaches the detachment zone with low but finite energy levels. The shear layer is therefore forced into the separation zone by a signal of finite energy and quasi-continuous spectrum in the frequency range $St_{L_{int}}\in [0.02,0.1]$. However, figure \ref{fig:DUPONTRAMA_U_y=0.3} shows a high selectivity of the mixing layer. Indeed, the quasi-continuous spectrum becomes strongly peaked in the shear layer at the frequencies already highlighted in figure \ref{fig:probe_1}, i.e. $St_{L_{int}}=0.04,0.0667$ and $0.1$. 

\begin{figure}
 	\begin{subfigure}{.5\textwidth}
 		\centerline{\includegraphics[width=1.0\linewidth]{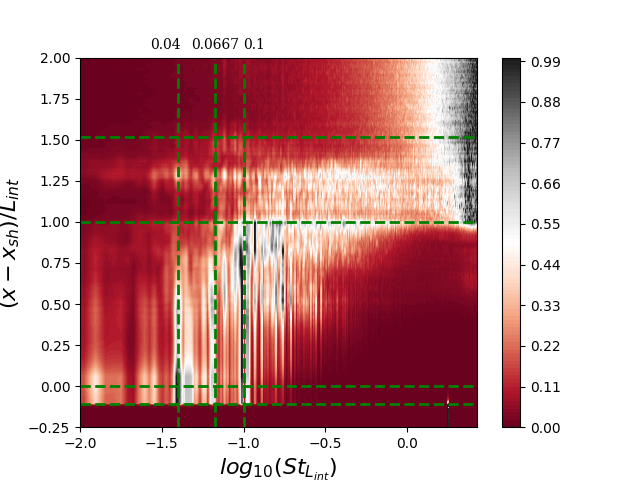}}
 		\caption{Distribution along the flat plate of the spanwise averaged premultiplied and normalised power spectral density of the wall pressure. The vertical dashed lines indicate significant frequency peaks in the low and medium-frequency range. The horizontal dashed lines indicate, from bottom to top: the reflected shock foot, the separation point, the incident shock-wave impingement location and the reattachment point.}
 		\label{fig:DUPONTRAMA_Pw}
 	\end{subfigure}%
 	\begin{subfigure}{.5\textwidth}
 		\centerline{\includegraphics[width=1.0\linewidth]{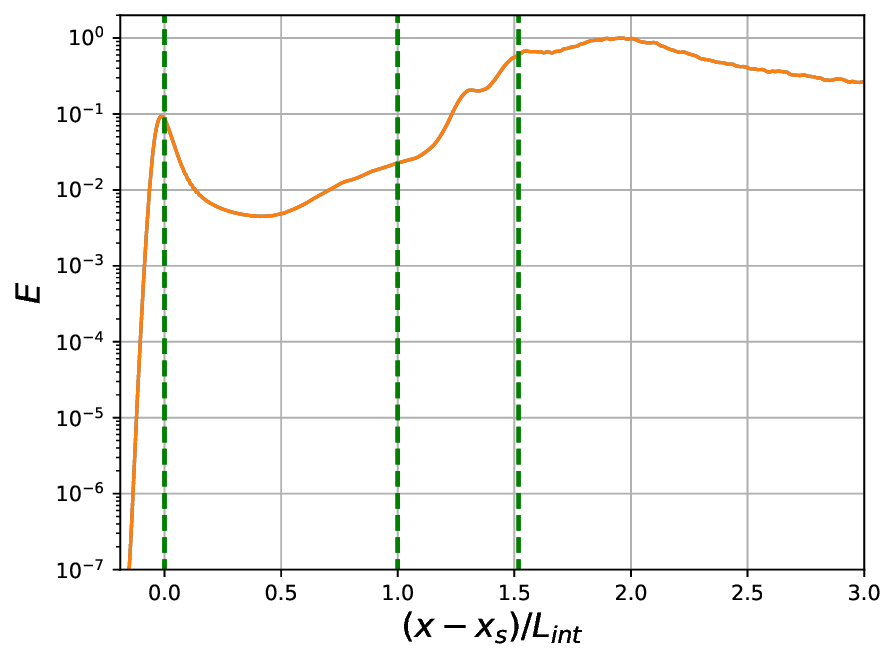}}
 		\caption{Distribution along the flat plate of the expected power in the range $St_{L_{int}}\in [0.02,0.1]$ of the wall pressure. The vertical dashed lines indicate, from left to right: separation point, the incident shock impingement location, and the reattachment point.}
 		\label{fig:Energie_Pw_St_0.01-0.1_y=0.3}\vspace{1cm}
 	\end{subfigure}%
 	\caption{Spanwise averaged power spectral density and power of the wall pressure along the flat plate.}
 	\label{fig:DUPONTRAMA}
 \end{figure}

 We are also interested in the pressure forces exerted on the wall in the interaction zone. To this end, we analyse the evolution in the streamwise direction of the premultiplied power spectral density normalised to the wall pressure shown in Figure \ref{fig:DUPONTRAMA_Pw}. A similar evolution of the spectrum can be observed between the points of separation and reattachment in the sense that we move from the low/medium frequency range to a wider range, extended to higher frequencies. However, we can clearly see that in the first part of the separation bubble, the wall pressure spectrum predominantly highlights the spectral signature of the shear layer, which is peaked at frequencies $St_{L_{int}}=0.04,0.0667$ and $0.1$. In both respects, the wall pressure spectrum is very similar to the spectra presented in \citet{Mauriello_2022}. In this simulation, we therefore record pressure oscillations at the foot of the reflected shock dominated by the low and medium frequency range $St_{L_{int}}\in [0.02,0.1]$ which characterises the SWBLI unsteadiness. The evolution in the streamwise direction of the power of the parietal pressure signal in this frequency range is shown in figure \ref{fig:Energie_Pw_St_0.01-0.1_y=0.3}. As with the near-wall velocity signal, there is a quasi-exponential attenuation of the wall pressure perturbations as we move upwards from the reattachment point. We then see an increase in the first quarter of the bubble as we get closer to the separation point and the shear layer becomes closer to the wall. The signal power in the separation zone is an order of magnitude lower than the power in the reattachment zone.

\begin{figure}
  \begin{subfigure}{.5\textwidth}
  \centering
    \includegraphics[width=.9\linewidth]{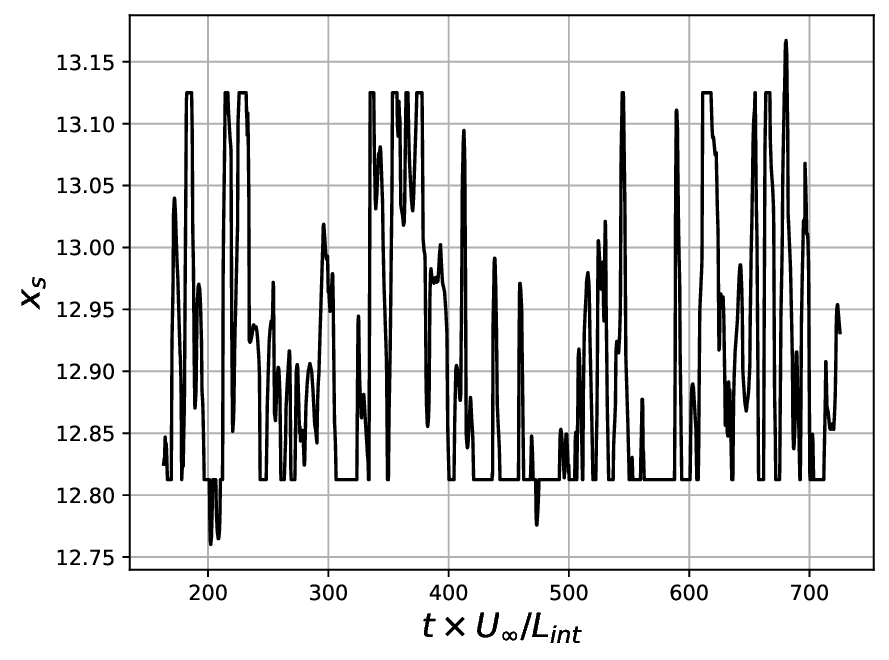}
    \caption{$x_S$}
  \end{subfigure}%
  \begin{subfigure}{.5\textwidth}
  \centering
    \includegraphics[width=.9\linewidth]{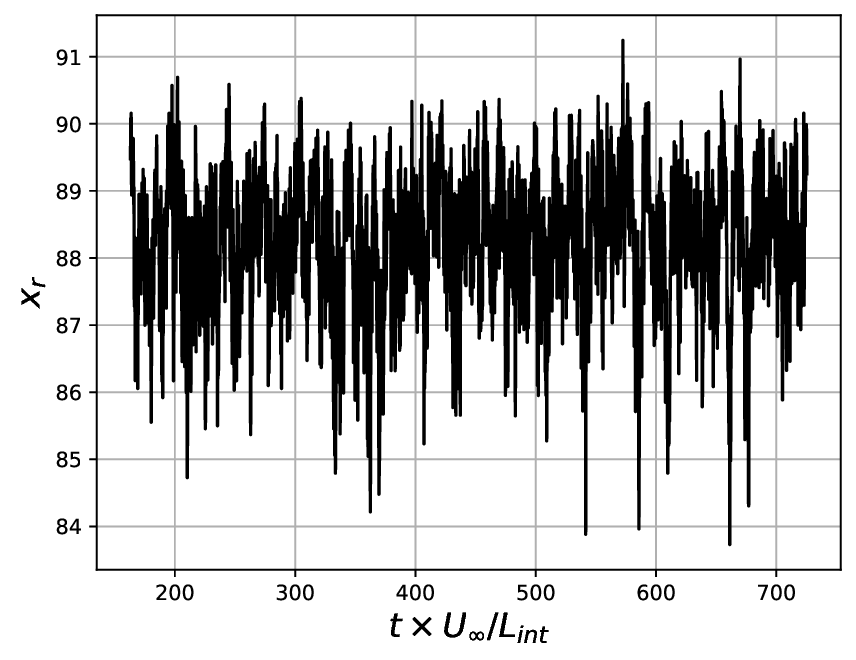}
    \caption{$x_R$}
  \end{subfigure}
  \caption{History of the spanwise averaged abscissa along the flat plate of the separation point, on the left, and the reattachment point, on the right.}
 \label{fig:separation_reattachement_abscissa_history}
\end{figure}

\begin{figure}
  \begin{subfigure}{.5\textwidth}
  \centering
    \includegraphics[width=.9\linewidth]{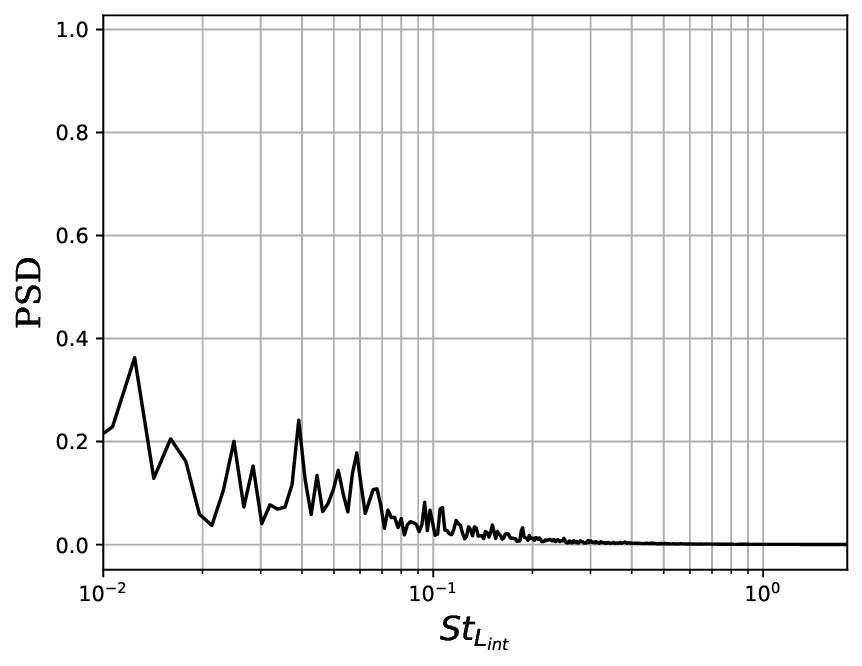}
    \caption{$x_S$}
  \end{subfigure}%
  \begin{subfigure}{.5\textwidth}
  \centering
    \includegraphics[width=.9\linewidth]{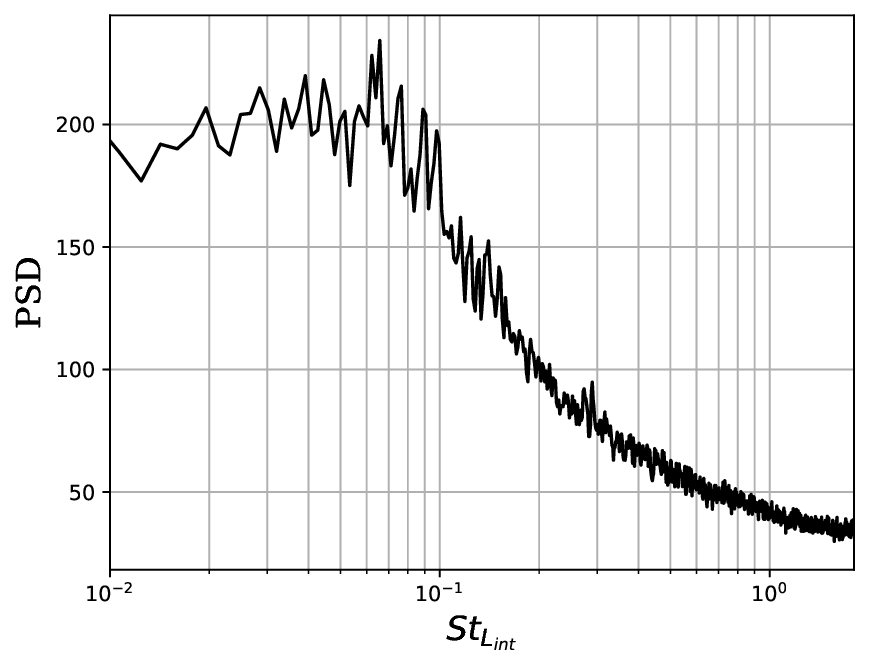}
    \caption{$x_R$}
  \end{subfigure}
  \caption{Spanwise averaged power spectral densities of the separation $x_S$ and reattachment point locations $x_R$.}
 \label{fig:PSD_xs_xr}
\end{figure}

The complex dynamics described above are also reflected in the breathing and flapping cycles of enlargement and shrinkage of the separation bubble, which will be described in section \ref{sec:SPOD_Analysis} documenting the SPOD analysis we carried out. This low- and medium-frequency dynamics of the bubble can also be seen by analysing the temporal evolution of the position of the separation and reattachment points, respectively $x_S$ and $x_R$, shown in Figure \ref{fig:separation_reattachement_abscissa_history}. It can be seen that the reattachment point is subject to oscillations whose maximum extrusions correspond to about $7\delta$ while the separation point is subject to very weak oscillations with maximum amplitudes of the order of $0.4 \delta$. This order-of-magnitude difference in the amplitudes of the oscillations of $x_S$ and $x_R$ is consistent with the order-of-magnitude difference observed in the power of the oscillations of the parietal pressure and velocity signals in these two regions. Similarly, in agreement with the wall pressure and velocity signals in these regions of the flow, the spectral content of $x_S(t)$ oscillations is mainly located in the low and medium frequency range with significant peaks at frequencies $St_{L_{int}}=0.04,0.0667$ and $0.1$, while the oscillations of $x_R(t)$ present a wider band spectrum with peaks also localised at these frequencies, as shown in Figure \ref{fig:PSD_xs_xr}. 

In the present flow configuration, the low energy level of perturbations coming from the reattachment point reaching the separation point explains the low amplitude of the separation point and associated reflected shock foot (not shown here) oscillations. This point is a salient feature of the flow studied here, which makes it different from configurations with a forced laminar or fully turbulent boundary layer. As already stated in the introduction of this paper, in this respect, the present case study represents a limit case in which the mechanisms at the origin of the low-frequency dynamics appear to be present at low intensity and are not disturbed by the natural frequencies of the incident turbulence what, we believe, makes it a good candidate for identifying the intrinsic underlying mechanisms of this dynamics. In the cases of forced or turbulent incoming boundary layers, the more intense dynamics of the mixing layer (due to upstream forcing) has two consequences: (i) the energy of the perturbations at reattachment, and therefore the energy of the signal emitted from reattachment is much higher, (ii) the separation length, i.e. the distance between the zone of signal emission and the separation point, is smaller. The combination of these two characteristics means that the disturbances received at the separation point are more energetic than in the case studied in this article. In the forced and turbulent cases, disturbances are then more likely (for a given strength of interaction) sufficiently energetic to induce higher amplitude oscillations at the separation point.

   \subsection{Database}
   \label{subsec:Database}
   
   After the initial transient, corresponding to the initial propagation of the incident shock wave in the domain and the creation of the separation bubble and subsequent shock wave system, $3 \times 10^6$ iterations have been simulated with a time step of $\Delta t \simeq 3 \times 10^{-4} \times L_{int} / U_\infty$. A snapshot has been recorded every $1000$ time steps so that a total of 3000 snapshots are available for analysis. Each snapshot consists in the state vector of primitive variables $\boldsymbol{q}(\boldsymbol{x}_i,t_l)=[\rho,u,v,w,T]^T$ evaluated at each mesh point $\boldsymbol{x}_i$ (with $ 1 \leq i \leq M$, $M$ being the number of mesh points considered) in the 3D domain containing the interaction region and at each time of snapshot recording $t_l$ (with $1 \leq l \leq N_t$, $N_t$ being the number of snapshots recorded). All data are gathered in the snapshot matrix 
   
   \begin{equation}
   	\mathsfbi{Q}=[\boldsymbol{q}_1~...~\boldsymbol{q}_{N_t}]=\begin{pmatrix}
   		\boldsymbol{q}(\boldsymbol{x}_1,t_1) &  \cdots & \boldsymbol{q}(\boldsymbol{x}_1,t_{N_t}) \\
   		\vdots  &    & \vdots  \\
   		\boldsymbol{q}(\boldsymbol{x}_{M},t_1) &  \cdots & \boldsymbol{q}(\boldsymbol{x}_{M},t_{N_t}) 
   	\end{pmatrix}.
   	\label{eq:snapshot_matrix}
   \end{equation}

For both SPOD and BSMD analysis algorithms introduced in sections \ref{sec:SPOD_Analysis} and \ref{sec:BSMD_Analysis}, the Welch's method  \citep{Welch_1967} is used to compute the discrete Fourier transform (DFT) of the state vector.  The matrix of snapshots is split into $N_b= 10$ with an overlapping of $67 \%$ between the blocks. With this slicing, the total length of the signal considered for the analysis is around 30 events at $St=0.02$ and 60 events at $St=0.04$.

In the following, for both SPOD and BSMD analyses, we consider the centered snapshot matrix  

   \begin{equation}
	\mathsfbi{Q}^c=[\boldsymbol{q}^c_1~...~\boldsymbol{q}^c_{N_t}]=\begin{pmatrix}
	\boldsymbol{q}^c(\boldsymbol{x}_1,t_1) &  \cdots & \boldsymbol{q}^c(\boldsymbol{x}_1,t_{N_t}) \\
		\vdots  &    & \vdots  \\
		\boldsymbol{q}^c(\boldsymbol{x}_{M},t_1) &  \cdots & \boldsymbol{q}^c(\boldsymbol{x}_{M},t_{N_t}) 
	\end{pmatrix},
	\label{eq:centered_snapshot_matrix}
\end{equation}
with $\boldsymbol{q}^c(\boldsymbol{x}_i,t_l)=\boldsymbol{q}(\boldsymbol{x}_i,t_l)-\boldsymbol{\overline{q}}(\boldsymbol{x}_i)$, where $\boldsymbol{\overline{q}}(\boldsymbol{x}_i)$ is the vector of primitive variables averages in time and spanwise direction.

\subsubsection{Remark on the signal length}

For the following SPOD and BSMD analyses. A convergence study has been performed. Indeed, these analyses have been performed with different numbers of snapshots (i.e. different length of the signal and frequency resolution) leading to the same conclusions than those presented below. 
   
   
\section{Spectral Proper Orthogonal Decomposition (SPOD) analysis}
\label{sec:SPOD_Analysis}

In order to get more insight in the low/medium-frequency dynamics of the flow, we perform a modal decomposition of the database formed by the centered snapshot matrix introduced in the previous section. In particular, we use the SPOD, introduced in \citet{Towne_Spectral_2018} and \citet{Schmidt_Guide_2020} (to which we refer for more details about the methodology), to highlight the physical structure of the flow at each frequency. Indeed, the SPOD can be schematically described as a POD applied in the frequency domain frequency by frequency. The first step of the algorithm is the computation of the discrete Fourier transform (DFT) of the state vector. To this end, the matrix of snapshots \eqref{eq:centered_snapshot_matrix} is split into $N_b$ segments of $N_f$ snapshots with an overlapping of $N_o$ snapshots ; thus $N_b$ snapshot matrices $	\mathsfbi{Q}^{c(n)}=[\boldsymbol{q}^{c(n)}_1~...~\boldsymbol{q}^{c(n)}_{N_f}]$ are obtained, with $ 1 \leq n \leq N_b$. The DFT is applied to each snapshot matrix to build $N_b$  snapshot matrices in the frequency domain $\mathsfbi{\hat{Q}}^{c(n)}=[\boldsymbol{\hat{q}}^{c(n)}_1~...~\boldsymbol{\hat{q}}^{c(n)}_{N_w}]$, with $N_w=N_f/2+1$ the number of resolved frequencies. As a second step of the algorithm, the snapshot matrices in the spectral domain $\mathsfbi{\hat{Q}}^{(n)}$ are sorted by frequency to extract the matrices of realisations by frequency $\mathsfbi{\hat{Q}}^c_{k}=\left[\boldsymbol{\hat{q}}^{c(1)}_k~...~\boldsymbol{\hat{q}}^{c(N_b)}_k\right]/\sqrt{N_b}$, with $ 1 \leq k \leq N_w$. As already stated in paragraph \ref{subsec:Database}, in our case, $N_b = 10$ and the overlapping between blocks is $67 \%$. The third step consists in performing a POD to each matrix of realisations $\mathsfbi{\hat{Q}}_{k}$. In our implementation, we use the snapshot POD. In the frequency domain, it consists in solving the following eigenvalue problem 
\begin{equation}
	\mathsfbi{C}_k \mathbf{\Theta}^H_k = \mathbf{\Theta}^H_k \Lambda_k,
\end{equation}
where $\mathsfbi{C}_k=\mathsfbi{\hat{Q}}^{cH}_{k} \mathsfbi{W} \mathsfbi{\hat{Q}}^{c}_{k}$ is the matrix of spectral densities, $\mathbf{\Theta}_k \in \mathbb{C}^{N_b \times N_b}$ is the projection matrix in which each column contains the projection coefficients of one realisation over the spatial modes, $(.)^H$ is the conjugate transpose and $\Lambda_k \in \mathbb{R}^{N_b \times N_b}$ is the diagonal matrix containing the eigenvalue associated to each spatial mode with $\lambda_1 \geqslant \lambda_2 \geqslant ... \geqslant \lambda_{N_b} \geqslant 0$.  In our implementation, the inner product is weighted using the diagonal matrix of spatial quadrature weights $\mathsfbi{W}$ introduced by \citet{Chu_1965}.
The matrix of spatial modes is then obtained by projection
\begin{equation}
 \mathbf{\Psi}_k = \mathsfbi{\hat{Q}}^{c}_{k} \mathbf{\Theta}^H_k.
\end{equation}

Each column of $\mathbf{\Psi}_k$ contains the $m^{th}$ spatial mode $\boldsymbol{\psi}_m(f_k)$ associated to $\lambda_m$.

As a result, each spectral realisation $\boldsymbol{\hat{q}}^{c(n)}_k$ at frequency $f_k$, is decomposed into a linear combination of $N_b$ spatial modes $\boldsymbol{\psi}_m$ as

  \begin{equation}	            \boldsymbol{\hat{q}}^{c(n)}(\boldsymbol{x}_i,f_k)=\sum_{m=1}^{N_b}a_m^{(n)}(f_k)\boldsymbol{\psi}_m(\boldsymbol{x}_i,f_k),
  \label{eq:SPOD_decomposition}	
  \end{equation}
where the $a_m^{(n)}$ coefficients are the components of the projection coefficients at each frequency.

The spatial modes correlate space and time and they are ordered by their probability of existence in the flow quantified by their associated eigenvalue.

The knowledge of the $m^{th}$ spatial mode at frequency $f_k$, $\boldsymbol{\psi}_m(\boldsymbol{x}_i,f_k)$, allows the construction of the associated $m^{th}$ spatio-temporal mode
by inverse Fourier transform
\begin{equation}
	\boldsymbol{\mathcal{\Phi}}_m(\boldsymbol{x}_i,t)=\boldsymbol{\psi}_m(\boldsymbol{x}_i,f_k) e^{2i\pi f_kt}.
	\label{eq:spatio_temporal_evolution}
\end{equation}
 The spatio-temporal modes represent structures that evolve coherently in space and time and are ordered according to the most likely dynamics for a given frequency (mode 1 being the most probable one, then mode 2, etc.).
 
 \subsection{Results}
 \label{subsec:SPOD_results}

\begin{figure}
     \centering
      \includegraphics[width=.9\linewidth]{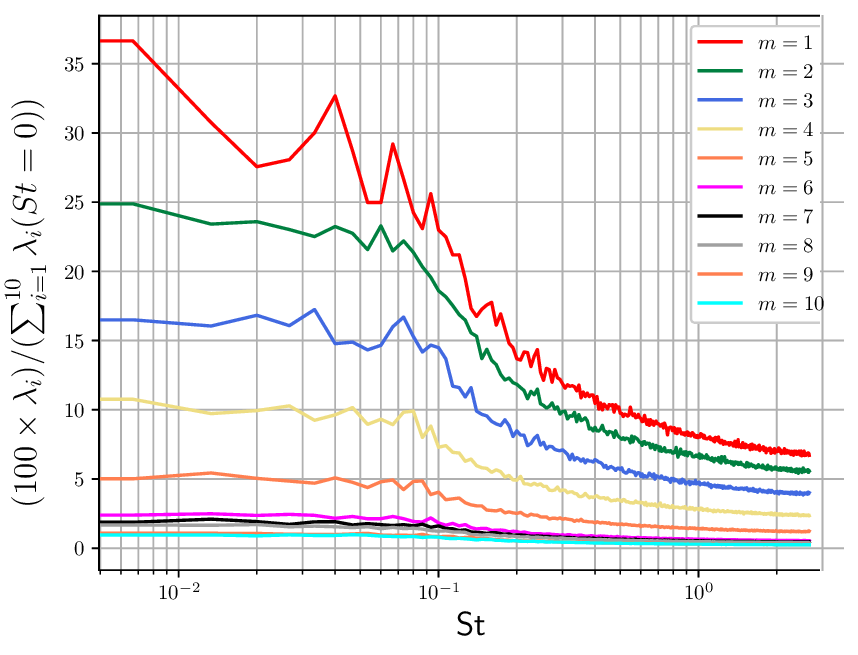}
 	\caption{SPOD spectrum of the state vector $\boldsymbol{\hat{q}}$ for each mode ($m_j,\>\>j=1,...,10$) expressed as a relative energy given in $\%$.}
 	\label{fig:SPOD_spectrum}
 \end{figure}

The SPOD spectrum of eigenvalues $\lambda_m$, is shown in figure \ref{fig:SPOD_spectrum}. We can see that the SPOD spectrum describes a marked dynamic in the low and medium frequency range, with a clear drop-off beyond that. We can clearly see that the energy decreases with the number of the mode, so that the first mode contains about twice as much energy as the third. In the low and medium frequency range, almost $60 \%$ of the flow energy is contained in the first two modes. Interestingly, the 3 main frequencies for mode 1 are the frequencies found in local DNS measurements (see Figure \ref{fig:PSD_prob}): $St_{L_{int}}=0.04,0.0667$ and $0.1$. In particular, we have already shown that these three frequencies, which are present throughout the shear layer, are largely dominant in the rising shear layer.  We can also see that mode 2 signs the frequency $St_{L_{int}}=0.04$ and its first sub-harmonic $St_{L_{int}}=0.02$ as well as the sum of the two ($St_{L_{int}}=0.06$) and another frequency at $St_{L_{int}}=0.073$. The three frequencies highlighted by the second mode (i.e. $St_{L_{int}}\simeq 0.02,0.06$ and $0.073$) are well present in the spectra of the DNS signals measured locally in the shear layer shown in Figure \ref{fig:PSD_prob}. However, unlike the frequencies highlighted by the first mode, these frequencies reach energy levels comparable to the former in the descending shear layer, downstream of the impact of the incident shock, and are comparatively less energetic in the upstream part of the separation bubble.  The first mode therefore seems to bring out the dominant frequencies in the whole of the interaction zone, while the second mode brings out frequencies that are mainly present in the downstream part of the interaction zone.

\begin{figure}
  \centering
  \begin{subfigure}{1.0\textwidth}
    \includegraphics[width=1.0\linewidth,trim=0 1 4 0,clip]{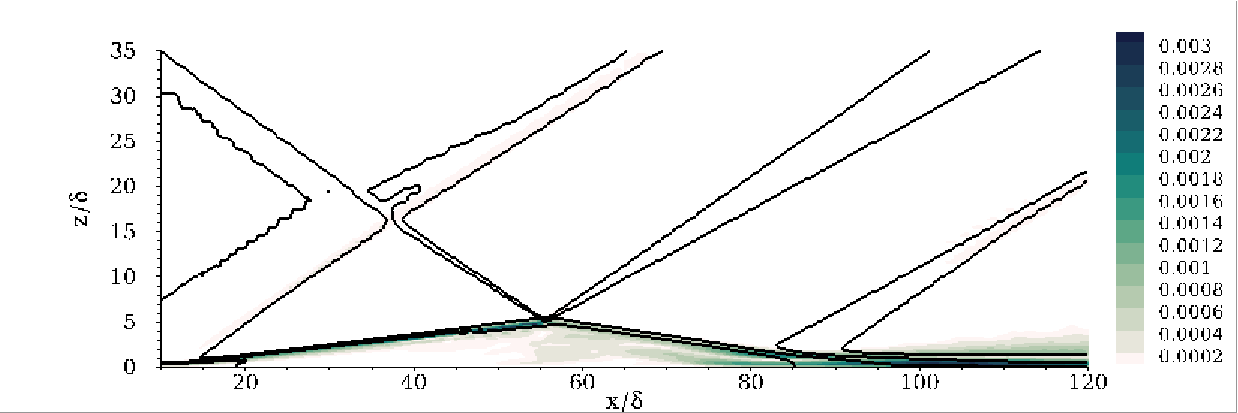}
    \caption{Longitudinal velocity $u$.}
  \end{subfigure}
  \begin{subfigure}{1.0\textwidth}
    \includegraphics[width=1.0\linewidth,trim=0 1.5 4 0,clip]{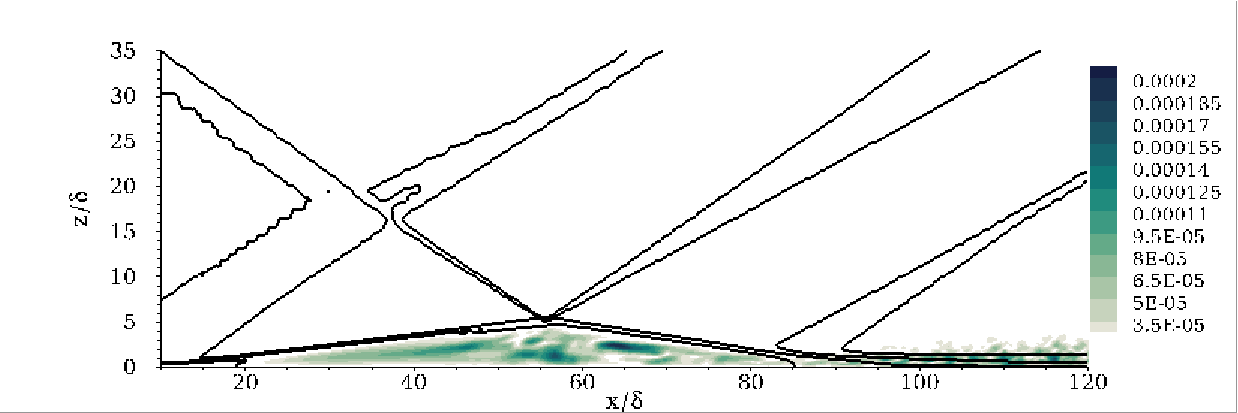}
    \caption{Spanwise velocity $v$.}
  \end{subfigure}
   \begin{subfigure}{1.0\textwidth}
  	\includegraphics[width=1.0\linewidth,trim=0 1.5 4 0,clip]{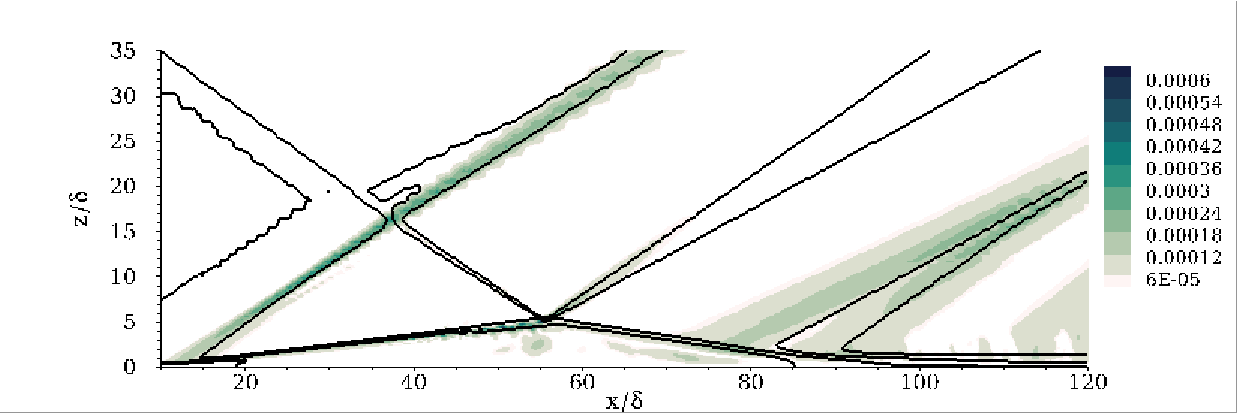}
  	\caption{Vertical velocity $w$.}
  \end{subfigure}
  \caption{First spatial SPOD mode $(m_1)$ averaged in the spanwise direction at $S_t=0.04$. The mean flow is indicated by isolines of the mean density field. }
  \label{fig:SPOD_spatial_mode_0.04}
\end{figure}

We are now interested in the shape of the dominant modes in the entire bubble. As an illustration, Figure \ref{fig:SPOD_spatial_mode_0.04} shows the modulus of the first mode at frequency $St=0.04$ (characteristic frequency of the separation bubble breathing) for the three velocity components $u$, $v$ and $w$. For the longitudinal component $u$, we can see that the activity of the mode is mainly localised in the shear layer. In particular, large amplitudes of the mode are observed in the rising shear layer. In the descending shear layer, the amplitude increases the closer we get to the abscissa of the reattachment point. In the attachment zone, strong amplitudes are also observed close to the wall and amplify downstream of the interaction. We also observe activity in the longitudinal velocity mode localised on the reflected shock as well as the reattachment shock, materialising the oscillations of these shocks at this low frequency, characteristic of SWBLI unsteadiness. These oscillations of the reflected and reattachment shocks are even more visible in the shape of the vertical velocity mode $w$, whose activity is localised in these zones as well as in the shear layer, with less intensification than for the longitudinal velocity in the reattachment zone. The activity of the mode for the spanwise velocity component $v$ is completely localised inside the separation bubble.

The spatial modes at frequencies $St_{L_{int}}=0.0667$ and $0.0934$ are qualitatively almost identical to those at frequency $St_{L_{int}}=0.04$. For the sake of brevity, they are shown in the appendix \ref{appA}.

In Figure \ref{fig:SPOD_spatio_temporal_St=0.04}, we plot the real part of the first spatio-temporal mode $\boldsymbol{\mathcal{\Phi}}_1(\boldsymbol{x}_i,t)$ of the three velocity components for the Strouhal $St = 0.04$, characteristic of the breathing of the recirculating bubble. We plot regularly spaced snapshots in a period $T=1/St$. 
For each speed component, a video of the animation of the spatio-temporal evolution of the associated mode is available in the supplementary material provided with the article.

\begin{figure}
 	 \centering
 	  \begin{subfigure}{1.0\textwidth}
     \includegraphics[width=1.0\linewidth]{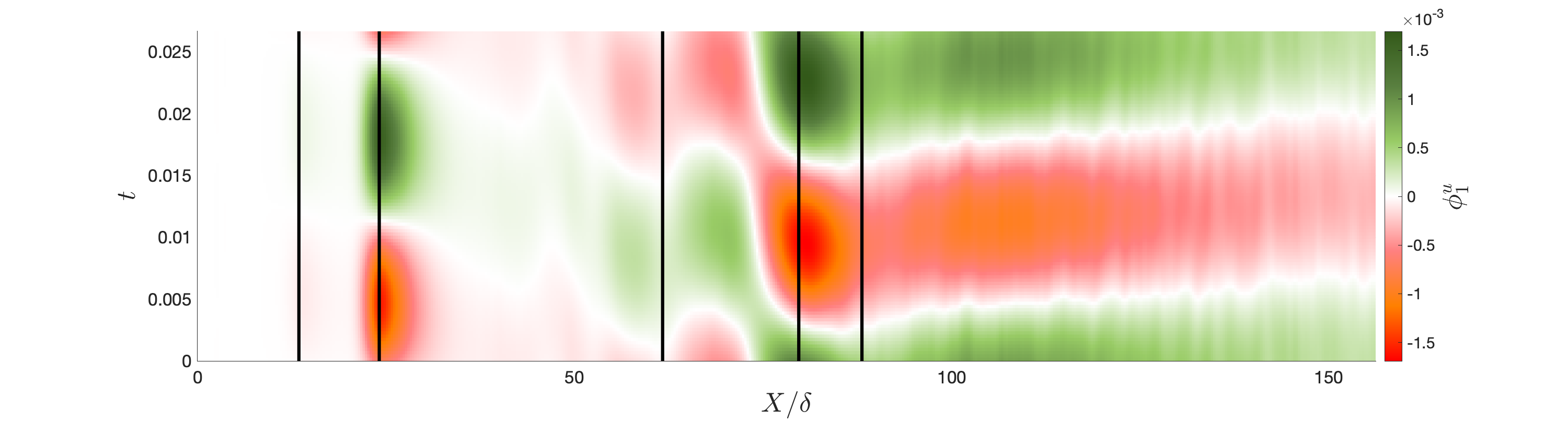}
     \caption{Velocity component $u$}
     \label{fig:SPOD_diagramme_x_t_U}
    \end{subfigure}
    \begin{subfigure}{1.0\textwidth}
     \includegraphics[width=1.0\linewidth]{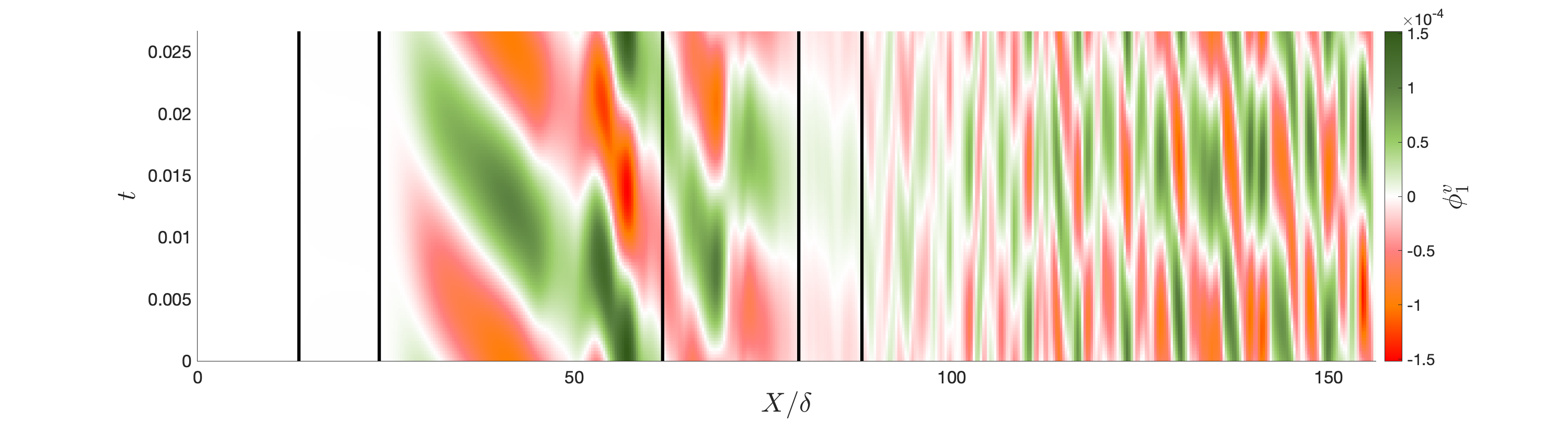}
     \caption{Velocity component $v$}
      	\label{fig:SPOD_diagramme_x_t_V}
    \end{subfigure}
 	\caption{Representation of $\boldsymbol{\mathcal{\Phi}}_1(x,z=\text{cste},t)$ for $z/\delta = 1.48$. The vertical black lines indicate, from left to right: the separation point, the crossing of the rising shear layer, the crossing of the incident shock-wave, the crossing of the descending shear layer and the reattachment point. }
 	\label{fig:SPOD_diagramme_x_t}
 \end{figure}

In the longitudinal velocity mode $u$, we can clearly see that the oscillations of the reflected shock in the longitudinal direction are in phase with the oscillations of the shear layer in this direction, and therefore in phase with the oscillations of the separation point. We can also see that the breathing cycles follow the following sequence: the value of the mode changes in the downstream part of the bubble before being propagated upstream in the first part of the bubble, eventually contaminating the entire shear layer. The upstream propagation of information in the bubble is shown in figure \ref{fig:SPOD_diagramme_x_t}, where we plotted versus time (for a single breathing period) the value of $\boldsymbol{\mathcal{\Phi}}^u_1$ along the flat plate measured at a constant height above it $z/\delta=1.48$. In the first part of the bubble, we clearly distinguish regions of same sign oriented with a negative slope in the $(x,t)$ frame, denoting an upstream propagation of this sign. This behaviour is qualitatively observed at all altitudes in the separation bubble. The upstream propagation from the latter part of the separation bubble is also visible in the spanwise velocity mode as shown in Figures         \ref{fig:SPOD_spatio_temporal_St=0.04_V} and       	\ref{fig:SPOD_diagramme_x_t_V}. This result seems to confirm that the downstream part of the separation plays a decisive role in the breathing dynamics of the separation bubble. More precisely, it confirms the analysis of section \ref{subsec:Mean_flow_and_frequency}: the low-frequency oscillations of the separation bubble are driven by oscillations at this frequency in its latter part (materialised by the change in sign of the mode in the reattachment zone) propagated upstream in the separation bubble.

A similar qualitative behavior is observed in the results obtained for the frequencies $St = 0.0667$ and $St=0.0934$, the latter being characteristic of the flapping of the separation bubble. The figures spatial evolution of the first mode at these frequencies can be found in appendix \ref{appA}. The similarity of the spatio-temporal modes for breathing and flapping may suggest a link to be determined between these two oscillation modes.

\begin{figure}
\begin{minipage}[c][\textheight]{\textwidth}
  \begin{subfigure}{0.333\textwidth}
    \includegraphics[width=4.8\linewidth, angle=90]{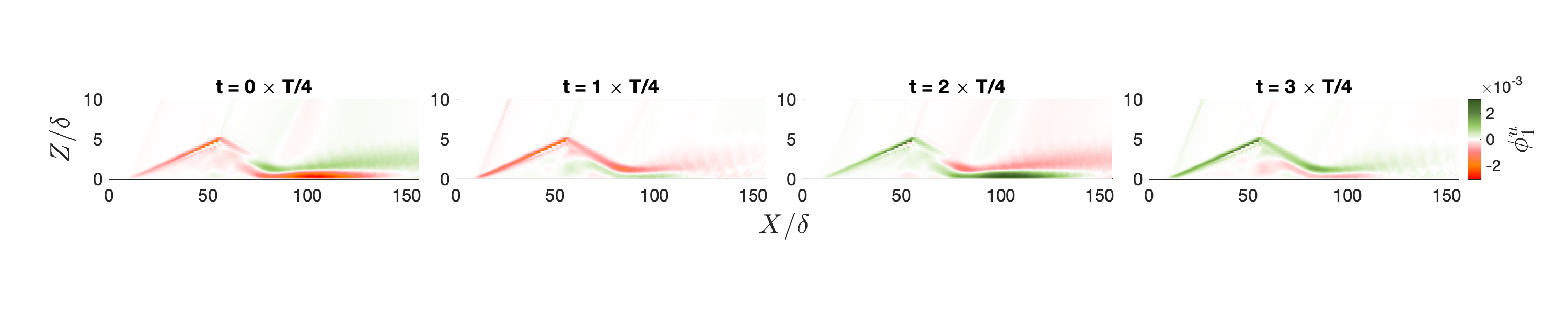}
    \caption{Velocity component $u$.}
  \end{subfigure}
  \begin{subfigure}{0.333\textwidth}
    \includegraphics[width=4.8\linewidth, angle=90]{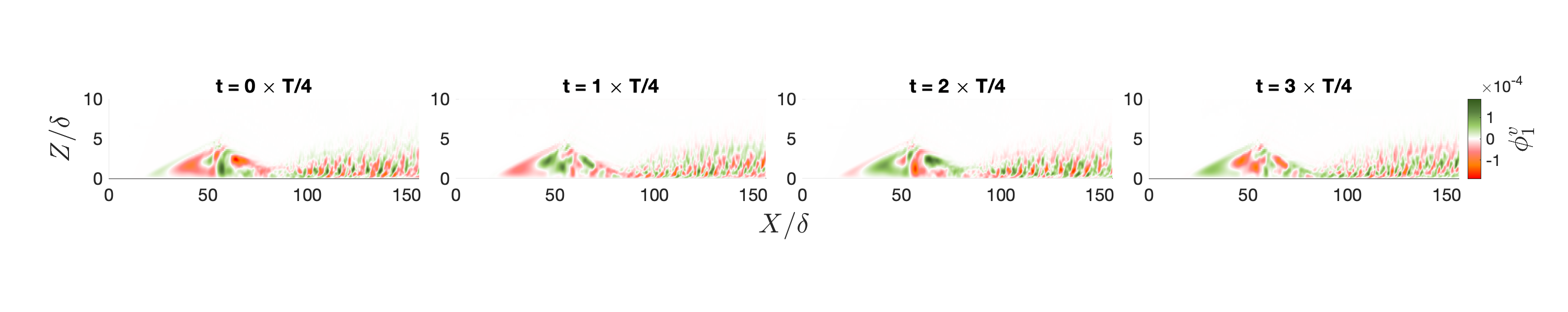}
        \caption{Velocity component $v$.}
        \label{fig:SPOD_spatio_temporal_St=0.04_V}
  \end{subfigure}
  \begin{subfigure}{0.333\textwidth}
    \includegraphics[width=4.8\linewidth, angle=90]{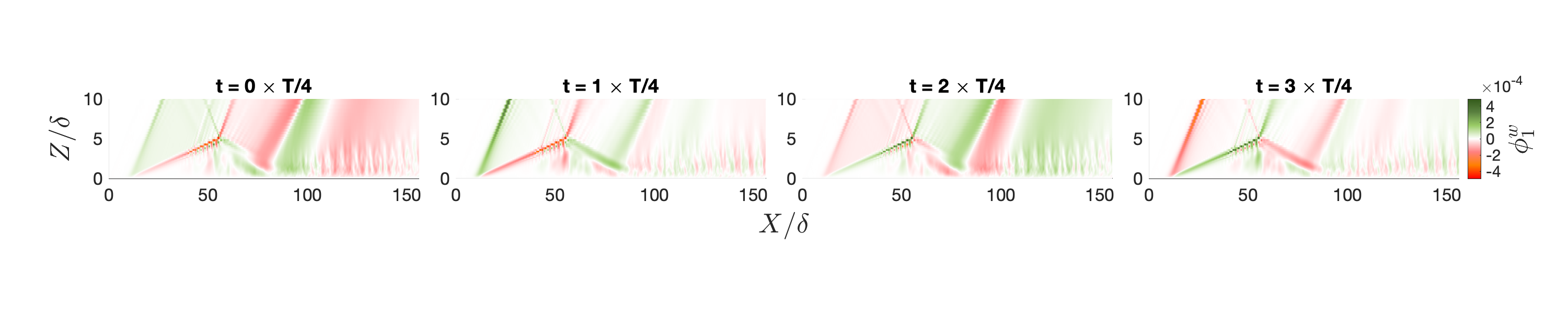}
        \caption{Velocity component $w$.}
  \end{subfigure}
 	\caption{Visualisation of $\boldsymbol{\mathcal{\Phi}}_1(\boldsymbol{x},t)$ velocity u at $St = 0.04$, averaged in the spanwise direction. }
 	\label{fig:SPOD_spatio_temporal_St=0.04}
\end{minipage}
\end{figure}

\section{BiSpectral Mode Decomposition (BSMD) analysis}
\label{sec:BSMD_Analysis}

In order to study possible non-linear links between modes at different frequencies,
the BSMD algorithm has been applied to highlight possible triadic interactions taking place in the interaction zone. Triadic interactions result from the quadratic non-linearities of the Navier-Stokes equations. They are the fundamental mechanism of energy transfer in fluid flows, and manifest themselves in the frequency domain as triplets of three frequencies $(f_1,f_2,f_3)$, the combination of which is zero: $f_1 \pm f_2 \pm f_3 =0$. Such interactions can be identified from triple correlations of frequency components, i.e., from the bispectrum, which is a measure of the quadratic non–linearities at the bifrequency $(f_1,f_2)$. For a scalar signal $q$, it is defined by 
\begin{equation}
	S_{qqq}(f_1,f_2)=\lim\limits_{T \rightarrow \infty} \frac{1}{T} E [\hat{q} (f_1)^*\hat{q} (f_2)^* \hat{q} (f_1+f_2) ],
	\label{eq:bispectrum}
\end{equation}
where $(.)^*$ denotes the complex conjugate and $E$ the expectation operator.

  \subsection{BSMD methodology}
  \label{subsec:BSMD_Methodology} 
  
 The BSMD algorithm is a modal decomposition that reveals the presence of triadic non-linear interactions from multidimensional data. To this end, an integral measure of the point-wise bispectral density between frequency components of the vector of primitive variables is introduced as:
  
  \begin{equation}
  	b(f_k,f_l)=E\Bigr[ \int_{\Omega}^{} \hat{\boldsymbol{q}}_k^* \circ \hat{\boldsymbol{q}}_l^* \circ \hat{\boldsymbol{q}}_{k+l} ~ d\boldsymbol{x} \Bigr]=E \Bigr[\hat{\boldsymbol{q}}_{k \circ l}^H \mathsfbi{\mathcal{W}} \hat{\boldsymbol{q}}_{k + l} \Bigr] = E\Bigr[ < \hat{\boldsymbol{q}}_{(k \circ l)} , \hat{\boldsymbol{q}}_{(k + l)} > \Bigr],
  	\label{eq:integral_bispectrum}
  \end{equation}
  where  $\hat{\boldsymbol{q}}_{(k \circ l)}=\hat{\boldsymbol{q}}(\boldsymbol{x},f_k) \circ \hat{\boldsymbol{q}}(\boldsymbol{x},f_l)$ (with $\circ$ is the Hadamard product), $\hat{\boldsymbol{q}}_{(k + l)} = \hat{\boldsymbol{q}}(\boldsymbol{x},f_{k+l})$ and $(.)^H$ is the conjugate transpose. $\mathsfbi{\mathcal{W}}$ is
  the diagonal matrix of spatial quadrature weights and
  $\Omega$ is the spatial domain over which the flow is defined.
  
  Using the Welch's slicing of the snapshot matrix, two modal projections are constructed: the cross-frequency fields
  \begin{equation}
     \boldsymbol{\Phi}^{[i]}_{k \circ l} (\boldsymbol{x},f_k, f_l) = \sum_{j=1}^{N_b} a_{i}^{[j]}(f_{k+l}) \hat{\boldsymbol{q}}_{(k \circ l)}^{[j]} ,
     \label{eq:cross-frequency_fields}
  \end{equation}
  and the bispectral modes 
  \begin{equation}
  	\boldsymbol{\Phi}^{[i]}_{k + l} (\boldsymbol{x},f_{k+l}) = \sum_{j=1}^{N_b} a_{i}^{[j]}(f_{k+l}) \hat{\boldsymbol{q}}_{(k + l)}^{[j]},
  	\label{eq:bispectral_modes}
  \end{equation}
 that share a common set of expansion coefficients $a_{i}^{[j]}$ with $ 1 \leqslant j \leqslant N_b$.
  
  The goal of bispectral mode decomposition is to compute modes that optimally represent the data in terms of the integral bispectral density defined in equation \eqref{eq:integral_bispectrum}. That is, we seek the set of expansion coefficients $a_1$ that maximises the absolute value of $b(f_k,f_l)$ as defined in Equation \eqref{eq:integral_bispectrum} with $\hat{\boldsymbol{q}}_{(k \circ l)}$ and $\hat{\boldsymbol{q}}_{(k + l)}$ replaced by their modal expansion. We therefore target the set of expansion coefficients that verifies
\begin{equation}
	\boldsymbol{a}_1 = \text{arg} \max_{\parallel \boldsymbol{a}_1 \parallel = 1} \mid E \Bigr[ \boldsymbol{\Phi}^{[1]}_{k \circ l} \mathsfbi{\mathcal{W}} \boldsymbol{\Phi}^{[i]}_{k + l} \Bigr] \mid,
	\label{eq:BSMD_optimization_problem}
\end{equation}
where the coefficient vector is required to be a unit vector in order to guarantee boundedness of the expansion.

More details about the BSMD methodology can be found in \citet{Schmidt_Bispectral_2020} especially about the construction and solving of this optimisation problem.

As a result of the BSMD analysis, we obtain optimal cross-frequency fields and bispectral modes, respectively $\mathbf{\Phi}^{[1]}_{k \circ l}$ and $\mathbf{\Phi}^{[1]}_{k + l}$, as well as the mode bispectrum 
 \begin{equation}
 	\lambda_1(f_k,f_l) = E[\mathbf{\Phi}^{[1]H}_{k \circ l} \mathbf{W} \mathbf{\Phi}^{[1]}_{k + l} ] \in \mathbb{C}.
 	\label{eq:mode_bispectrum}
 \end{equation}
 
 The bispectrum mode is the fundamental result of the BSMD analysis. Indeed, significant triadic interactions are indicated by local maxima of the modulus of this quantity, also called bispectrum mode amplitude.
 Bispectral modes $\mathbf{\Phi}^{[1]}_{k + l}$ are linear combinations of Fourier modes and can be interpreted as observable physical structures resulting from quadratic interactions between frequencies $f_k$ and $f_l$. The multiplicative cross-frequency fields $\mathbf{\Phi}^{[1]}_{k \circ l}$, on the contrary, are maps of phase-alignment between two frequency components that may not directly be observed. Once significant triadic interactions are identified in the map of bispectrum mode amplitude, interaction maps can be computed as:
 \begin{equation}
 	\varPsi_{k,l} (\mathbf{x},f_k,f_l)= | \mathbf{\Phi}^{[1]}_{k \circ l} \circ \mathbf{\Phi}^{[1]}_{k+l} |.
 	\label{eq:interaction_map}
 \end{equation}
 This field indicates the location of the triadic interaction involving the triad $(f_k,f_l,f_{k+l})$ in the flow field, as it quantified the average local bicorrelation between this frequencies in the domain.

To sum up, the BSMD analysis process involves calculating the map of mode bispectrum amplitude $\lVert \lambda_1(f_k,f_l) \rVert$ to identify significant quadratic interactions. These are the interactions that contribute most to the flow dynamics. Once these interactions have been identified, the associated interaction map $\varPsi_{k,l} (\mathbf{x},f_k,f_l)$ is studied. This allows us to identify the regions of the flow where this triadic interaction occurs. Finally, the associated bispectral mode $\mathbf{\Phi}^{[1]}_{k+l}$, which is the physical mode resulting from this interaction is visualised and interpreted.

In the following, we will present the BSMD analysis of the database introduced in paragraph \ref{subsec:Database}. The interpretation of the results will be performed following the process that has just been outlined above, focusing on the low/medium-frequency dynamics of the interaction. In particular we will be interested in identifying and documenting non linear links between medium-frequencies and low-frequencies.
 

  \subsection{Mode bispectrum amplitude}
  \label{subsec:Mode_bispectrum_amplitude}

The modulus of the mode bispectrum $\lVert \lambda_1(St_k,St_l) \rVert$ obtained when applying BSMD to the snapshot matrix \eqref{eq:centered_snapshot_matrix}, is shown in Figure \ref{fig:bispectre}. Only the region of non redundant information, as demonstrated in \cite{Schmidt_Bispectral_2020} is shown. Moreover, the interactions between under resolved frequencies for which 2 or less periods are resolved through the Welch Fourier Transform estimation have been greyed out on the graph. The region above the horizontal axis of the graph represents sum interactions $St_1+St_2=St_3$, the region below the horizontal axis represents difference interactions $St_1-St_2=St_3$. The interactions located on the horizontal axis correspond to interactions for which $St_2=0$ resulting in $St_3=St_1$ and therefore correspond to the linear evolution of the flow around its mean. In our analysis, we do not consider local maxima located on the line $St_1=-St_2$ as they result in modes of frequency $St_3=0$ corresponding to the mean flow distortion. 

 We can clearly identify strong quadratic couplings (highlighted by the high value of $\lVert \lambda_1(St_k,St_l) \rVert $) for $(St_1,St_2) \in [0,0.20]^2$, namely in the low and medium-frequency ranges. The intensity of interactions decreases sharply at higher frequencies. In the low and medium-frequency range, three types of interesting couplings (highlighted by the high value of $\lVert \lambda_1(St_k,St_l) \rVert $) can be identified, that will be analysed in more details in the following paragraphs:

\begin{figure}
    \centering
 	\includegraphics[width=14cm,trim=4 4 4 4,clip]{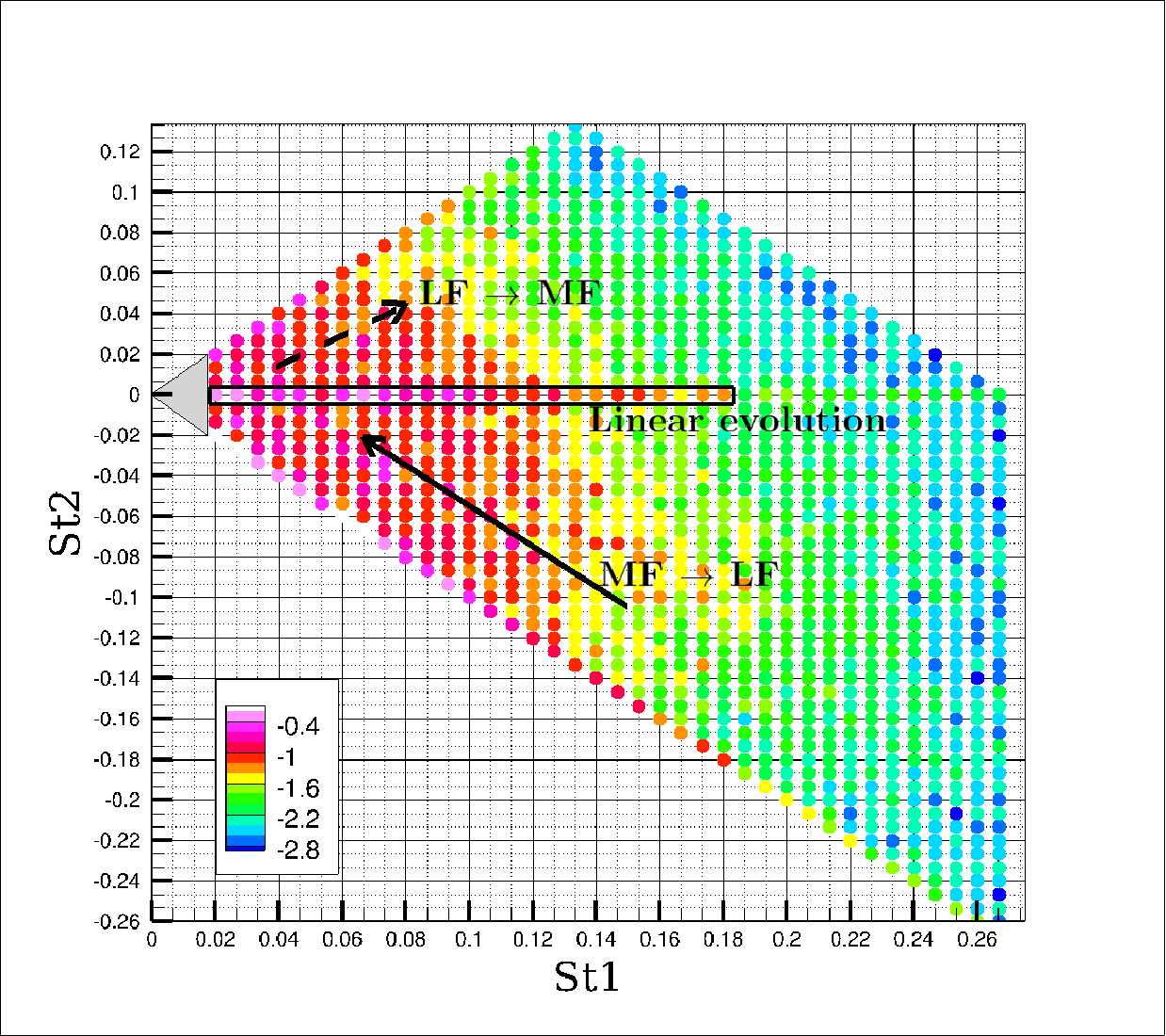}
 	\caption{Modulus of the complex mode bispectrum of the centered state vector $\boldsymbol{q}^c$ in the $(St_1 , St_2 )$ plane. Relevant types of interactions are circled or indicated by arrows: interactions between medium-frequencies creating low-frequencies (arrow) ; interactions between low-frequencies creating medium-frequencies (dashed arrow) ; interactions reflecting the linear evolution of the flow around its mean. Interactions between unresolved frequencies are greyed.}
 	\label{fig:bispectre}
 \end{figure}

\begin{enumerate}
    \item The interactions in the difference interaction region correspond to a cascade of interactions generating low-frequencies (LF) from medium frequencies (MF). This cascade is indicated by the solid black arrow. 
    \item The interactions in the sum interaction region correspond to a cascade of sum interactions between generate medium frequencies (MF) from low frequencies. This cascade is indicated by the dashed arrow.
    \item The interactions circled correspond to interactions associated to the linear evolution of the flow around the mean flow that contribute significantly to the dynamics of the flow.
    \end{enumerate}

In the following, in order to further interpret the physical role of the triadic interactions that we have highlighted, for each type of interaction, we analyse in more details the involved frequencies, as well as their characteristic interaction maps and the shape of the associated bispectral modes.

\subsection{Cascade of interactions generating low-frequencies (LF) from medium frequencies (MF) (type i))}
\label{subsec:MF->LF}

\begin{figure}
    \centering
 	\includegraphics[width=14cm,trim=4 4 4 4,clip]{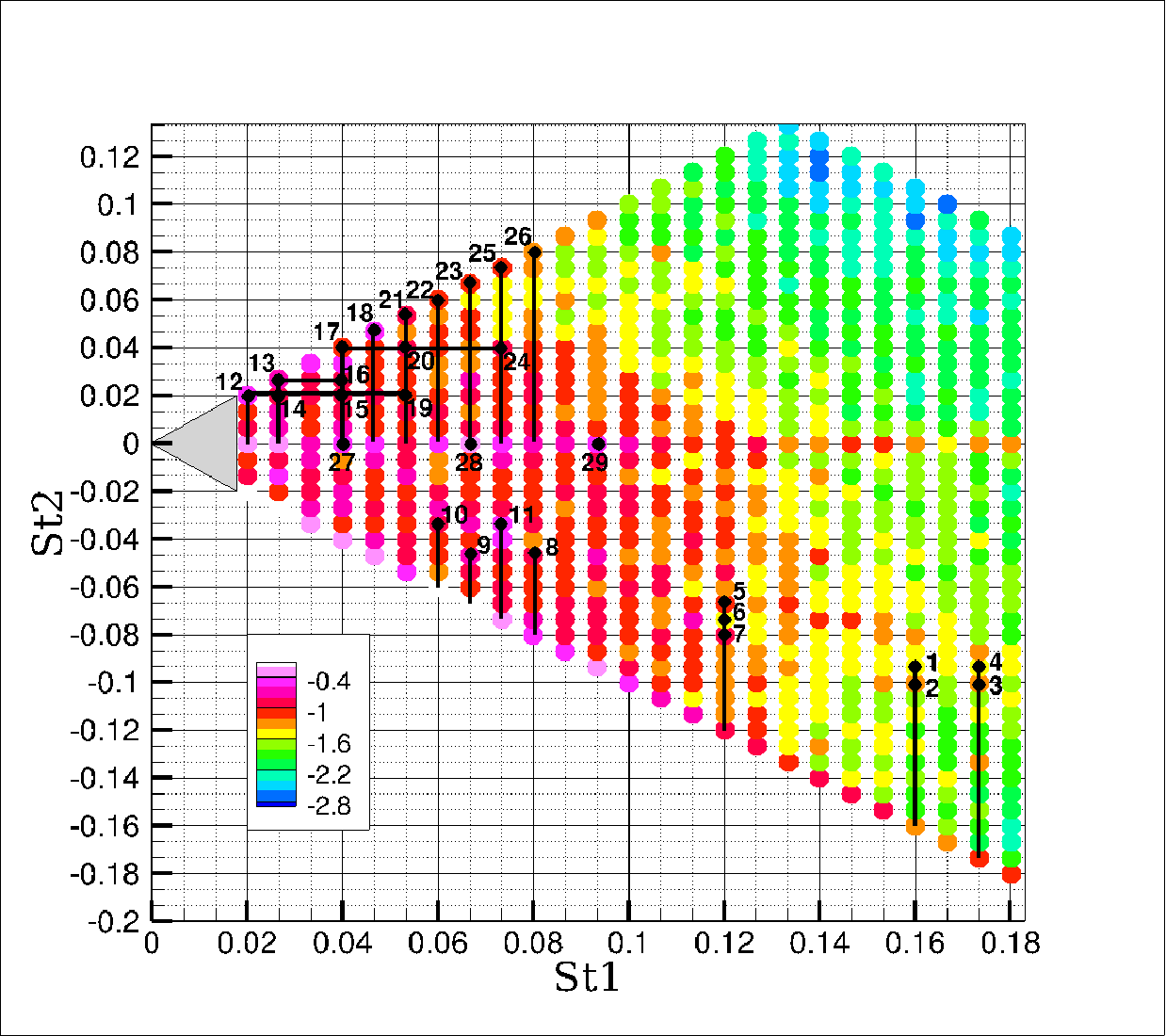}
 	\caption{Modulus of the complex mode bispectrum of the centered state vector $\boldsymbol{q}^c$ in the $(St_1 , St_2 )$ plane: zoom on medium/low-frequencies. Sequences of interactions of each type are highlighted and numbered. Interactions between unresolved frequencies are greyed.}
 	\label{fig:bispectre_zoom}
 \end{figure}

In order to get more insight in the low and medium frequency range, the map of mode bispectrum amplitude is reproduced in figure \ref{fig:bispectre_zoom} by zooming in these ranges.

In the region of the map below the horizontal axis (difference interaction region), there is a large continuous zone of interaction of significant amplitude defined by $(St_1,St_2)\in ]0,0.18]\times ]0,-0.18]$. These interactions are difference interactions, each interaction therefore result in the creation of a lower frequency $St_3$ from two interacting frequencies $St_1$ and $St_2$. As this zone is continuous, a multitude of cascades of successive interactions exist, creating low frequencies from interactions in the mid-frequency range characteristic of the well documented flapping of the shear layer $St \in [0.1,0.18]$.

\begin{figure}
    \centering
 	\includegraphics[width=14cm,trim=4 4 4 4,clip]{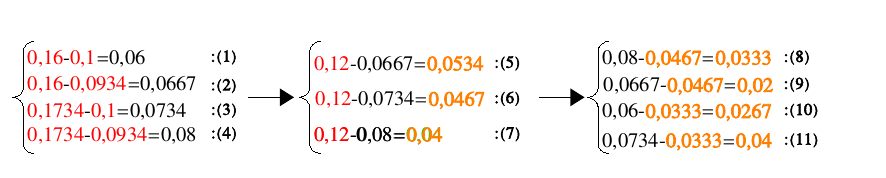}
 	\caption{Example of triadic interactions sequence between medium-frequencies eventually creating low-frequencies. The frequencies are expressed as $St_{int}$. Medium-frequencies at the origin of the sequence are written in red. The resulting low-frequencies are written in orange. Each interaction has a number between parentheses reported in figure \ref{fig:bispectre_zoom}.}
 	\label{fig:triades_descendantes}
 \end{figure}

As an illustration, an example of such a cascade is shown in Figure \ref{fig:triades_descendantes}. Each interaction is given a number (between brackets) which is reported on the mode bispectrum amplitude map (Figure \ref{fig:bispectre_zoom}). We would stress that the cascade considered here is just one example of a multitude of other possible paths. The first column (interactions of numbers $(1)$, $(2)$, $(3)$ and $(4)$) contains interactions between medium frequencies present in the flow (typical frequencies of the well documented flapping mode) that are written in red. The presence of these medium-frequencies in the flow is sufficient to initiate a cascade of interactions that eventually creates low-frequencies. Indeed the interactions of the first column generate lower frequencies. In column 2, we report interactions of these generated lower frequencies with medium frequencies (interactions of numbers $(5)$, $(6)$, and $(7)$). These interactions generate low-frequencies typical of the breathing mode, written in orange. We show in the third column that interactions between frequencies generated from interactions of the first and second columns also contribute to the generation of low-frequencies characteristic of the breathing (interactions of numbers $(8)$, $(9)$, $(10)$ and $(11)$). 

This example of cascade shows how the presence in the flow of the well documented medium-frequencies characteristic of the flapping generates low-frequencies through a multitude of cascades of difference interactions. 

\begin{figure}
  \centering
    \begin{subfigure}{1.0\textwidth}
     \includegraphics[width=1.0\linewidth,trim=0 1 4 0,clip]{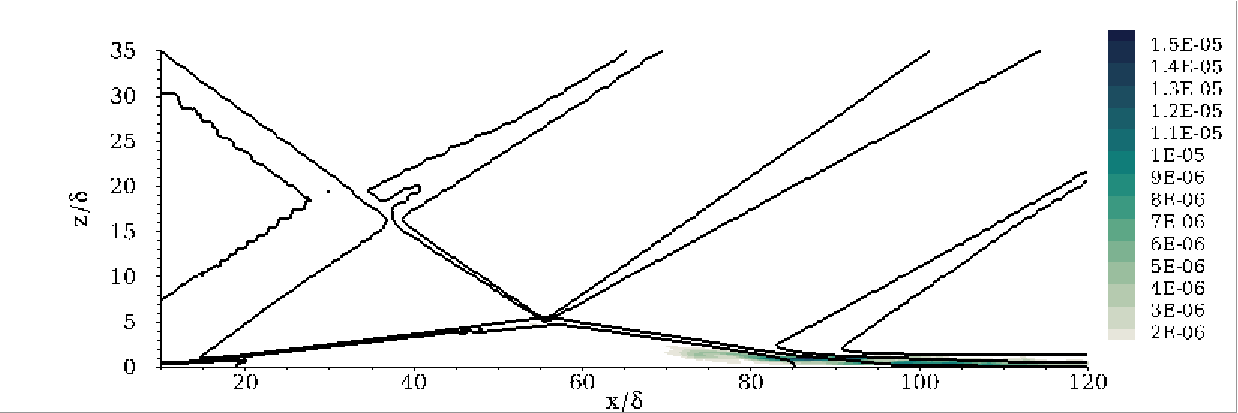}
     \caption{Longitudinal velocity $u$.}
    \end{subfigure}
    \begin{subfigure}{1.0\textwidth}
    \includegraphics[width=1.0\linewidth,trim=0 1 4 0,clip]{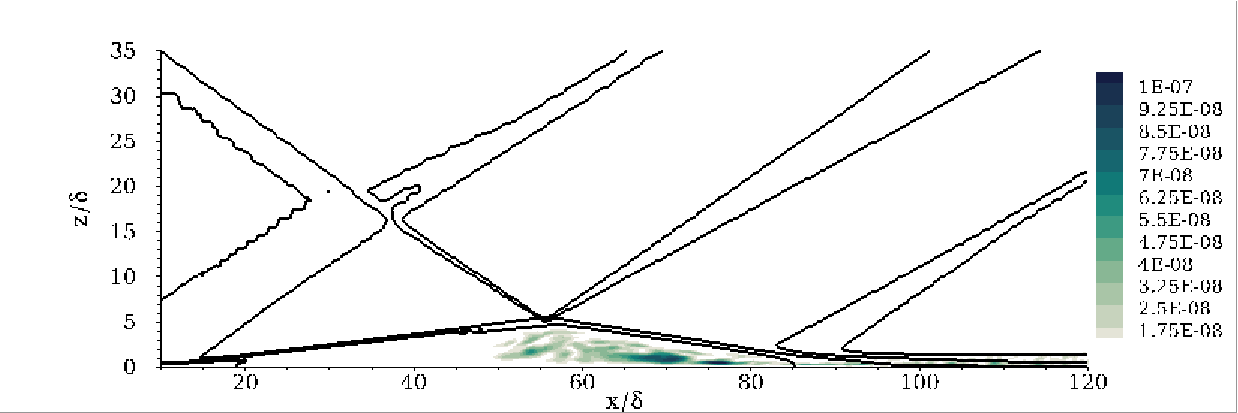}
        \caption{Spanwise velocity $v$.}
    \end{subfigure}
    \begin{subfigure}{1.0\textwidth}
     \includegraphics[width=1.0\linewidth,trim=0 1 4 0,clip]{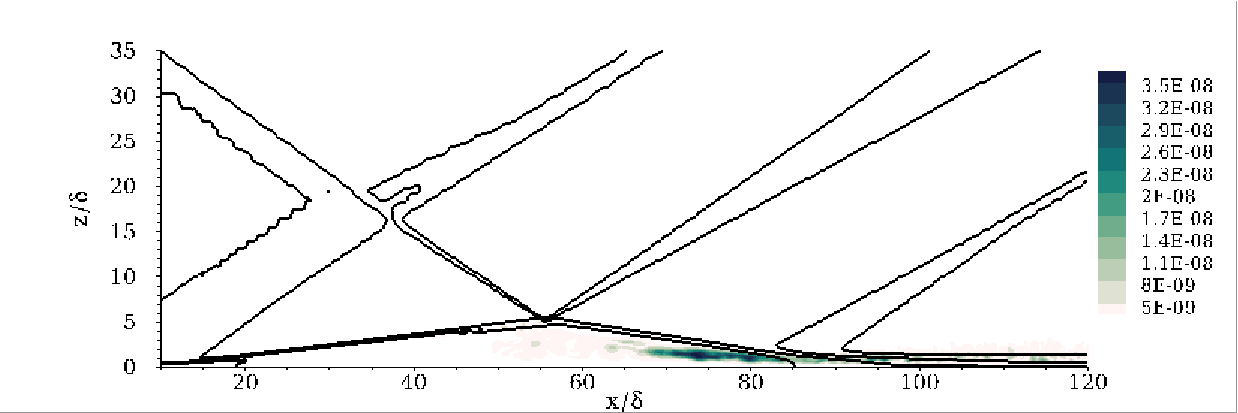}
         \caption{Vertical velocity $w$.}
    \end{subfigure}
    \caption{Spanwise averaged interaction maps for a triadic interaction (interaction (7) in figures \ref{fig:bispectre_zoom} and \ref{fig:triades_descendantes}) illustrating the sequence of interactions creating low-frequencies from medium-frequencies (interactions of type i). Frequencies involved: $St_1=0.12$; $St_2=-0.08$; $St_3=0.04$. The mean flow is indicated by isolines of the mean density field.}
   \label{fig:InterMaps_Mf-LF}
\end{figure}

The interaction map of an emblematic triadic interaction of type i) is shown in figure \ref{fig:InterMaps_Mf-LF} for the three components of velocity. Other interactions of type i) are not shown here for the sake of brevity, but they exhibit the same qualitative behavior. We clearly identify that the non-linear interactions constituting the cascade creating energy at LFs from MFs are mainly localised in the second part of the separation bubble and especially in the region of reattachment. This point is consistent with the analyses and interpretations exposed in previous sections, namely the direct analysis of the DNS data base (section \ref{subsec:Database}) and the SPOD analysis (section \ref{sec:SPOD_Analysis}), in which the low-frequency breathing dynamics of the separation bubble has been shown to be driven by dynamical activity taking place in the downstream part of the separation bubble. Moreover, this result confirms and refines the results of \citet{Sansica_2014} and \citet{Mauriello_2022}. Indeed, both works suspected a role of non-linear interactions in generating low-frequencies in the downstream part of the interaction zone, although the precise determination of the location of significant triadic interactions were difficult to the local nature of the methods employed.

\begin{figure}
  \centering
  \begin{subfigure}{1.0\textwidth}
    \includegraphics[width=1.0\linewidth,trim=0 1 4 0,clip]{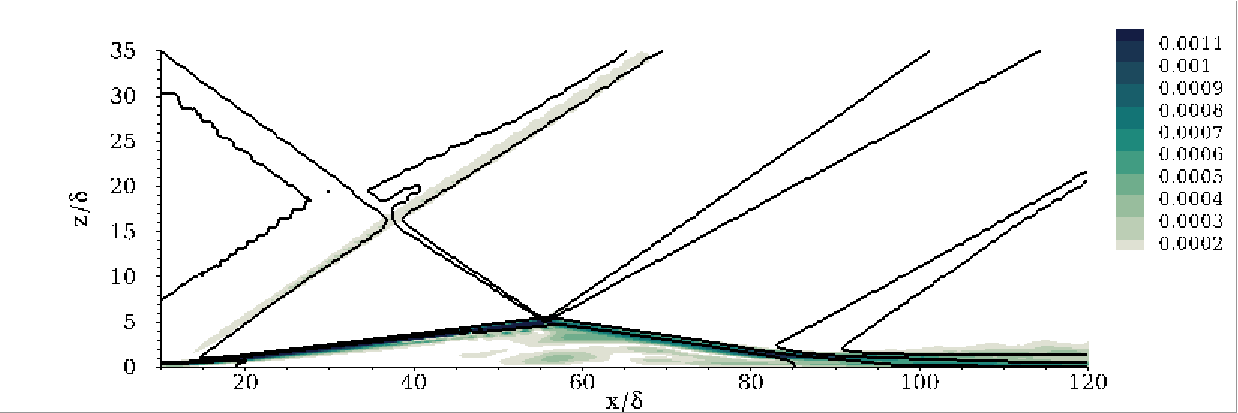}
    \caption{Longitudinal velocity $u$.}
  \end{subfigure}
  \begin{subfigure}{1.0\textwidth}
    \includegraphics[width=1.0\linewidth,trim=0 1.5 4 0,clip]{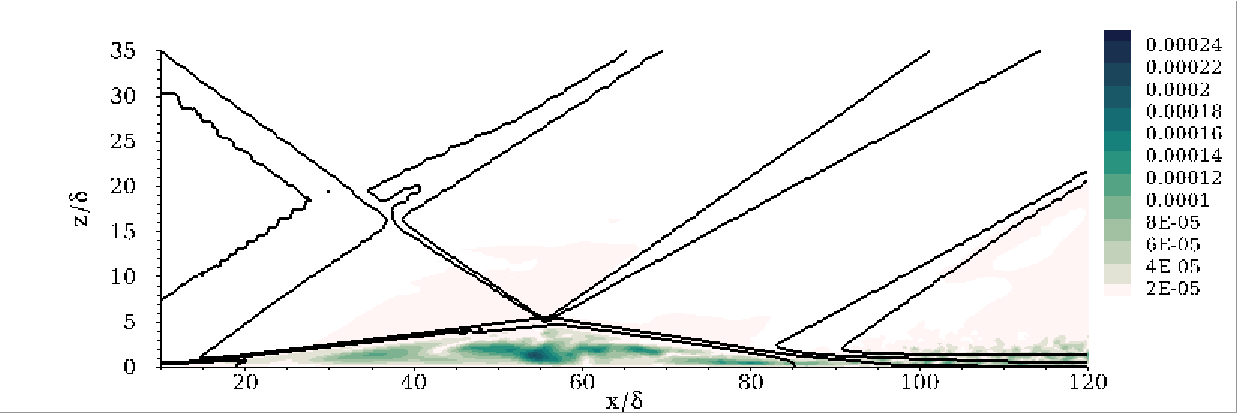}
    \caption{Spanwise velocity $v$.}
  \end{subfigure}
    \begin{subfigure}{1.0\textwidth}
    \includegraphics[width=1.0\linewidth,trim=0 1.5 4 0,clip]{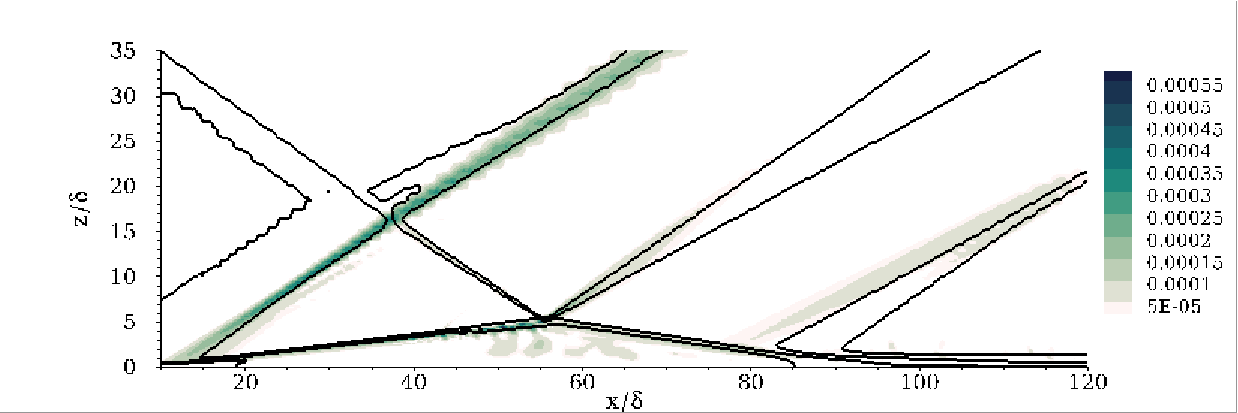}
    \caption{Vertical velocity $w$.}
  \end{subfigure}
  \caption{spanwise averaged bispectral mode for a triadic interaction (interaction (7) in figures \ref{fig:bispectre_zoom} and \ref{fig:triades_descendantes}) illustrating the sequence of interactions creating low-frequencies from medium-frequencies (interactions of type i). Frequencies involved: $St_1=0.12$; $St_2=-0.08$; $St_3=0.04$. The mean flow is indicated by isolines of the mean density field.}
  \label{fig:BispectralModes_Mf-LF}
\end{figure}

The low-frequency bispectral mode resulting from the triadic interactions of type i) evoked in figure \ref{fig:InterMaps_Mf-LF} is shown in figure \ref{fig:BispectralModes_Mf-LF} for the 3 components of velocity. For all components of velocity, we clearly see that the bispectral mode is very similar  to the first SPOD mode at the same frequency ($St=0.04$, characteristic of the breathing) shown in figure \ref{fig:SPOD_spatial_mode_0.04}. This result strongly suggests that the low-frequency breathing dynamics is fuelled by interactions between medium-frequency modes characteristic of flapping.

\subsection{Cascade of interactions creating medium-frequencies modes from low-frequencies (type ii)}
\label{subsec:cascade:LF->MF}

By analysing Figure \ref{fig:bispectre_zoom}, we identify a region of high amplitudes of the mode bispectrum above the horizontal axis (sum interactions), defined by $(St_1,St_2)\in [0,0.15]\times ]0,0.1]$. This region is symmetrical to the region of type i) interactions analysed in the previous paragraph. It has an analog effect on flow dynamics. However, as it is populated by sum interactions, it consists of a multitude of cascades of interactions generating medium-frequencies from interactions between low-frequencies instead of the opposite.

\begin{figure}
       \centering
 	   \hspace{-2cm}\includegraphics[width=15.5cm, trim=4 8 4 4,clip]{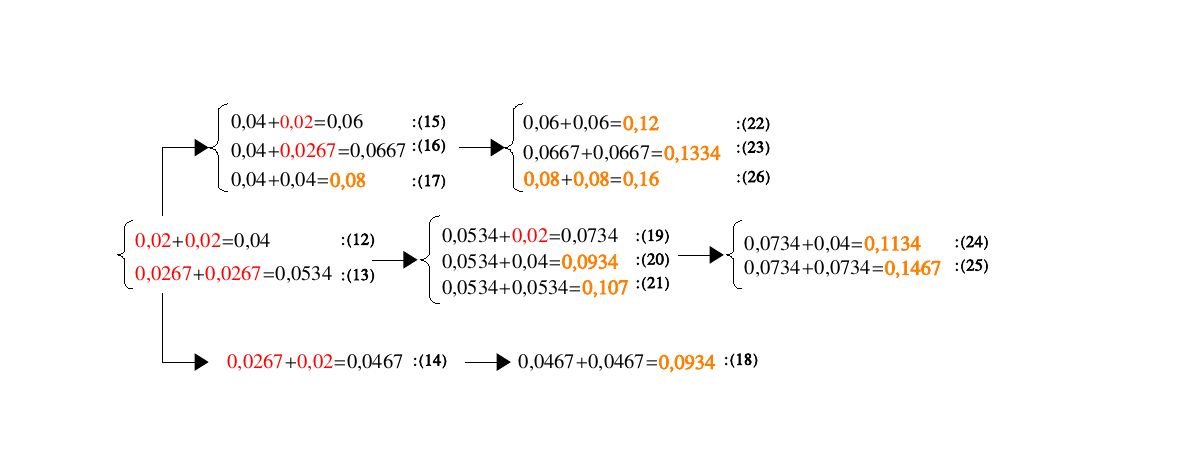}
 	   \caption{Example of a cascade of triadic interactions creating medium-frequencies from low-frequencies. The frequencies are expressed as $St_{int}$. Low-frequencies at the origin of the cascade are written in red. The resulting medium-frequencies are written in orange. Each interaction has a number between parentheses reported in figure \ref{fig:bispectre_zoom}.}
 	   \label{fig:triades_montantes}
    \end{figure}

In the same way as for interactions of type i), we show in Figure  \ref{fig:triades_montantes} a particular path for a cascade of such interactions. Each interaction is given a number (between brackets) which is reported on the mode bispectrum amplitude map (Figure \ref{fig:bispectre_zoom}). To form this cascade, we only assume the existence of two low-frequencies marked in red, i.e. here: $St=0.02$ and $St=0.0267$. The first stage of the cascade consists of the self-interactions of these frequencies and the interactions of these frequencies with each other, creating higher frequencies. We then unfold a cascade of interactions between the frequencies created and show how medium-frequencies (marked in orange) can be created in one or two stages. Again, as the region of high amplitude sum interactions is continuous, a multitude of cascade exist that has the same effect on the flow: creating medium-frequencies, characteristic of the flapping, from low-frequencies, characteristic of the breathing of the separation bubble.

\begin{figure}
  \centering
    \begin{subfigure}{1.0\textwidth}
    \includegraphics[width=1.0\linewidth,trim=0 4 4 0,clip]{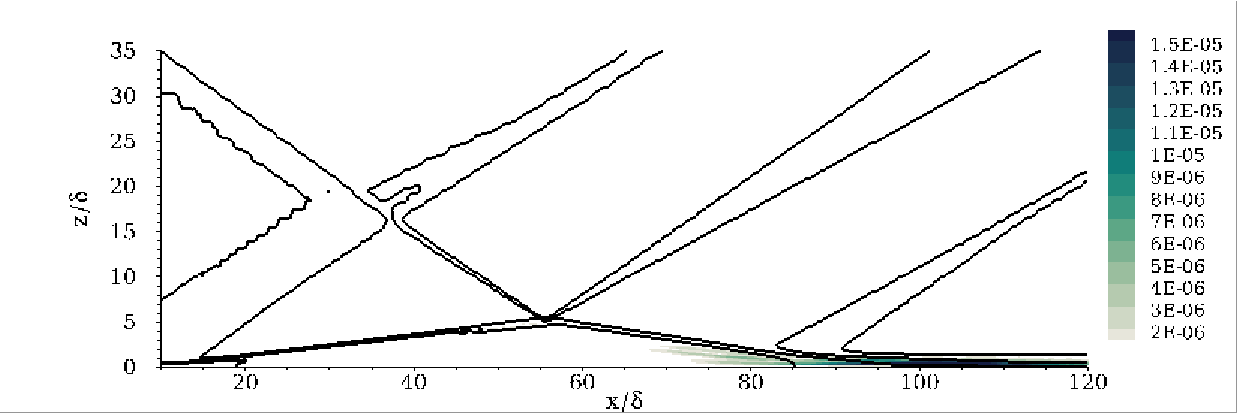}
   \caption{Longitudinal velocity $u$.}
    \end{subfigure}
    \begin{subfigure}{1.0\textwidth}
    \includegraphics[width=1.0\linewidth,trim=2 4 8 0,clip]{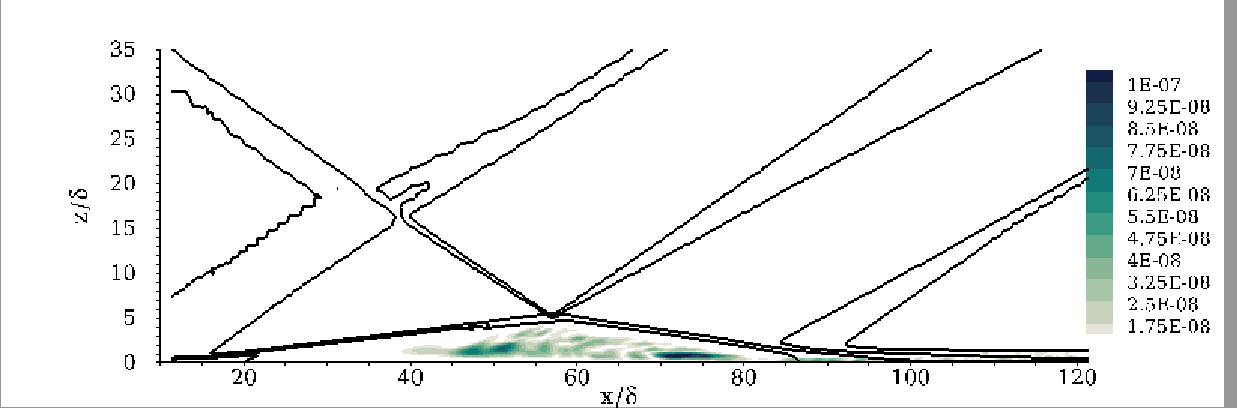}
    \caption{Spanwise velocity $v$.}
    \end{subfigure}
    \begin{subfigure}{1.0\textwidth}
    \includegraphics[width=1.0\linewidth,trim=0 4 4 0,clip]{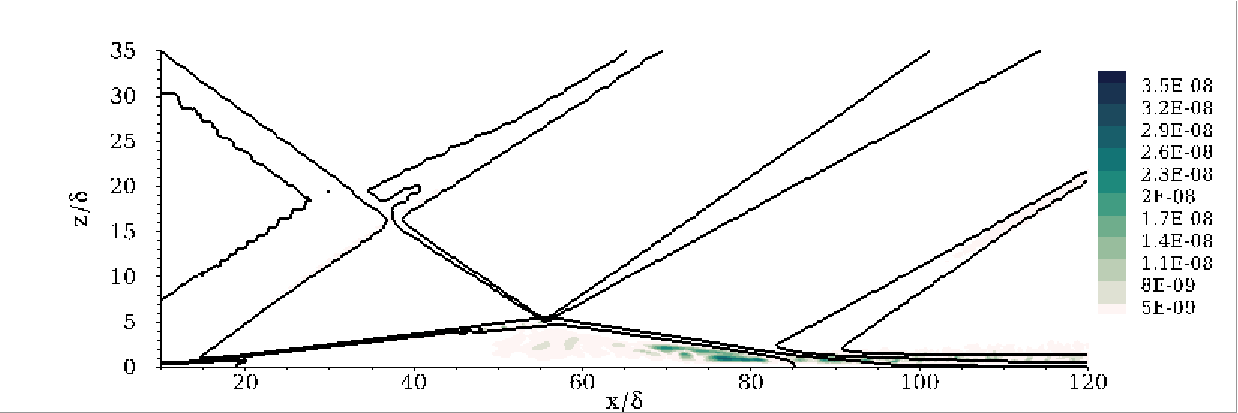}
    \caption{Vertical velocity $w$.}
    \end{subfigure}
    \caption{Spanwise averaged interaction map for a triadic interaction (interaction (18) in figures \ref{fig:bispectre_zoom} and \ref{fig:triades_montantes}) illustrating the cascade of interactions creating medium frequency modes from low frequencies (interactions of type ii). Frequencies involved: $St_1=0.04667$; $St_2=0.04667$; $St_3=0.0933$. The mean flow is indicated by isolines of the mean density field.}
  \label{fig:InterMaps_cascade_Lf-MF}
\end{figure}

The interaction map of an emblematic triadic interaction of type ii) is shown in figure  \ref{fig:InterMaps_cascade_Lf-MF} for the three components of velocity. Other interactions of type ii) are not shown here for the sake of brevity, but the exhibit the same qualitative behavior. Again, these interaction maps clearly shows that the cascade of interactions creating medium frequency modes from low frequencies mainly take place in the second part of the separation bubble and especially in the region of reattachment. This result is consistent with the SPOD analysis of the dominant dynamics at medium frequency (mode 1 at $St=0.0934$), that shows a driving role of the reattachment region's dynamics in the flapping's dynamics. The direct analysis of the DNS database also revealed a strong intensification of the medium-frequency dynamics in the second part of the separation bubble.

\begin{figure}
  \centering
  \begin{subfigure}{1.0\textwidth}
    \includegraphics[width=1.0\linewidth,trim=0 1 4 0,clip]{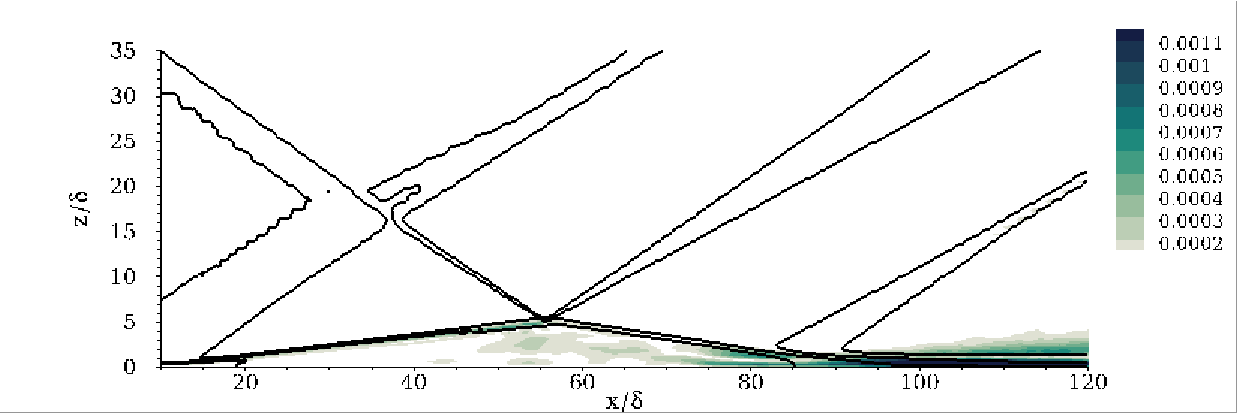}
    \caption{Longitudinal velocity $u$.}
  \end{subfigure}
  \begin{subfigure}{1.0\textwidth}
    \includegraphics[width=1.0\linewidth,trim=0 1.5 4 0,clip]{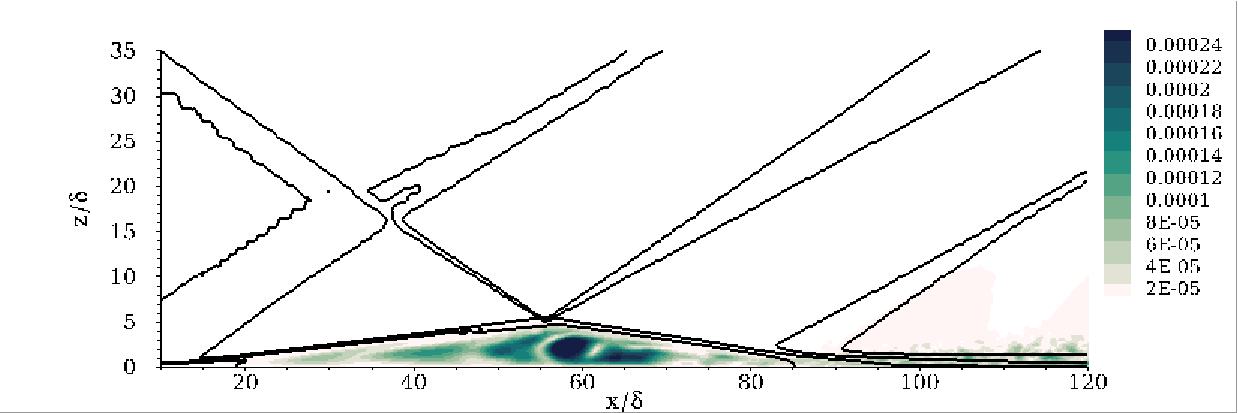}
    \caption{Spanwise velocity $v$.}
  \end{subfigure}
    \begin{subfigure}{1.0\textwidth}
    \includegraphics[width=1.0\linewidth,trim=0 1.5 4 0,clip]{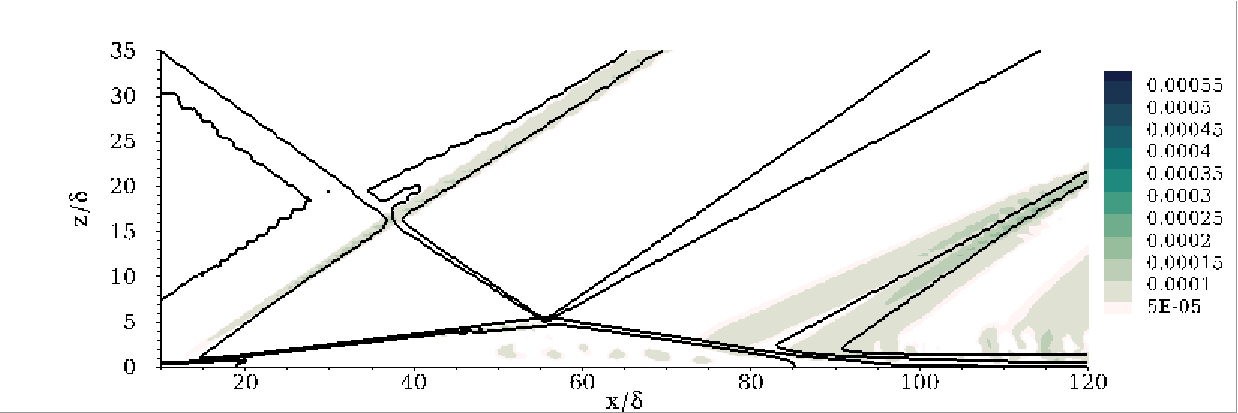}
    \caption{Vertical velocity $w$.}
  \end{subfigure}
  \caption{Spanwise averaged bispectral mode for a triadic interaction (interaction (18) in figures \ref{fig:bispectre_zoom} and \ref{fig:triades_montantes}) illustrating the cascade of interactions creating medium frequency modes from low frequencies (interactions of type ii). Frequencies involved: $St_1=0.04667$; $St_2=0.04667$; $St_3=0.0933$. The mean flow is indicated by isolines of the mean density field.}
  \label{fig:BispectralModes_cascade_Lf-MF}
\end{figure}

The medium-frequency bispectral mode resulting from the triadic interactions of type ii) evoked in figure \ref{fig:InterMaps_cascade_Lf-MF} is shown in figure \ref{fig:BispectralModes_cascade_Lf-MF} for the 3 components of velocity. The medium-frequency bispectral modes resulting from the cascade of interactions of type ii) are very similar to the medium-frequency flapping modes highlighted by the SPOD analysis and shown in appendix \ref{appA}. This result suggests a feedback loop between the medium and low frequencies. Indeed, the analysis of the triadic interactions of type i) (undertaken in paragraph \ref{subsec:MF->LF}) suggest that low-frequencies modes (breathing) are fuelled by triadic interactions between medium-frequencies characteristic of flapping. In turn the current analysis of interactions of type ii) suggest that the medium-frequency flapping dynamics is fuelled by a cascade of triadic interactions between low-frequency modes characteristic of the breathing of the separation bubble.

\subsubsection{Modes arising from the linear evolution of the flow (type iii)}
\label{subsec:linear_evolution_modes}

We clearly see a range of high amplitudes along the horizontal axis of the map of the mode bispectrum amplitude. These points, circled in Figure \ref{fig:bispectre}, correspond to interactions for which $St_1 \in [0.02,0.18]$ and $St_2=0$. Hence, $St_3=St_1$ and the corresponding bispectral modes can be interpreted as the linear evolution of the flow around its mean, according to \citet{Schmidt_Bispectral_2020}. Interestingly, the range of frequency of these modes corresponds to the range of low and medium-frequencies characteristic of the SWBLI unsteadiness, including breathing dynamics and flapping dynamics. The flow therefore shows a strong linear tendency of the flow to oscillate around it's mean field at medium and low-frequencies characterising respectively the flapping and the breathing of the bubble. In particular, we observe high intensities at the frequencies identified by the direct analysis of the DNS data and the SPOD analysis (peak frequencies of the first mode), namely $St_1=0.04$, $St_1=0.0667$ and $St_1=0.0934$. These three modes are reported on Figure \ref{fig:bispectre_zoom} as numbers $(27)$, $(28)$ and $(29)$.

\begin{figure}
  \centering
  \begin{subfigure}{1.0\textwidth}
    \includegraphics[width=1.0\linewidth,trim=0 1 4 0,clip]{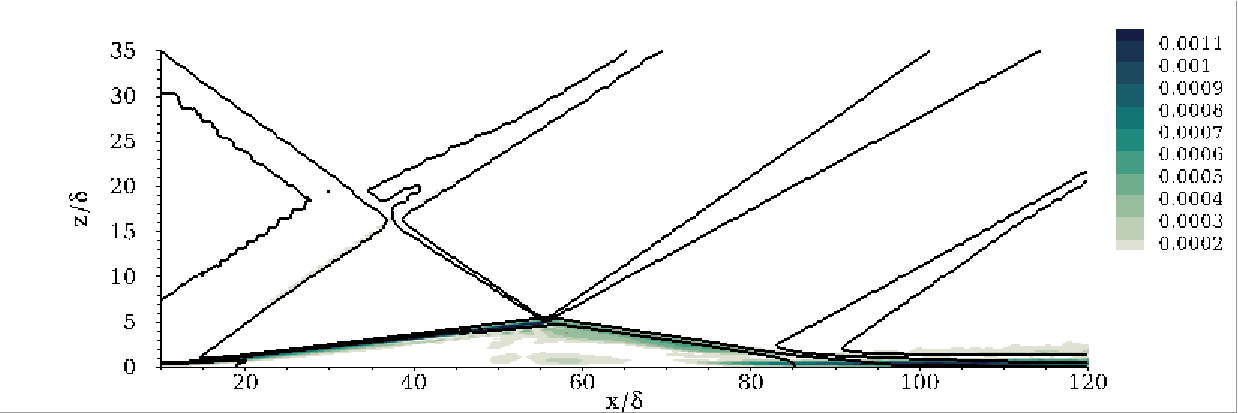}
    \caption{Longitudinal velocity $u$.}
  \end{subfigure}
  \begin{subfigure}{1.0\textwidth}
    \includegraphics[width=1.0\linewidth,trim=0 1.5 4 0,clip]{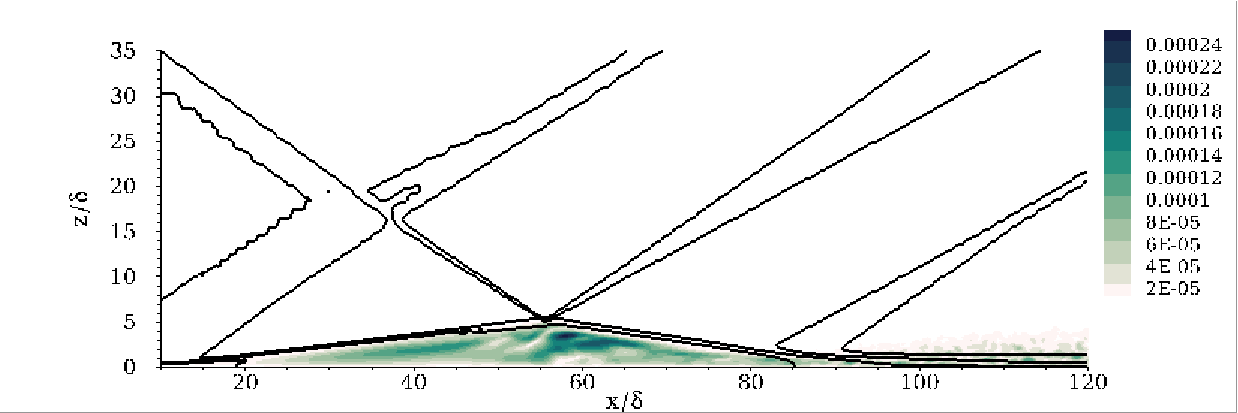}
    \caption{Spanwise velocity $v$.}
  \end{subfigure}
\begin{subfigure}{1.0\textwidth}
    \includegraphics[width=1.0\linewidth,trim=0 1.5 4 0,clip]{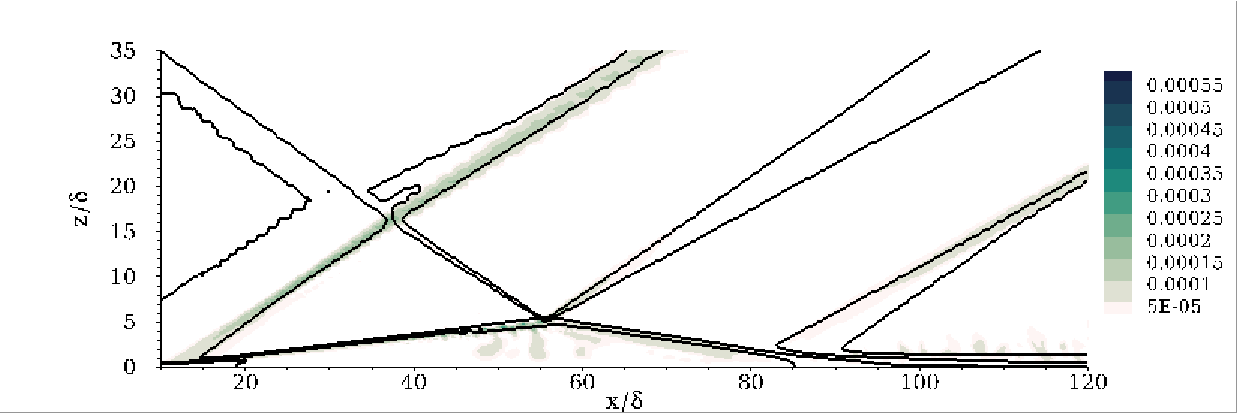}
    \caption{Vertical velocity $w$.}
  \end{subfigure}
  \caption{Spanwise averaged bispectral mode for a triadic interaction (interaction (28) in figures \ref{fig:bispectre_zoom}) illustrating the interactions expressing the linear evolution of the flow around its mean field (interactions of type iii). Frequencies involved: $St_1=0.04$; $St_2=0$; $St_3=0.04$. The mean flow is indicated by isolines of the mean density field.}
  \label{fig:BispectralModes_linear}
\end{figure}

\begin{figure}
  \centering
  \begin{subfigure}{1.0\textwidth}
    \includegraphics[width=1.0\linewidth,trim=0 1 4 0,clip]{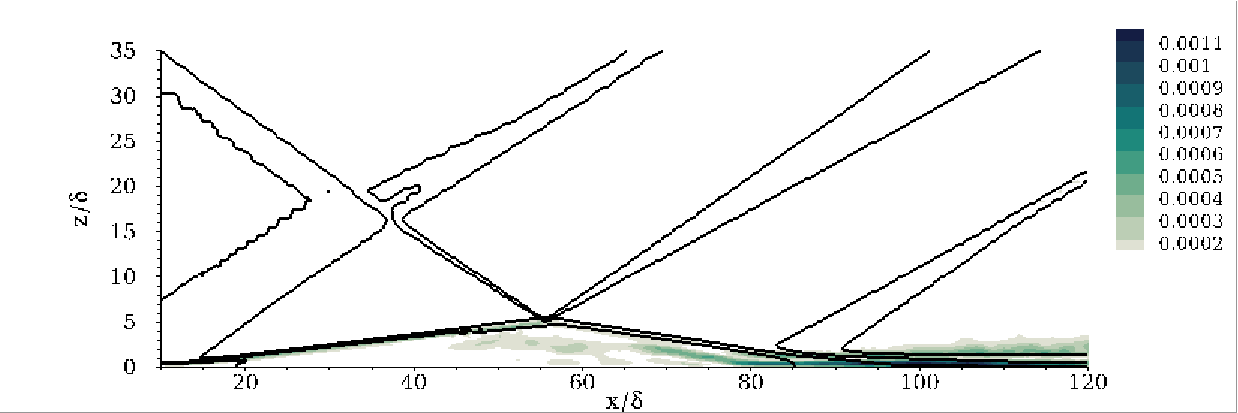}
    \caption{Longitudinal velocity $u$.}
  \end{subfigure}
  \begin{subfigure}{1.0\textwidth}
    \includegraphics[width=1.0\linewidth,trim=0 1.5 4 0,clip]{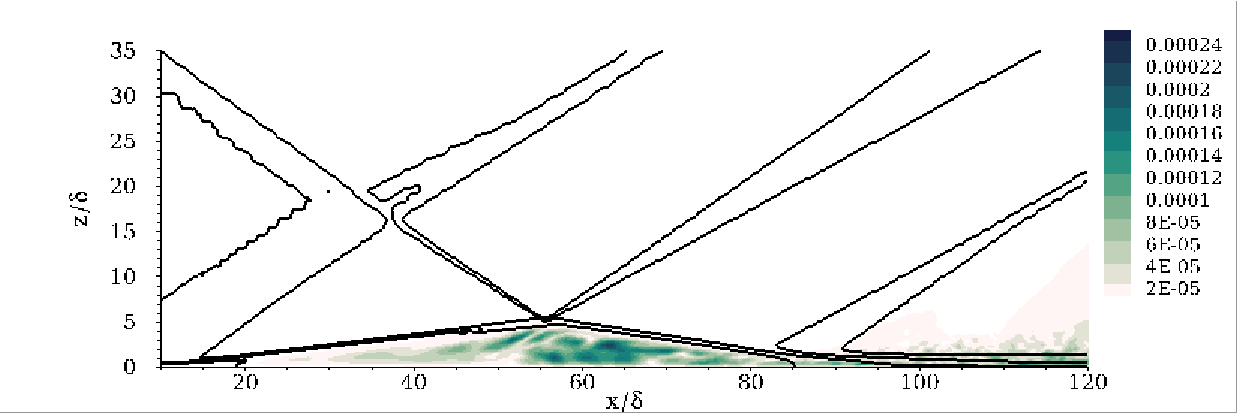}
    \caption{Spanwise velocity $v$.}
  \end{subfigure}
\begin{subfigure}{1.0\textwidth}
    \includegraphics[width=1.0\linewidth,trim=0 1.5 4 0,clip]{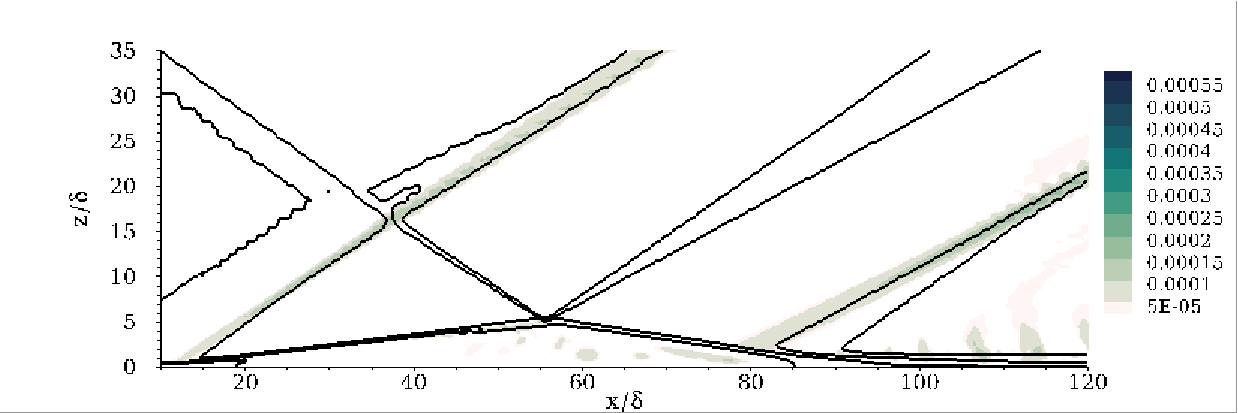}
    \caption{Vertical velocity $w$.}
  \end{subfigure}
  \caption{Spanwise averaged bispectral mode for a triadic interaction (interaction (28) in figures \ref{fig:bispectre_zoom}) illustrating the interactions expressing the linear evolution of the flow around its mean field (interactions of type iii). Frequencies involved: $St_1=0.0934$; $St_2=0$; $St_3=0.0934$. The mean flow is indicated by isolines of the mean density field.}
  \label{fig:BispectralModes_linear_0.0934}
\end{figure}

We show bellow the bispectral modes for the modes oscillating at $St_1=0.04$ and $St_1=0.0934$, respectively on figures \ref{fig:BispectralModes_linear} and \ref{fig:BispectralModes_linear_0.0934}. These modes are qualitatively very similar to the low-frequency breathing and medium-frequency flapping modes obtained by the SPOD analysis. It shows that these linear modes are contributing to the breathing and flapping modes of the separation bubble.

\begin{figure}
  \centering
    \begin{subfigure}{1.0\textwidth}
    \includegraphics[width=1.0\linewidth,trim=0 2 4 0,clip]{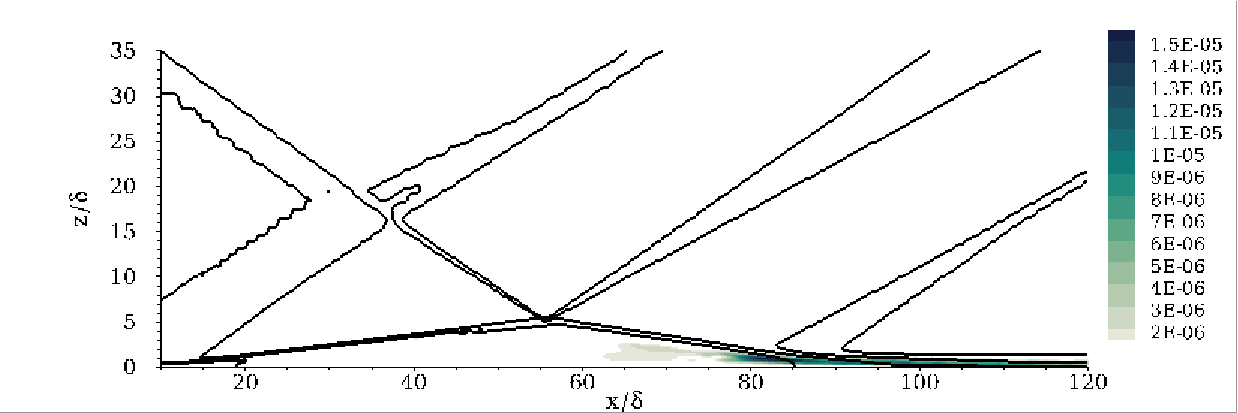}
    \caption{Longitudinal velocity $u$.}
   \end{subfigure}
    \begin{subfigure}{1.0\textwidth}
    \includegraphics[width=1.0\linewidth,trim=2 2 6 0,clip]{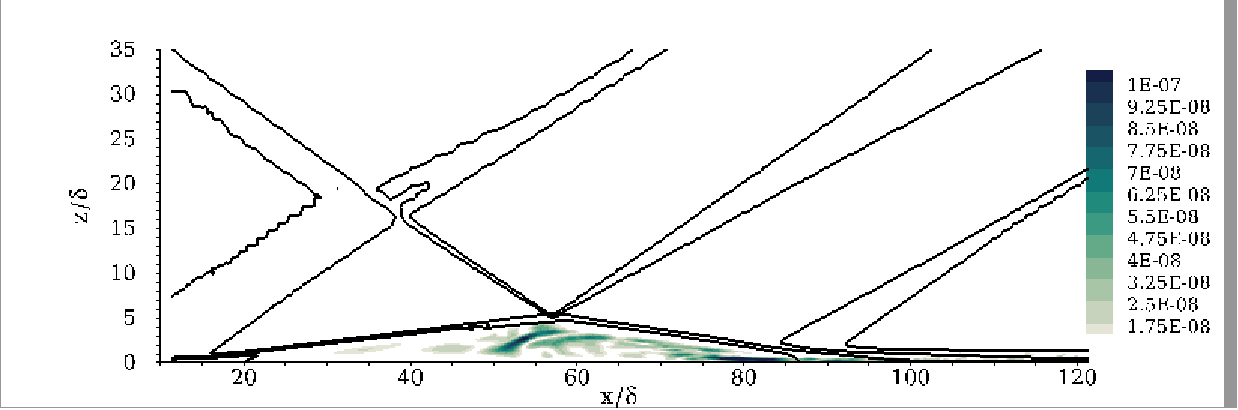}
    \caption{Longitudinal velocity $v$.}
   \end{subfigure}
       \begin{subfigure}{1.0\textwidth}
    \includegraphics[width=1.0\linewidth,trim=0 2 4 0,clip]{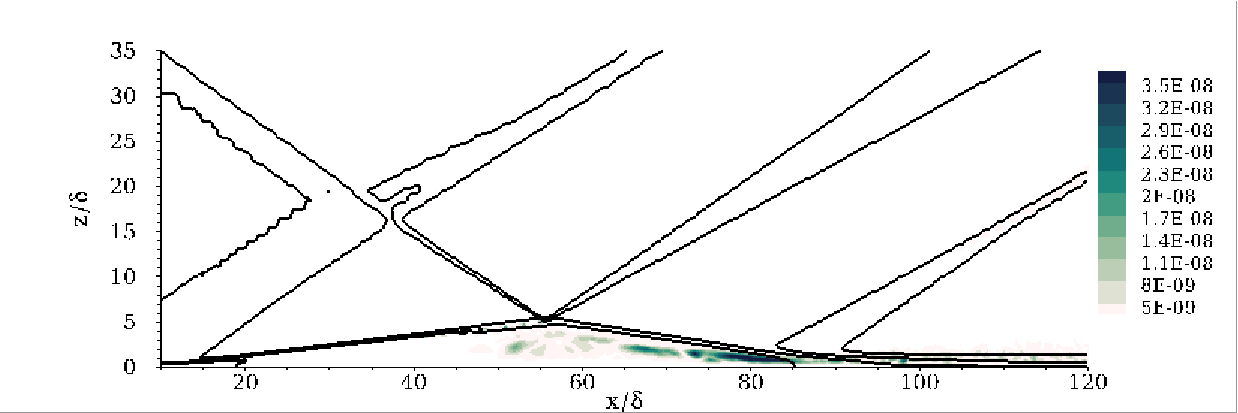}
    \caption{Longitudinal velocity $w$.}
   \end{subfigure}
  \caption{Spanwise averaged interaction map for a triadic interaction (interaction (28) in figures \ref{fig:bispectre_zoom}) illustrating the interactions expressing the linear evolution of the flow around its mean field (interactions of type iii). Frequencies involved: $St_1=0.04$; $St_2=0$; $St_3=0.04$. The mean flow is indicated by isolines of the mean density field.}
    \label{fig:InterMaps_linear}
\end{figure}

\begin{figure}
  \centering
    \begin{subfigure}{1.0\textwidth}
    \includegraphics[width=1.0\linewidth,trim=0 2 4 0,clip]{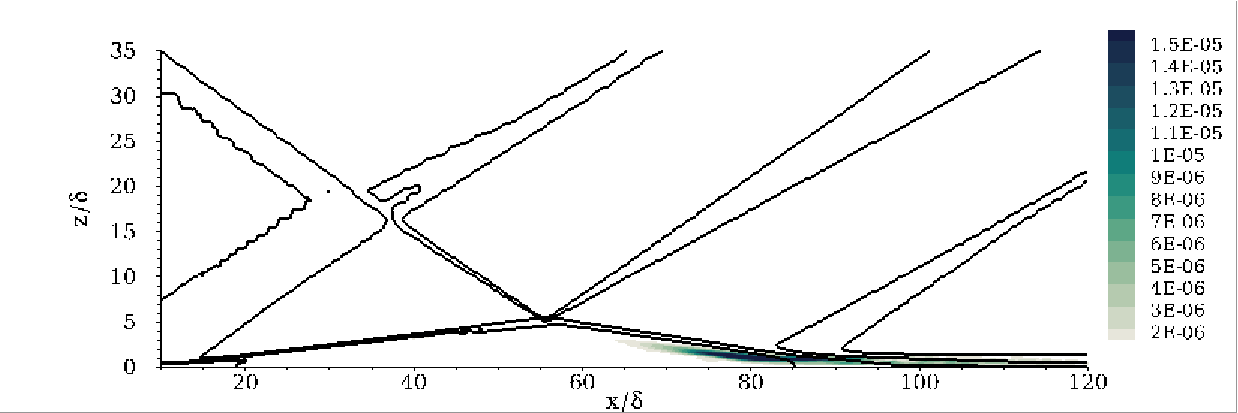}
    \caption{Longitudinal velocity $u$.}
   \end{subfigure}
    \begin{subfigure}{1.0\textwidth}
    \includegraphics[width=1.0\linewidth,trim=2 2 6 0,clip]{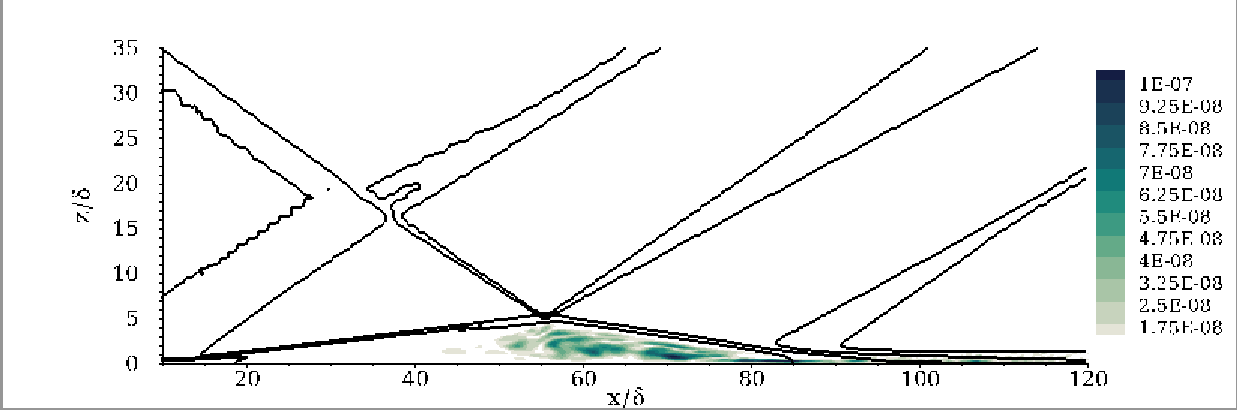}
    \caption{Longitudinal velocity $v$.}
   \end{subfigure}
       \begin{subfigure}{1.0\textwidth}
    \includegraphics[width=1.0\linewidth,trim=0 2 4 0,clip]{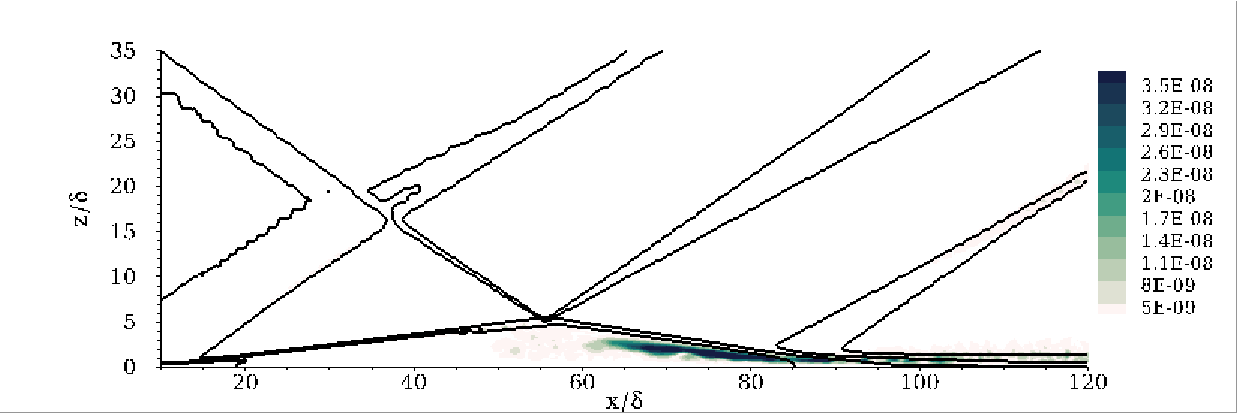}
    \caption{Longitudinal velocity $w$.}
   \end{subfigure}
  \caption{Spanwise averaged interaction map for a triadic interaction (interaction (28) in figures \ref{fig:bispectre_zoom}) illustrating the interactions expressing the linear evolution of the flow around its mean field (interactions of type iii). Frequencies involved: $St_1=0.0934$; $St_2=0$; $St_3=0.0934$. The mean flow is indicated by isolines of the mean density field.}
    \label{fig:InterMaps_linear_0.0934}
\end{figure}

The interaction maps corresponding to these modes are shown in Figures \ref{fig:InterMaps_linear} and \ref{fig:InterMaps_linear_0.0934}. The amplitude of these maps are mainly significant in the second part of the separation bubble and especially in the reattachment region. It clearly shows that these medium and low-frequency modes highlighted are driven by the dynamical activity in the downstream part of the interaction, similarly to the modes resulting from interactions of type i) and ii). 

\section{Discussion}
\label{sec:Discussion}

At this point, it is important to remember that the medium-frequency dynamics of separation bubbles is a well-documented phenomenon on which there is a consensus. The motivation of the preceding analyses was the documentation of the lower-frequency breathing-type dynamics that is associated with reflected shock oscillations in the case of SWBLIs, and whose mechanisms are still under debate. In this respect we are particularly interested in highlighting the non linear link between these two ranges of frequencies.


The analyses conducted in the preceding sections lead to a number of factual conclusions:

\begin{enumerate}
    \item the dominant dynamics of this flow are in the low and medium frequency range, with breathing and flapping modes associated with reflected shock oscillations at these frequencies,
    \item the dynamics of these modes are qualitatively similar. The associated frequencies develop in the mixing layer to reach maximum amplitudes in the downstream zone of the separation bubble close to the reattachment,
    \item in addition, for both modes, there is an upstream propagation of information inside the bubble that forces the separation zone.
    \item the downstream zone of the separation bubble, is the site of significant triadic interaction cascades providing energy to the breathing dynamics from interactions between existing frequencies of the shear layer flapping. In return, these low-frequency modes interact nonlinearly, also in the downstream part of the bubble, and form cascades of triadic interactions feeding energy into the flapping modes of the shear layer.  
\end{enumerate}

\begin{figure}
    \centering
 	\includegraphics[width=14cm,trim=4 4 4 4,clip]{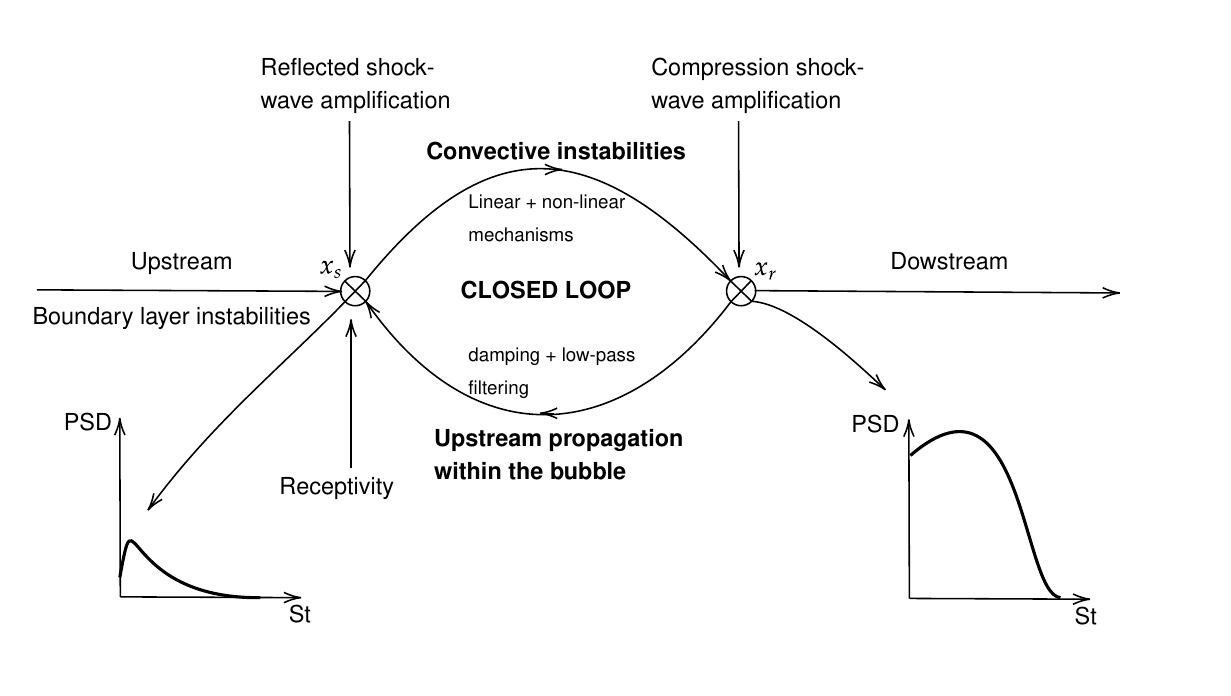}
 	\caption{Diagram explaining the suspected mechanism underlying the low-frequency unsteadiness of the transitional SWBLI.}
 	\label{fig:discussion_physique}
 \end{figure}

In the light of these elements, we can propose a mechanism underlying the SWBLI unsteadiness. This mechanism is illustrated in Figure \ref{fig:discussion_physique}.

For the description, the flow is divided in three zones: zone 1 before the interaction zone and zone 3 downstream of the interaction, zone 2 being the interaction zone. The first zone consists in the boundary layer upstream of the interaction zone that amplifies free stream perturbations. The flow then enters zone 2 by passing through the reflected shock-wave. For given Mach and Reynolds numbers of the freestream flow, the reflected shock strength does not depend on the incident shock wave strength as predicted by the free interaction theory. Through the shock, the disturbances are slightly amplified. At this stage, for sufficiently small freestream upstream disturbances, it is reasonable to assume that the disturbances are still in the linear regime. These perturbations act as a forcing of the shear layer bounding the separation bubble in the separation region. These perturbations are then amplified in the shear layer in which convective instabilities also develop. If the interaction is strong enough, the separation bubble is large enough and the shear layer is long enough so that the linear shear layer modes saturate and exhibit a non linear dynamics with cascades of triadic interactions generating the low-frequency activity. In the present case, after the reattachment, the flow enters in zone 3 (downstream of reattachment), where the vortical structures created in the interaction zone are propagated and finally lead the boundary layer to transition to turbulence, which is not the subject of the present study. Back in zone 2, there is an upstream transport of a part of the dynamical activity at medium and low-frequency from the reattachment zone until the separation zone. This upstream propagation have been shown to be damped and selective, as it damps high frequencies and select low and medium-frequencies. These upstream transported perturbations act as finite amplitude perturbations in the receptivity process of the shear layer, along with the upstream boundary layer perturbations. The shear layer is therefore forced by perturbations at low and medium-frequencies characteristic of the breathing and flapping. 

The mechanism thus described is a feedback loop whose key elements are the generation of low frequencies by the non-linear dynamics of the shear layer and the upstream propagation of these disturbances within the mixing layer.





\section{Conclusions and perspectives}
\label{sec:Conclusion}    

\subsection{Conclusions}

The aim of this paper was to improve our physical understanding of the low-frequency dynamics of SWBLIs, and more specifically the phenomenon of SWBLI unsteadiness. Pursuing this objective, we turned our attention to the case of strong interactions, creating large separation bubbles, for which the consensus is that the low-frequency dynamics of the interaction is an intrinsic feature of the recirculation bubble. The specific interest of our study was to document the possible role of triadic interactions in the mechanism behind low-frequency instationnarities.

A DNS of an strong impinging shock wave SWBLI at $M=2.15$ on a flat plate have therefore been performed and analysed. In order to avoid the possible influence of the upstream boundary layer dynamics on the interaction dynamics, we have chosen to simulate an interaction with a laminar incident boundary layer, without forcing.

First, the DNS database was analysed to determine the average flow topology, its characteristic frequencies and the nature of the vortex structures within the interaction. This analysis showed that the flow exhibit a large, quasi-symmetric mean separation bubble. Instantaneous fields analysis showed that the vortical activity is mainly concentrated in the shear layer and in the downstream part of the separation bubble. Görtler types structures have been identified inside the separation bubble downstream of its apex, whereas the downstream part of the shear layer (after the incident impingement) have been shown to be populated by longitudinal streaks. This dynamical activity of the late part of the interaction leads to transition after the reattachment point. The analysis of the frequency content across the interaction zone showed that the dominant dynamical activity of this flow is in the low and medium-frequency range characterising respectively the breathing of the separation bubble and the flapping of the shear layer. The associated frequencies develop in the mixing layer to reach maximum amplitudes in the downstream zone of the separation bubble close to the reattachment. Part of this signal is emitted from the downstream part of the separation bubble and travel upstream until eventually forcing the shear layer in the separation region. This uspstream propagation has also been shown to be damped and selective, as it selects low and medium-frequencies.   

To get more insight in the physical structure of the flow at these frequencies, a SPOD analysis have been performed. Interestingly, the peak frequencies of the dominant mode have been found to match the frequencies highlighted by the direct analysis of the DNS database. The corresponding spatial modes consist in separation bubble oscillations typical of the breathing and flapping modes accompanied by oscillations of the reflected shock. Moreover, the SPOD modes at these frequencies were found to be piloted the downstream part of the interaction and the upstream propagation of perturbations originating from this zone inside the separation has been highlighted in the space-time evolution of these dominant modes eventually reaching the separation point. 

A BSMD analysis has then been performed in order to characterise the triadic interactions at play in the interaction, especially in the downstream part of the shear layer, and to which existent such interactions could play a role in the onset of the breathing dynamics, given that the flapping dynamics of the shear layer is now well documented and is a consensus. This analysis has shown that the strong dynamical activity in the downstream part of the interaction is largely fuelled by strong triadic interactions in this zone. Indeed, this region is the seat of strong triadic interactions between frequencies characterising the flapping, contributing to the breathing activity. In turn, in the same region of the flow, triadic interactions between frequencies in the range characterising the breathing. 

All these results strongly suggest a mechanism underlying the breathing motion of the separation bubble involving the non linear saturation of the convective instabilities of the shear layer, creating energy at low and medium-frequencies in the downstream part of the separation bubble. These perturbations are then transported upstream inside the bubble and force the shear layer in the upstream part of the separation bubble. Such a feedback loop could participate in the onset of the low frequency breathing of the bubble and therefore the low-frequency unsteadiness of the SWBLI.

\subsection{Perspectives}

Another important feature of the flow, highlighted by the BSMD analysis, is the existence of linear modes oscillating around the mean flow in the low and medium frequency range characterising breathing and flapping (type iii) interactions). This suggests possible other contributions to the low-frequency unsteadiness that were not further investigated as it is out of the scope of the present paper that was focused on the role of triadic interactions. Indeed, these results indicate the possibility of the existence of unstable or slightly stable global modes around the mean flow field in the low and medium-frequency range. In the case of slightly stable modes, the perturbations arising from the non-linear dynamics of the downstream part of the bubble could then act as a finite amplitude forcing of these modes at the appropriate frequency thus sustaining their oscillations. Such a scenario is not inconsistent with the fact that globally unstable modes are not documented in the literature for stability analyses around the fixed point. Indeed, these analyses around the fixed point neglect to take into account at least part (the distortion) of the non-linear effects linked to the convective dynamics responsible for the transition. A stability analysis around the mean field, on the other hand, takes partial account of the effect of non-linearities, which play a major role in the described scenario. Such a stability study could be carried out in future work to confirm the presumptions based on the results of the present BSMD analysis. It should be noted, however, that such a neutral mode has been highlighted by a linear stability analysis around the mean flow of an axi-symmetrical compression ramp in \citet{Lugrin_2022}. Moreover, in a recent work \citep{cura2023lowfrequency}, the authors performed a global stability analysis around the mean flow of an incompressible separation bubble (with non-fixed separation point), whose results suggest that the breathing motion is driven by such a forced modal mechanism. This result lends further credence to our hypothesis that calls for a future work in which the global stability analysis of the flow around its mean field should be undertaken.


Finally, in this work, we considered an idealised flow in which the upstream boundary layer is a laminar unforced boundary layer. This case is a limit case of more realistic flows in which the incoming boundary layer is forced by environmental disturbances and has been chosen to serve as a reference. As perspectives of this work, we will perform analog analyses for interaction with forced incident laminar boundary layer, in order to characterise the effect of this forcing on the low-frequency dynamics of transitional SWBLIs. Future studies should also be undertaken involving turbulent incident boundary layers to study the fate of the mechanisms highlighted in transitional interactions when submitted to an intense turbulent forcing.

\section*{Acknowledgements}
The authors thank the “Compressible Flows” team at IUSTI Marseille and more particularly L. Larchevèque and P. Dupont for their discussions around the dynamics of compressible separation.

This work was granted access to the HPC resources of IDRIS under the allocation 2024- A0152A07195 made by GENCI.
\appendix

\section{SPOD spatial modes}\label{appA}

\subsection{Spatial mode 1 at $St=0.0667$ and its spatio-temporal evolution}
\begin{figure}
  \centering
  \begin{subfigure}{1.0\textwidth}
    \includegraphics[width=1.0\linewidth,trim=0 2 4 0,clip]{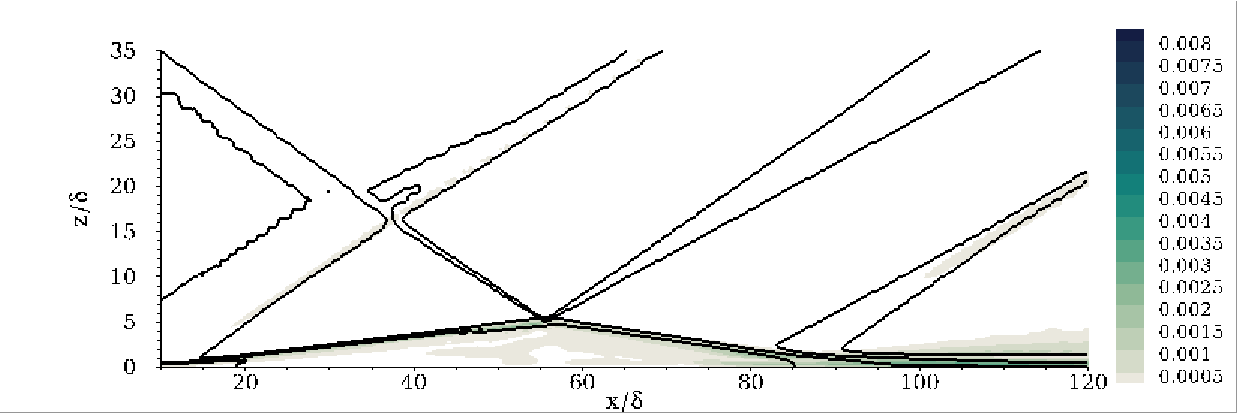}
    \caption{Longitudinal velocity $u$.}
  \end{subfigure}
  \begin{subfigure}{1.0\textwidth}
    \includegraphics[width=1.0\linewidth,trim=0 2 4 0,clip]{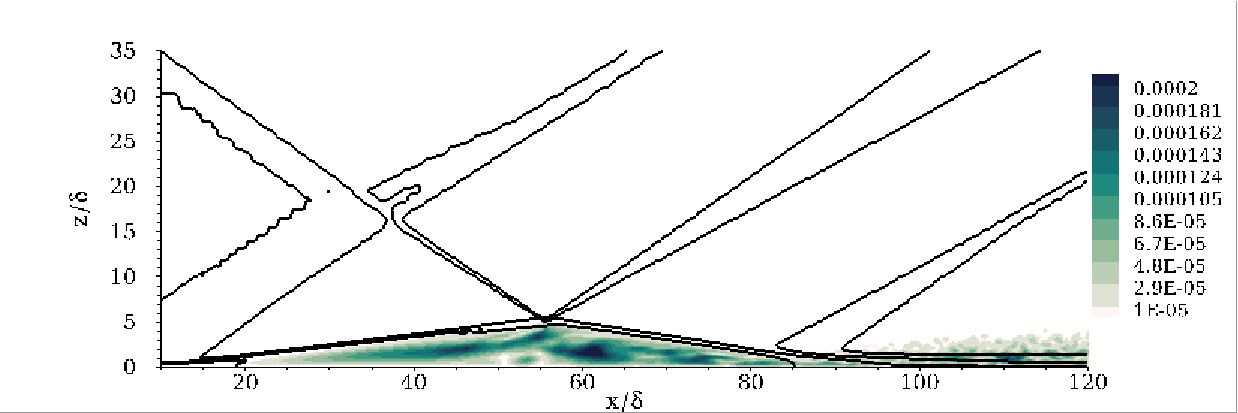}
    \caption{Spanwise velocity $v$.}
  \end{subfigure}
  \begin{subfigure}{1.0\textwidth}
    \includegraphics[width=1.0\linewidth,trim=0 2 4 0,clip]{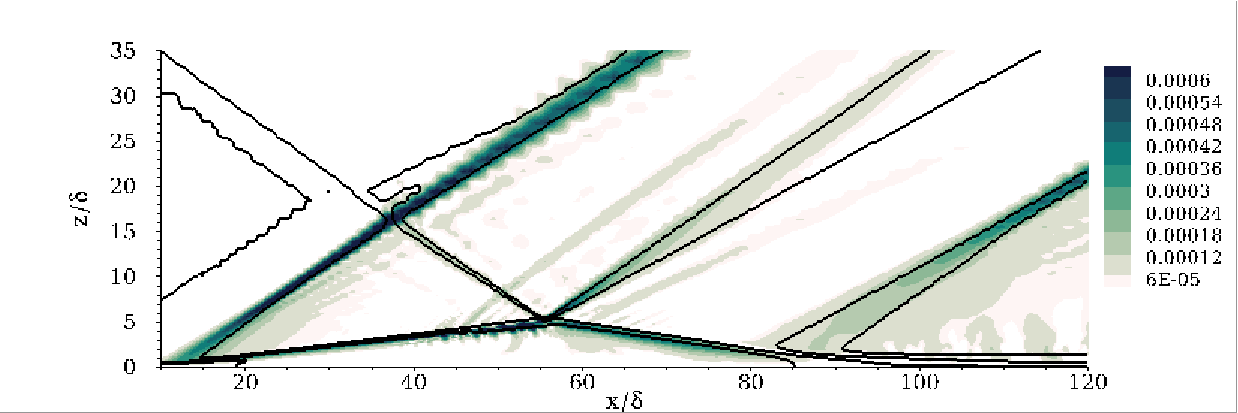}
    \caption{Vertical velocity $w$.}
  \end{subfigure}
  \caption{First spatial SPOD mode $(m_1)$ averaged in the spanwise direction at $S_t=0.0667$. The mean flow is indicated by isolines of the mean density field. }
  \label{fig:SPOD_spatial_mode_0.0667}
\end{figure}

\begin{figure}
\begin{minipage}[c][\textheight]{\textwidth}
  \begin{subfigure}{0.333\textwidth}
    \includegraphics[width=4.8\linewidth, angle=90]{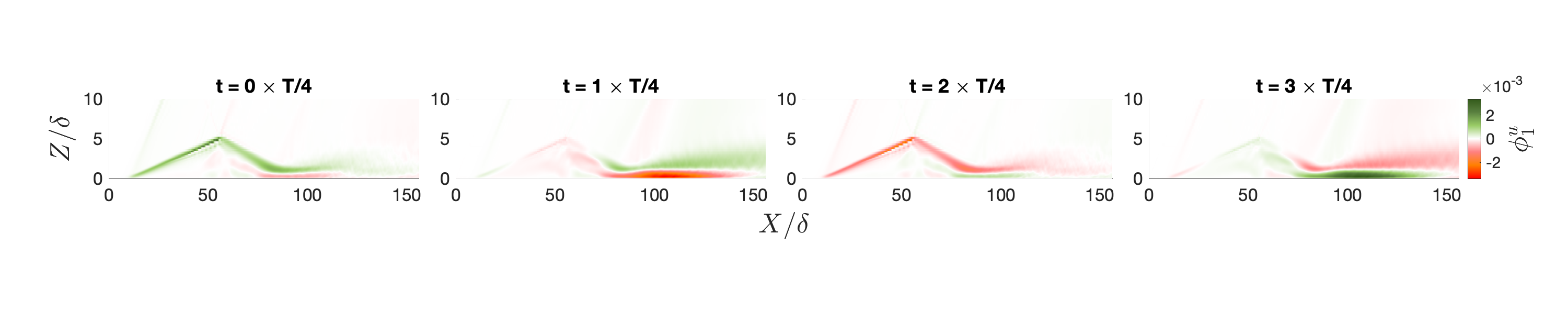}
    \caption{Velocity component $u$.}
  \end{subfigure}
  \begin{subfigure}{0.333\textwidth}
    \includegraphics[width=4.8\linewidth, angle=90]{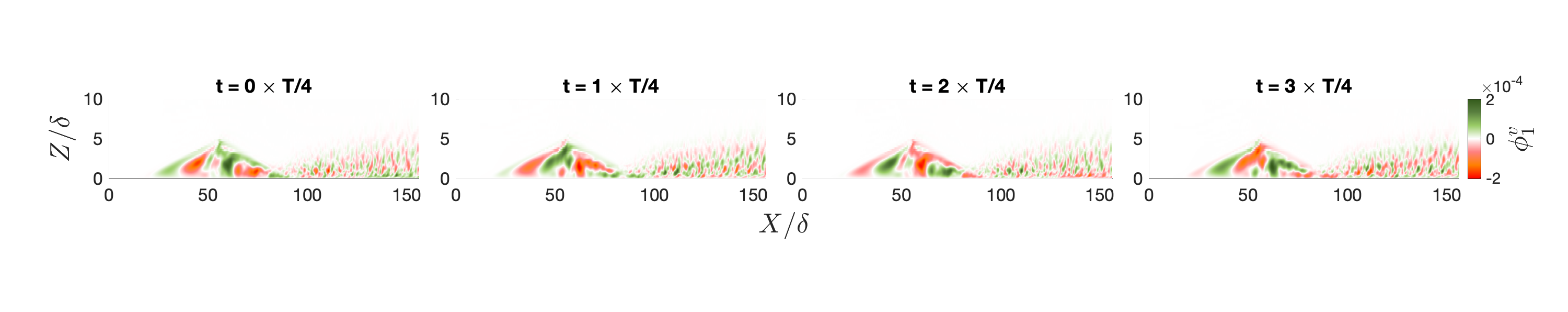}
        \caption{Velocity component $v$.}
        \label{fig:SPOD_spatio_temporal_St=0.0667_V}
  \end{subfigure}
  \begin{subfigure}{0.333\textwidth}
    \includegraphics[width=4.8\linewidth, angle=90]{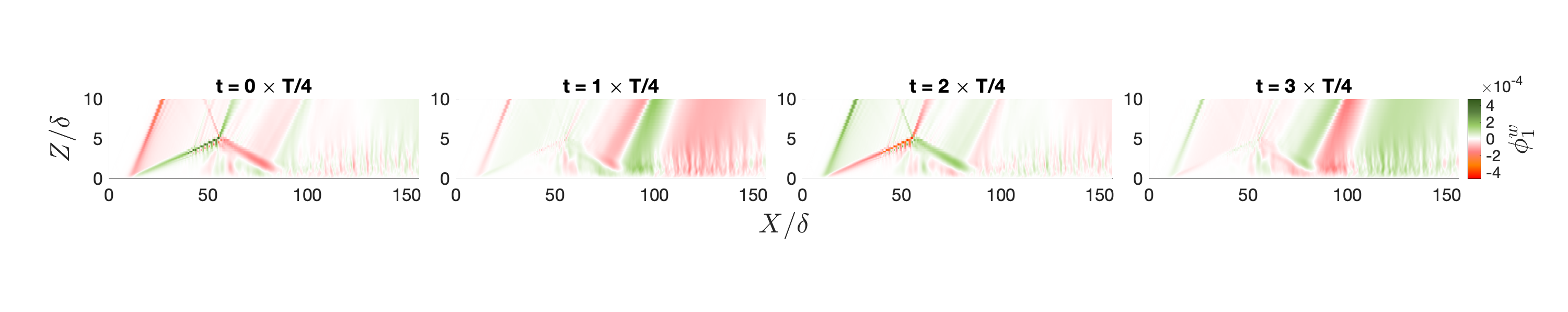}
        \caption{Velocity component $w$.}
  \end{subfigure}
 	\caption{Visualisation of $\boldsymbol{\mathcal{\Phi}}_1(\boldsymbol{x},t)$ velocity u at $St = 0.0667$, averaged in the spanwise direction.}
 	\label{fig:SPOD_spatio_temporal_St=0.0667}
\end{minipage}
\end{figure}

\begin{figure}
 	 \centering
 	  \begin{subfigure}{1.0\textwidth}
     \includegraphics[width=1.0\linewidth]{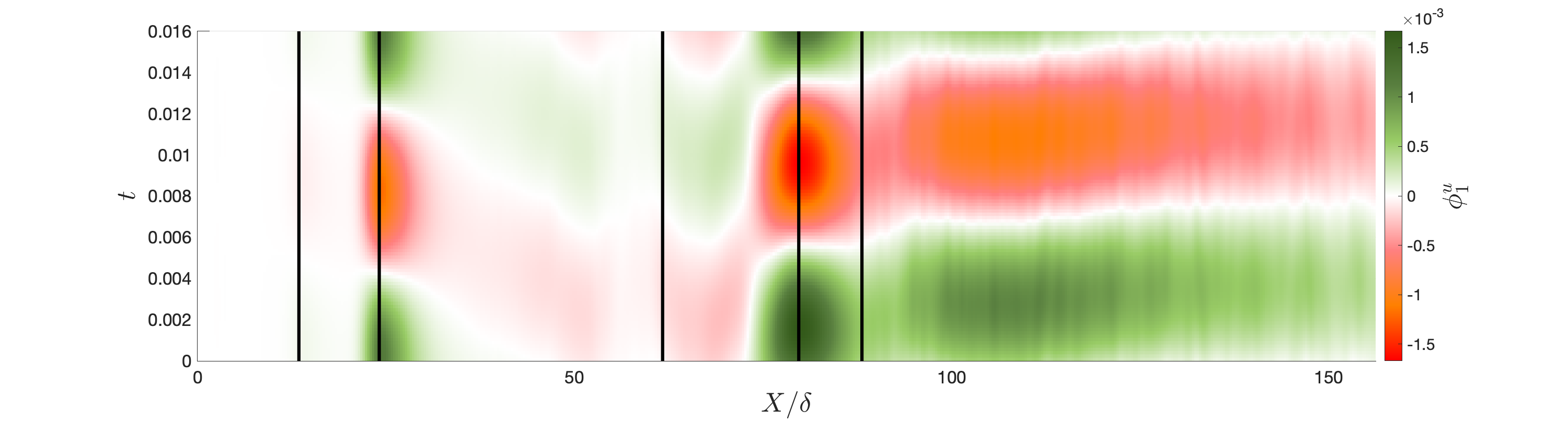}
     \caption{Velocity component $u$}
     \label{fig:SPOD_diagramme_x_t_U_0v0667}
    \end{subfigure}
    \begin{subfigure}{1.0\textwidth}
     \includegraphics[width=1.0\linewidth]{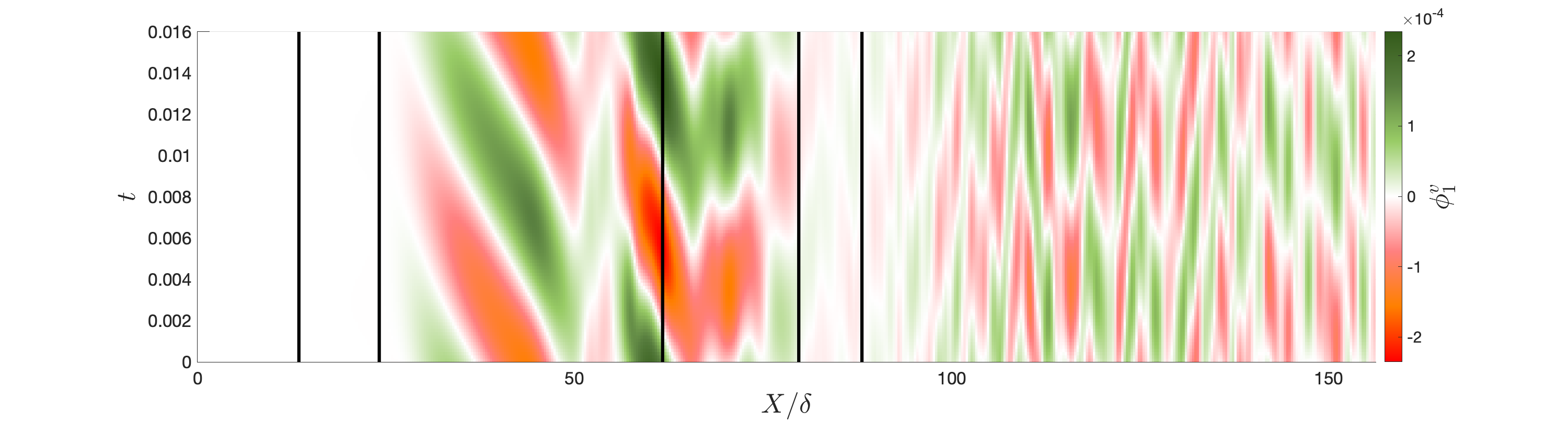}
     \caption{Velocity component $v$}
      	\label{fig:SPOD_diagramme_x_t_V_0v0667}
    \end{subfigure}
 	\caption{Representation of $\boldsymbol{\mathcal{\Phi}}_1(x,z=\text{cste},t)$ at $St=0.0667$ for $z/\delta = 1.48$. The vertical black lines indicate, from left to right: the separation point, the crossing of the rising shear layer, the crossing of the incident shock-wave, the crossing of the descending shear layer and the reattachment point.}
 	\label{fig:SPOD_diagramme_x_t_0v0667}
 \end{figure}

\subsection{Spatial mode 1 at $St=0.0934$ and its spatio-temporal evolution}

\begin{figure}
  \centering
  \begin{subfigure}{1.0\textwidth}
    \includegraphics[width=1.0\linewidth,trim=0 1 4 0,clip]{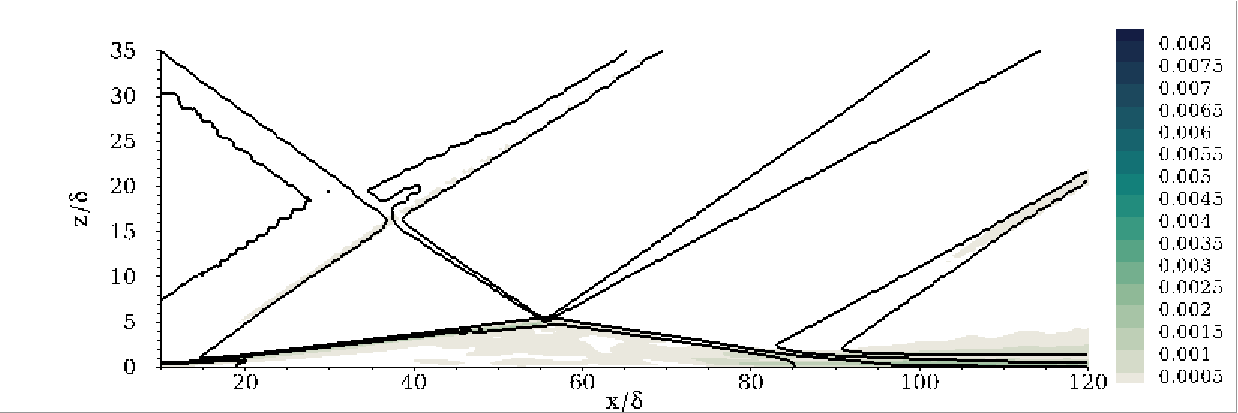}
    \caption{Longitudinal velocity $u$.}
  \end{subfigure}
  \begin{subfigure}{1.0\textwidth}
    \includegraphics[width=1.0\linewidth,trim=0 1.5 4 0,clip]{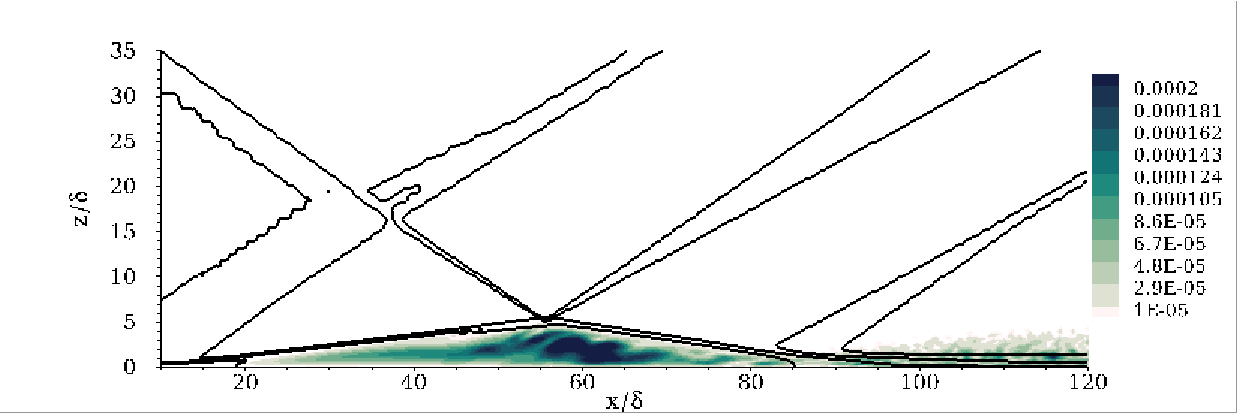}
    \caption{Spanwise velocity $v$.}
  \end{subfigure}
  \begin{subfigure}{1.0\textwidth}
    \includegraphics[width=1.0\linewidth,trim=0 1.5 4 0,clip]{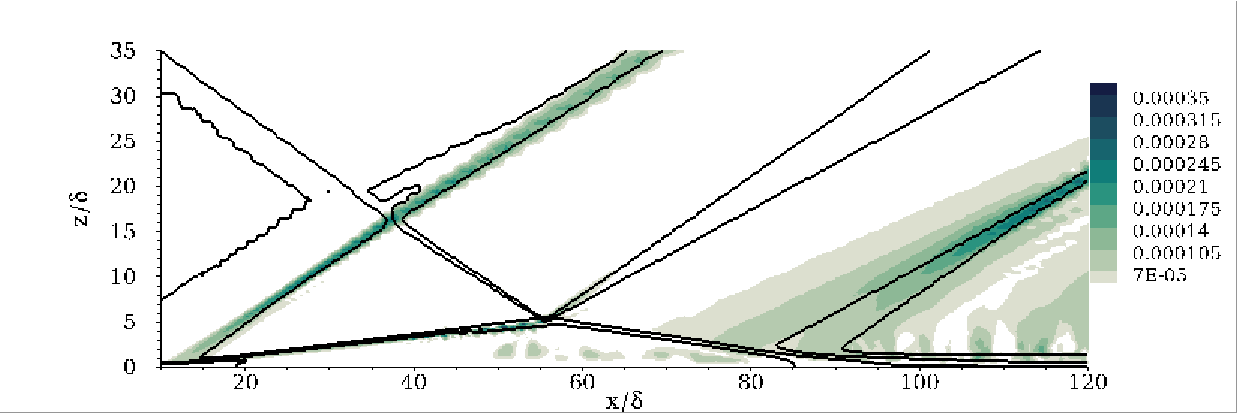}
    \caption{Vertical velocity $w$.}
  \end{subfigure}
  \caption{First spatial SPOD mode $(m_1)$ averaged in the spanwise direction at $S_t=0.0934$. The mean flow is indicated by isolines of the mean density field. }
  \label{fig:SPOD_spatial_mode_0.0934}
\end{figure}

\begin{figure}
\begin{minipage}[c][\textheight]{\textwidth}
  \begin{subfigure}{0.333\textwidth}
    \includegraphics[width=4.8\linewidth, angle=90]{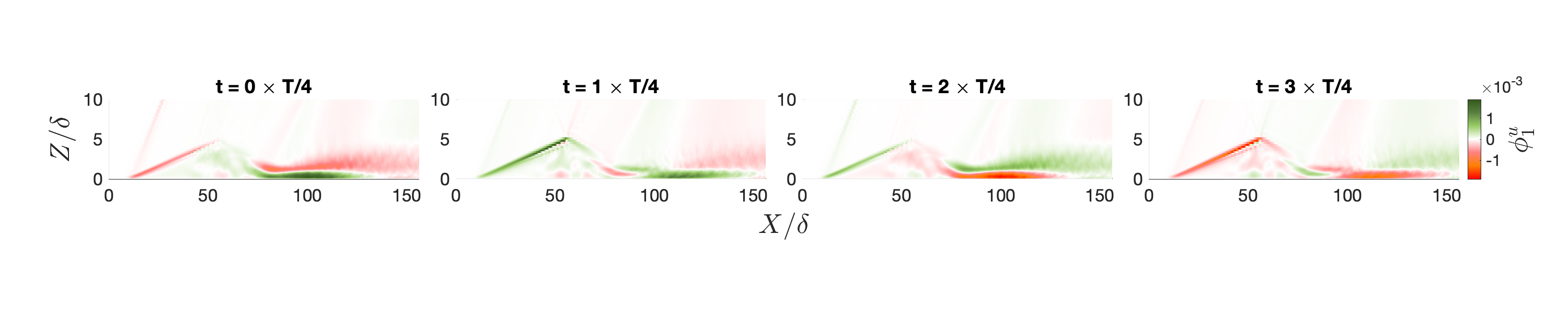}
    \caption{Velocity component $u$.}
  \end{subfigure}
  \begin{subfigure}{0.333\textwidth}
    \includegraphics[width=4.8\linewidth, angle=90]{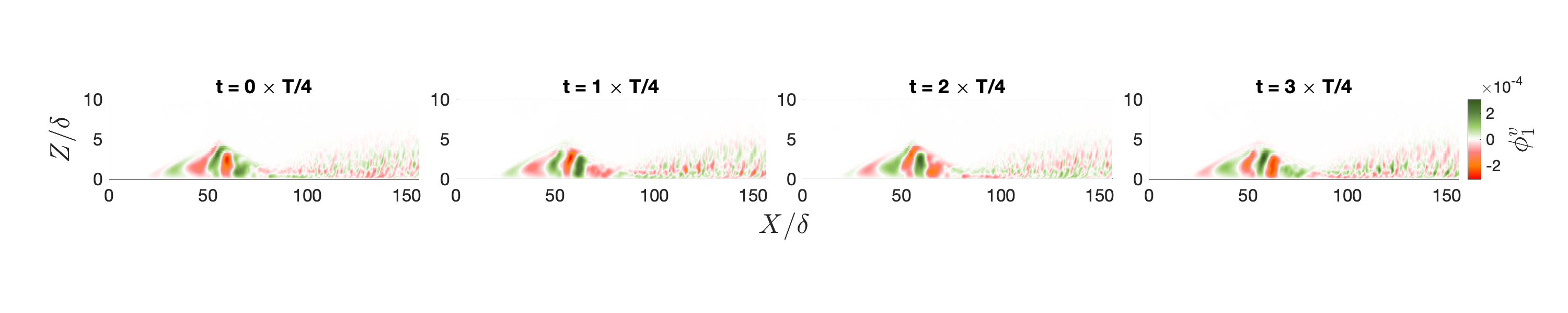}
        \caption{Velocity component $v$.}
        \label{fig:SPOD_spatio_temporal_St=0.0934_V}
  \end{subfigure}
  \begin{subfigure}{0.333\textwidth}
    \includegraphics[width=4.8\linewidth, angle=90]{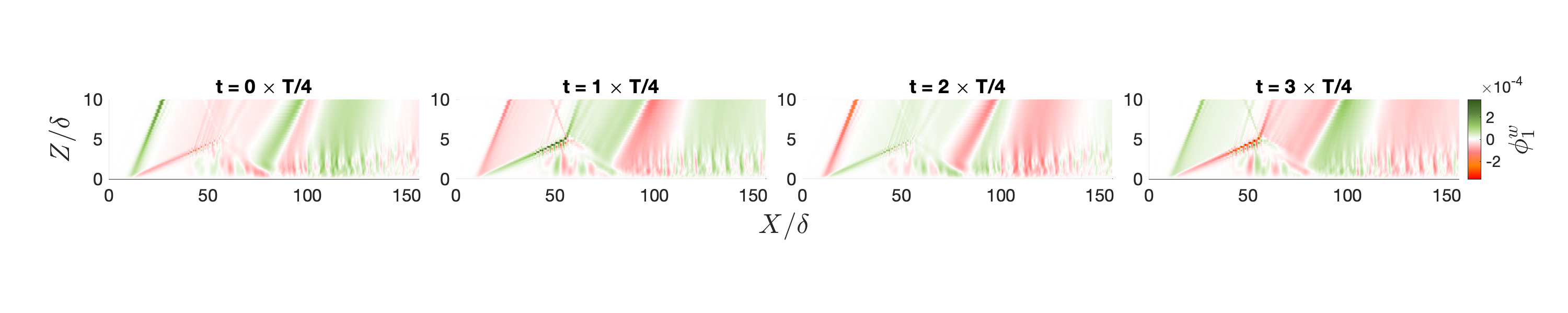}
        \caption{Velocity component $w$.}
  \end{subfigure}
 	\caption{Visualisation of $\boldsymbol{\mathcal{\Phi}}_1(\boldsymbol{x},t)$ velocity u at $St = 0.0934$, averaged in the spanwise direction. }
 	\label{fig:SPOD_spatio_temporal_St=0.0934}
\end{minipage}
\end{figure}
 
 \begin{figure}
 	 \centering
 	  \begin{subfigure}{1.0\textwidth}
     \includegraphics[width=1.0\linewidth]{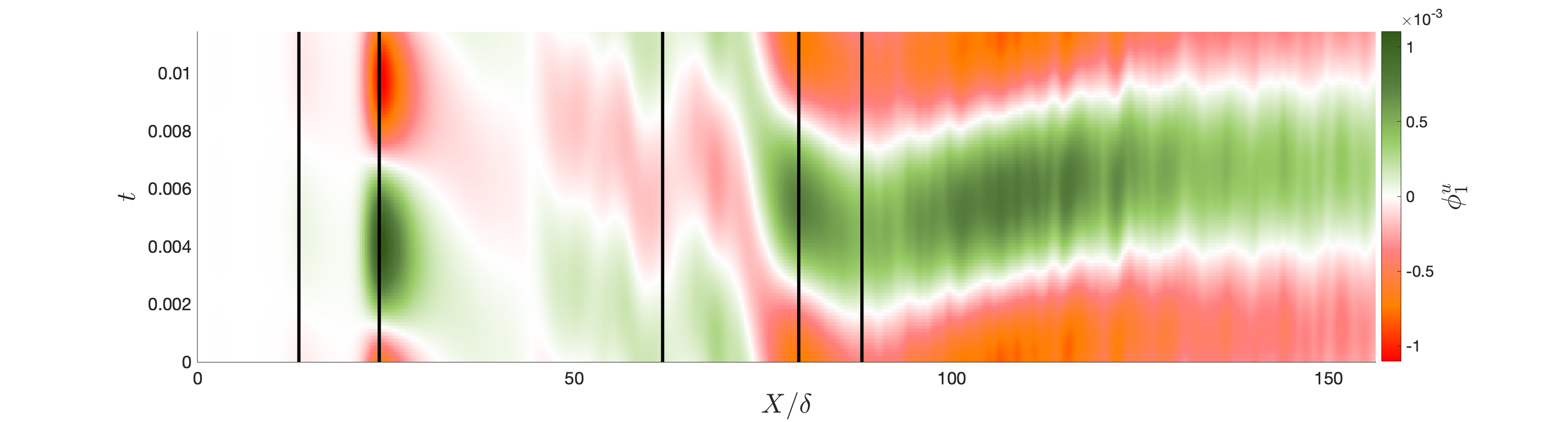}
     \caption{Velocity component $u$}
     \label{fig:SPOD_diagramme_x_t_U_0v0934}
    \end{subfigure}
    \begin{subfigure}{1.0\textwidth}
     \includegraphics[width=1.0\linewidth]{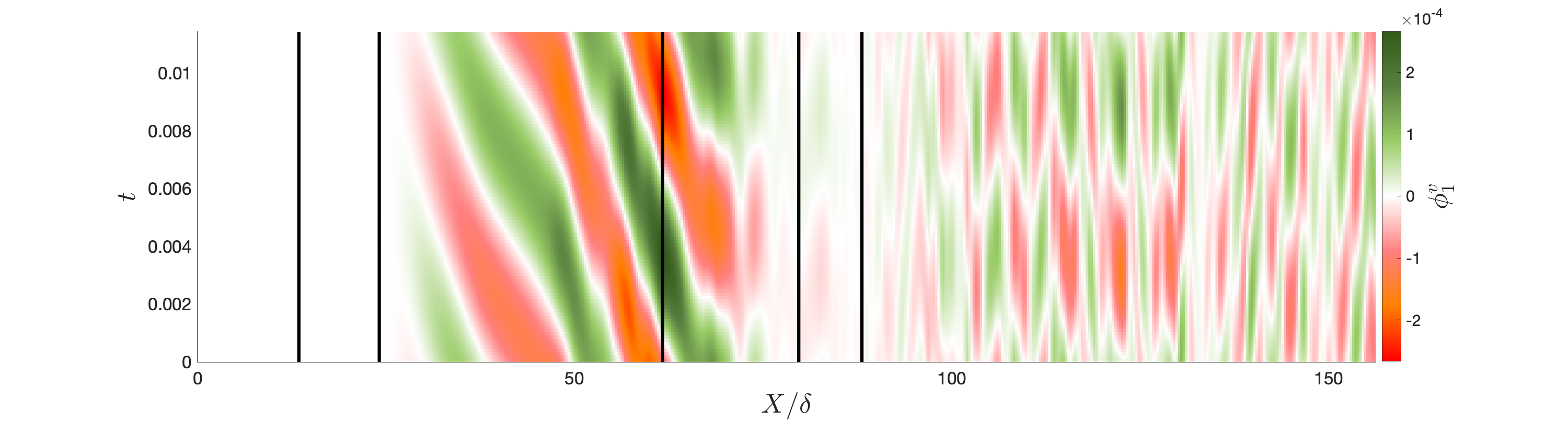}
     \caption{Velocity component $v$}
      	\label{fig:SPOD_diagramme_x_t_V_0v0934}
    \end{subfigure}
 	\caption{Representation of $\boldsymbol{\mathcal{\Phi}}_1(x,z=\text{cste},t)$ at $St=0.0934$ for $z/\delta = 1.48$. The vertical black lines indicate, from left to right: the separation point, the crossing of the rising shear layer, the crossing of the incident shock-wave, the crossing of the descending shear layer and the reattachment point.}
 	\label{fig:SPOD_diagramme_x_t_0v0934}
 \end{figure}



\bibliographystyle{jfm}
\bibliography{jfm}

\end{document}